\documentclass[10pt]{article}
\usepackage{epsf}
\usepackage{color}
\usepackage{amsmath,amsfonts,amsbsy,amssymb}
\usepackage{indentfirst}
\usepackage{mathtools}
\usepackage{arydshln}

\newcommand{\fsl}[1]{\ensuremath{\mathrlap{\not{\phantom{#1}}}#1}}

\newcommand{\nn}{\nonumber}

\def\be{\begin{equation}}
\def\ee{\end{equation}}
\def\bse{\begin{subequations}}
\def\ese{\end{subequations}}
\def\bal{\begin{align}}
\def\ealn{\end{align}}
\def\tr{\text{tr}}
\def\bs{\boldsymbol}

\begin{document}

\begin{titlepage}

\def\slash#1{{\rlap{$#1$} \thinspace/}}

\begin{flushright} 

\end{flushright} 

\vspace{0.1cm}

\begin{Large}
%\vspace{1cm}
\begin{center}

%%%%%%%%%%%%%%%%%%%%%%%%%%%%%%%%%%%%%%%%%%%%%%%%%%%%%%%%%%%%%%%%%%%%%
{\bf   
  $SO(4)$ Landau Models and Matrix Geometry}
\end{center}
\end{Large}
%%%%%%%%%%%%%%%%%%%%%%%%%%%%%%%%%%%%%%%%%%%%%%%%%%%%%%%%%%%%%%%%%%%%%

\vspace{1cm}

%%%%%%%%%%%%%%%%%%%%%%%%%%%%%%%%%%%%%%%%%%%%%%%%%%%%%%%%%%%%%%%%%%%%%%%%%%%%%%%%%
%%%%%%%%%%%%%%%%%%%%%%%%%%%%%%%%%%%%%%%%%%%%%%%%%%%%%%%%%%%%%%%%%%%%%%%%%%%%%%%%%
\begin{center}
{\bf Kazuki Hasebe}   \\ 
%%%%%%%%%%%%%%%%%%%%%%%%%%%%%%%%%%%%%%%%%%%%%%%%%%%%%%%%%%%%%%%%%%%%%%%%%%%%%%%%%
%%%%%%%%%%%%%%%%%%%%%%%%%%%%%%%%%%%%%%%%%%%%%%%%%%%%%%%%%%%%%%%%%%%%%%%%%%%%%%%%%
%%%%%%%%%%%%%%%%%%%%%%%%%%%%%%%%%%%%%%%%%%%%%%%%%%%%%%%%%%%%%%%%%%%%%%%%%%%%%%%%%
%%%%%%%%%%%%%%%%%%%%%%%%%%%%%%%%%%%%%%%%%%%%%%%%%%%%%%%%%%%%%%%%%%%%%%%%%%%%%%%%%
\vspace{0.5cm} 
\it{
National Institute of Technology, Sendai College,  
Ayashi, Sendai, 989-3128, Japan} \\ 
%%%%%%%%%%%%%%%%%%%%%%%%%%%%%%%%%%%%%%%%%%%%%%%%%%%%%%%%%%%%%%%%%%%%%%%%%%%%%%%%%
%%%%%%%%%%%%%%%%%%%%%%%%%%%%%%%%%%%%%%%%%%%%%%%%%%%%%%%%%%%%%%%%%%%%%%%%%%%%%%%%%

\vspace{0.5cm} 

{\sf
khasebe@sendai-nct.ac.jp} 

\vspace{0.8cm} 

{\today} 

\end{center}

\vspace{1.0cm}

%%%%%%%%%%%%%%%%%%%%%%%%%%%%%%%%%%%%%%%%%%%%%%%%%%%%%%%%%%%%%%%%%%%%%%%%%%%%%%%%%
%%%%%%%%%%%%%%%%%%%%%%%%%%%%%%%%%%%%%%%%%%%%%%%%%%%%%%%%%%%%%%%%%%%%%%%%%%%%%%%%%
\begin{abstract}
\noindent

\baselineskip=18pt

We develop an in-depth analysis of the $SO(4)$ Landau models on $S^3$ in the  $SU(2)$ monopole background and their associated matrix geometry. The Schwinger and Dirac gauges for the $SU(2)$ monopole are introduced to provide a concrete coordinate representation of $SO(4)$ operators and wavefunctions. The gauge fixing enables us to demonstrate algebraic relations of the operators and  the $SO(4)$ covariance of the eigenfunctions.  With the spin connection of  $S^3$, we construct an $SO(4)$ invariant Weyl-Landau operator  and analyze its eigenvalue problem with explicit form of the eigenstates. The  obtained results include the known formulae of the free Weyl operator eigenstates in the free field limit.  
Other eigenvalue problems of variant relativistic Landau models, such as massive Dirac-Landau and  supersymmetric Landau models, are investigated too. 
With the developed $SO(4)$ technologies, we  derive the three-dimensional matrix geometry in  the Landau models. By applying the level projection method to the Landau models, we identify the  matrix elements of the $S^3$ coordinates as the fuzzy three-sphere.  For the non-relativistic model, it is shown that the  fuzzy three-sphere geometry emerges in each of the Landau levels and only in the degenerate lowest energy sub-bands.   We also point out that Dirac-Landau operator accommodates two fuzzy three-spheres in each Landau level and the mass term  induces interaction  between them.

\end{abstract}
%%%%%%%%%%%%%%%%%%%%%%%%%%%%%%%%%%%%%%%%%%%%%%%%%%%%%%%%%%%%%%%%%%%%%%%%%%%%%%%%%
%%%%%%%%%%%%%%%%%%%%%%%%%%%%%%%%%%%%%%%%%%%%%%%%%%%%%%%%%%%%%%%%%%%%%%%%%%%%%%%%%

\end{titlepage}

\newpage 

\tableofcontents

\newpage 

%%%%%%%%%%%%%%%%%%%%%%%%%%%%%%%%%%%%%%%%%%%%%%%%%%%%%%%%%%%%%%%
\section{Introduction}
%%%%%%%%%%%%%%%%%%%%%%%%%%%%%%%%%%%%%%%%%%%%%%%%%%%%%%%%%%%%%%%

The Landau models are  physical models that manifest the non-commutative geometry in a most obvious way.  It is well known  \cite{Hasebe-2015,Hatusda-Iso-Umetsu-2003} that the fuzzy two-sphere geometry \cite{berezin1975,Hoppe1982,madore1992} is realized in the $SO(3)$ Landau model \cite{Dirac-1931,Wu-Yang-1976}  that provides a set-up of the  2D quantum Hall effect \cite{Haldane-1983}. Similarly  
the set-up of the $SO(5)$ Landau model  \cite{Yang-1978-I,Yang-1978-II} is used for  the construction of the 4D quantum Hall effect \cite{Zhang-Hu-2001} whose underlying geometry is the fuzzy four-sphere \cite{Grosse-Klimcik-Presnajder-1996,Castelino-Lee-Taylor-1997, Kimura2002}. The correspondence was further explored on $S^{2k}$ \cite{Hasebe-2014-1, Hasebe-Kimura-2003} and  the  $SO(2k+1)$ Landau model was shown to realize the geometry of fuzzy $2k$-sphere \cite{Ho-Ramgoolam-2002,Kimura2003}. Besides spheres, there are many manifolds that incorporate  non-commutative geometry, and  Landau models have been constructed on various manifolds, $i.e.$ $\mathbb{C}P^n$, supermanifolds, hyperboloids, etc. \cite{Karabali-Nair-2002, Bernevig-Hu-Toumbas-Zhang-2003,Hasebe-2005, Jellal-2005,  Hasebe-2008, Daoud-Jellal-2008,   Hasebe-2010, Balli-Behtash-Kurkcuoglu-Unal-2014, Karabali-Nair-2016, Lapa-Jian-Ye-Hughes-2016,  Heckman-Tizzano-2017}.    The works have brought  deeper understanding of the Landau  physics and the associated fuzzy geometry as well.   
%All of such Landau models share common properties that the gauge groups of the monopoles are compatible with the holonomies of the base-manifolds (Table \ref{table:correspDNYM}). 
The magnetic field is the vital for the realization of the non-commutative geometry in the Landau model, and  for spheres, the magnetic field is brought by the monopole at the center of the spheres. Since the monopole charge mathematically corresponds  to the Chern number that is defined on even dimensional manifold,  
%, and the monopole charge quantization stems from the topological quantization of the Chern number. 
all of the manifolds used in the above works are even dimensional.  
% Therefore, it is quite reasonable to consider  Landau models on even dimensional manifolds. 
Also in the viewpoint of the non-commutative geometry, adoption of the  even dimensional manifolds is quite reasonable, because  the geometric quantization is performed by replacing the Poisson bracket with the commutator, and   even dimensional symplectic manifold generally accommodates non-commutative structure by such a quantization procedure.

%%%%%%%%%%%%%%%%%%%%%%%%%%%%%%%%%%%%%%%%%%%%%%%%%%%%%%%%%%%%%%%%%%%%%%%%%%%%%%%%
\begin{table}
\hspace{-0.8cm}
%\begin{center}
   \begin{tabular}{|c|c|c|c|}\hline
    /      &   2D  &  3D &  4D \\ \hline
Base-manifold           &  $S^{2}$      & $S^{3}$        & $S^4$        \\ \hline 
Holonomy               &  $SO(2)\simeq U(1)$    &  $SO(3)\simeq SU(2)/Z_2$   &   $SO(4)\simeq SU(2)\otimes SU(2)/Z_2$   \\ \hline  
Monopole gauge group              &  $U(1)$    &  $SU(2)$   &   $SU(2)$   \\ \hline  
Landau model              &  $SO(3)$ Landau model     &  $SO(4)$ Landau model   &   $SO(5)$ Landau model   \\ \hline
Quantum Hall effect              &  2D QHE     &  3D QHE   &   4D QHE   \\ \hline
    \end{tabular}       
\caption{Landau models on low dimensional spheres and associated monopoles}
\label{table:correspDNYM}
\end{table}
%%%%%%%%%%%%%%%%%%%%%%%%%%%%%%%%%%%%%%%%%%%%%%%%%%%%%%%%%%%%%%%%%%%%%%%%%%%%%%%%%

From  above point of view, the Landau model on $S^3$  which Nair and Randjbar-Daemi first proposed \cite{Nair-Daemi-2004} \footnote{We refer to the model  as the $SO(4)$  Landau model in this paper, since the model respects the $SO(4)$ global symmetry.} was rather exotic, though the model  nicely fits in between the $SO(3)$ and $SO(5)$ Landau models (see Table \ref{table:correspDNYM}).  In the model,  the quantization of the $SU(2)$ monopole charge was assumed, but there is no reason  to justify the assumption: The Chern number is not defined in odd dimensions, and so the monopole charge quantization is not guaranteed.    
Also for odd dimensional manifolds, the symplectic structure cannot be embedded and then the geometric quantization procedure mentioned above is useless. Even if we adopt the quantum Nambu three-bracket instead of the usual commutator \cite{Nambu1973}, we encounter other problems, such as the violation of the Jacobi identity \cite{CurtrightZachos2003}. It thus seemed to exist fundamental difficulties for  Landau models and non-commutative geometry in odd dimensional space. % and embed non-commutative structure to odd dimensional space. 
In Refs.\cite{JabbariTorabian2005,Jabbari2004, Ramgoolam2002} however, it was pointed out that the usage of the odd dimensional  bracket can be circumvented by treating the odd dimensional bracket as a sub-bracket of the one-dimension higher even bracket, which indicates that the odd dimensional non-commutative space is not apparently consistent by itself but   consistent as a subspace of  one-dimension higher even dimensional space.  Inspired by this observation, we proposed a resolution for the difficulty of  odd dimensional  Landau model.   We showed that the $SO(4)$ Landau model is naturally embedded in the $SO(5)$ Landau model, and  the monopole charge quantization  is accounted for by that on one-dimension higher space $S^4$ \cite{Hasebe-2014-2}. 
%the lowest Landau level degeneracies of $SO(4)$ Landau model is exactly equal to  that of the $SO(5)$ Landau model,  and  
%In the previous papers \cite{Hasebe-2014-2, Hasebe-2017},  we proposed a resolution to the problem.  
 We also demonstrated that similar relation holds for arbitrary odd and even dimensional Landau models \cite{Hasebe-2017} and the dimensional relation has its origin in differential topology;  the dimensional ladder of anomaly or the spectral flow of Atiyah-Patodi-Singer.    
%%%%%%%%%%%%%%%%%%%%%
%\be
%S^4_F ~\rightarrow~S^3_F ~\rightarrow~S^2_F. 
%\ee
%%%%%%%%%%%%%%%%%%%
Though the foundation of the odd dimensional Landau models was thus established, there are merely  a handful of  works about them up to the present \cite{Nair-Daemi-2004,Hasebe-2014-2, Hasebe-2017,Coskun-Kurkcuoglu-Toga-2017}. (See also \cite{LiWu2013,Li-I-Y-W-2012,LiZhangWu2012}  for odd dimensional topological insulator  Landau models based on the Dirac oscillator.)

%In the previous works \cite{Hasebe-2017, Hasebe-2014-2}, we mainly focused on the dimensional relationship of the Landau models.   
% and have not derived explicit forms of the eigenstates, especially for the relativistic case.   
In this paper, we revisit the  $SO(4)$ Landau model -- the minimal model of the odd dimensional Landau models.    Through a full investigation of the $SO(4)$ Landau model, we learn  properties  specific to the odd dimensional Landau model  and associated non-commutative geometry whose analyses are technically difficult in higher dimensions.  
%Furthermore,  since the spacial dimension of the $SO(4)$ Landau model is three, the model is related to Weyl/Dirac  semimetal, chiral topological insular, and  hopefully  experiments. 
%to derive the Landau level eigenstates and matrix geometries both for non-relativistic and relativistic cases.  
%give the explicit expression of the $SO(4)$ Landau level eigenstates  by fixing a gauge as either Schwinger gauge or Dirac gauge.   We also provide explicit coordinate representation of operators  to demonstrate algebraic relations that has not been clarified. 
%In particular, we give a precise meaning of the lowest Landau level basis states constructed in \cite{Hasebe-2014-2} and establish  relations to the results of  Nair and Randjbar-Daemi.  
The main achievements  are as follows: 
$(i)$ We introduce the Schwinger gauge and the Dirac gauge for $S^3$ and solve the Landau problem with  explicit form of the wavefunction and the operators.   The gauge fixing enables us to  demonstrate   important algebraic relations, such as the $SO(4)$ invariance of the Dirac-Landau operator and covariance of the $SO(4)$ Landau level eigenstates.    
%$(ii)$ We analyze the complete Dirac-Landau operator with the spin connection,\footnote{In the previous analysis \cite{Nair-Daemi-2004}, they used the Dirac-Landau operator without the spin connection which is not reduced to the free the free Dirac operator in the free limit.} and the spin connection part plays an important role in the analysis. We demonstrate the results are indeed reduced to the known free Dirac operator eigenvalues and eigenstates in the free limit.  
$(ii)$ We analyze relativistic Landau operators on $S^3$ with  spin connection.\footnote{The previous work \cite{Nair-Daemi-2004} adopted  the Dirac-Landau operator $\it{without}$ the spin connection, and so the role of the spin connection has not been taken into account.  We then investigate the eigenvalue problem of the  Dirac-Landau operator with the spin connection from the beginning.}   Besides the eigenvalues, we derive a concrete coordinate representation of the eigenstates. 
It is shown that the obtained results indeed include  the known formulae of the (free) relativistic operator  \cite{CamporesiHiguchi1996,Tranutman1995,Tranutman1993}  in the free background limit.   %Typical relativistic Landau operators, $i.e.$ Weyl, Dirac, supersymmetric, are investigated. 
$(iii)$ The matrix elements of the arbitrary Landau levels of the $SO(4)$ Landau models are  derived explicitly. We  demonstrate that the obtained matrix geometry is identical to that of the fuzzy three-sphere. This is the first  derivation of the odd dimensional matrix geometry in the context of  the Landau model.  
%We also clarify the matrix geometry of the relativistic Landau models, and 
Especially, we point out that the mass parameter of the Dirac-Landau model induces interaction between  two fuzzy three-spheres realized in each of the relativistic Landau levels.

%It is also confirmed that the known results  of the $SO(4)$ spherical harmonics   %\cite{Biedenharn1961,Domokos-1967,Hochstadt-book} and the free Dirac operator eigenstates on $S^3$  %\cite{CamporesiHiguchi1996,Tranutman1995,Tranutman1993} are reproduced in the free background limit.

 %We thus adopt the level projection method  to derive    fuzzy geometries from various non-relativistic and relativistic Landau levels.  
 
% It should also be added that the level projection method has the advantage that the obtained non-commutative %geometry automatically becomes  a mathematically consistent formulation, since the whole Hilbert space  of the %Landau model  is  well-defined, quantum mechanically \cite{Hasebe-2015}. 

This paper is organized as follows. 
In Sec.\ref{sec:three-sphereint}, we introduce the Schwinger and Dirac gauges  for geometric quantities of three-sphere. Sec.\ref{sec:nonrelaLandaumodel} discusses  the non-relativistic Landau model in the Dirac gauge. We analyze the eigenvalue problem of the spinor Landau model with synthesized connection in Sec.\ref{sec:spinorlandau}. Subsequently,  the eigenvalue problem of  relativistic Landau models is solved for Weyl-type, Dirac-type  and supersymmetric-type   in Sec.\ref{sec:relativisLandau}. The  matrix geometries of the Landau models are identified as fuzzy three-sphere in Sec.\ref{sec:matrixgeo}. Sec.\ref{sec:summary} is devoted to summary and discussions.

%%%%%%%%%%%%%%%%%%%%%%%%%%%%%%%%%%%%%%%%%%%%%%%
\section{Geometric Quantities of Three-sphere}\label{sec:three-sphereint}
%%%%%%%%%%%%%%%%%%%%%%%%%%%%%%%%%%%%%%%%%%%%%%%%%

Here, we summarize basic geometric quantities of $S^3$ based on the exterior derivative method \cite{Eguchi-Gilkey-Hanson-1980,Boulware-1975}. (For the component method, see Appendix \ref{append:threespheregeo}.)   

We first  parameterize the  coordinates on $S^3$ as 
%%%%%%%%%%%%%%%%%%%%%%%%%%%%%
\be
x_1=\sin\chi\sin\theta\cos\phi, ~~~x_2= \sin\chi\sin\theta\sin\phi, ~~~x_3= \sin\chi\cos\theta, ~~~x_4= \cos\chi, 
\label{fs3coordinatesspherical}
\ee
%%%%%%%%%%%%%%%%%%%%%%%%%%%%%
with the ranges 
%%%%%%%%%%%%%%%%%%
\be
0\le \chi \le \pi,~~0\le \theta\le  \pi,~~~0\le \phi < 2\pi. 
\ee
%%%%%%%%%%%%%%%%%%%
The world-line on $S^3$ is given by 
%%%%%%%%%%%%%%%%%%%%%%%%%%%%%%%%%%
\be
ds^2={dx_1}^2+{dx_2}^2+{dx_3}^2+{dx_4}^2=d\chi^2+\sin^2 \chi~d\theta^2+\sin^2\chi\sin^2\theta ~d\phi^2, 
\label{metricthreesphere}
\ee
%%%%%%%%%%%%%%%%%%%%%%%%%%%%%%%%%%%%
and the area  of $S^3$ is   
%%%%%%%%%%%%%%%%%%%%
\be
A(S^3)=\int_{S^3}d\Omega_3=\int_0^{2\pi}d\phi \int_0^{\pi}d\theta\sin\theta\int_0^{\pi}
d\chi\sin^2\chi=2\pi^2. 
\ee
%%%%%%%%%%%%%%%%%%
From the formula 
%%%%%%%%%%%%%%%%%%%%%%%
\be
ds^2=\delta_{ab}~e^a ~e^b, 
\label{ssquaredelta}
\ee
%%%%%%%%%%%%%%%%%%%%%%%%%
we can read off the dreibein as 
%%%%%%%%%%%%%%%%%%%%
\be
e^1_{\text{S}}=d\chi, ~~e^2_{\text{S}}=\sin\chi ~d\theta, ~~e^3_{\text{S}}=\sin\chi\sin\theta ~d\phi, 
\label{dreibeinschwing}
\ee
%%%%%%%%%%%%%%%%%%%%%
where 
%%%%%%%%%%%%%%%%%%%%%
%\be
$a, b=1,2, 3$ 
%\ee
%%%%%%%%%%%%%%%%%%%%%%
 denote locally flat space coordinate indices.  
Since $\delta_{ab}$ (\ref{ssquaredelta}) is the $SO(3)$ invariant tensor, the choice of the dreibein is not unique  due to the degrees of freedom of $SO(3)$ rotation. The choice (\ref{dreibeinschwing}) is the simplest one, and   we  refer to it as the Schwinger gauge.  Another useful  choice is the Dirac gauge 
%%%%%%%%%%%%%%%%%%%%%
\begin{align}
&e_{\text{D}}^1=\cos\theta d\chi-\sin\chi\sin\theta d\theta,~~~~e^2_{\text{D}}=\sin\theta \cos\phi d\chi+\sin\chi\cos\theta \cos \phi d\theta-\sin\chi\sin\theta \sin \phi d\phi, \nn\\
&e^3_{\text{D}}=\sin\theta \sin\phi d\chi+\sin\chi\cos\theta \sin \phi d\theta+\sin\chi\sin\theta \cos \phi d\phi, 
\label{diracgaugedreibein}
\end{align}
%%%%%%%%%%%%%%%%%%%%%%%
which is related to  the Schwinger gauge by the following $SO(3)$ rotation   
%%%%%%%%%%%%%%%%%%%%%%%
\be
\begin{pmatrix}
e^1_{\text{D}} \\
e^2_{\text{D}} \\
e^3_{\text{D}}
\end{pmatrix} =O(\theta, \phi)
\begin{pmatrix}
e^1_{\text{S}}\\
e^2_{\text{S}} \\
e^3_{\text{S}} 
\end{pmatrix} , 
\label{diracgaugedreibeinrel}
\ee
%%%%%%%%%%%%%%%%%%%%%%
where 
%%%%%%%%%%%%%%%%%%%
\be
O(\theta, \phi) = e^{-i\phi t_1}e^{-i\theta t_3}=
\begin{pmatrix}
\cos\theta & -\sin\theta & 0 \\
\sin\theta\cos\phi & \cos\theta\cos\phi & -\sin\phi\\
\sin\theta\sin\phi & \cos\theta\sin\phi & \cos\phi
\end{pmatrix}, \label{so3rotation}
\ee
%%%%%%%%%%%%%%%%%%%%
with $(t_a)_{bc}\equiv -i\epsilon_{abc}$ being  the $SO(3)$ generators of the adjoint representation.  
The spin connection can be found from the torsion free condition 
%%%%%%%%%%%%%%%%%
\be
de^a =-\omega^{a}_{~~b}e^b. 
\ee
%%%%%%%%%%%%%%%
For the Schwinger gauge (\ref{dreibeinschwing}), we have\footnote{Since the present dreibein (\ref{dreibeinschwing}) is not the Maurer-Cartan one-form, they do $\it{not}$ satisfy $\omega_{ab}=\epsilon_{abc}e_c$ 
%.   About the the Maurer-Cartan one-form of $SU(2)\simeq S^3$ 
(Appendix B of \cite{Hasebe-2015}). 
Non-zero components of the Riemann curvature 2 form, $R^a_{~~b}=d\omega^a_{~~b}+\omega^a_{~~c}\omega^c_{~~b}$, are derived as 
%%%%%%%%%%%%%%%%%%%%%%%%%%%%%
\begin{align}
&R_{12}=-R_{21}=\sin\chi ~d\chi \wedge d\theta =e_1\wedge e_2, ~~~~R_{31}=-R_{13}=-\sin\chi\sin\theta ~d\chi \wedge d\phi=e_3\wedge e_1, \nn\\
&R_{23}=-R_{32}=\sin^2\chi\sin\theta~ d\theta \wedge d\phi=e_2\wedge e_3. 
\label{curvaturetensors3}
\end{align}
%%%%%%%%%%%%%%%%%%%%%%%%%%%%%% 
 For spheres, a special relation, $R_{ab}=%\frac{1}{2}R_{mn \mu\nu}dx^{\mu}\wedge dx^{\nu}=
e_a\wedge e_b$, holds in arbitrary dimensions   
(p.378 in \cite{Eguchi-Gilkey-Hanson-1980}). 
Reading off the Riemann curvature $R^a_{~bcd}$ from $R^{a}_{~b}=\frac{1}{2}R^a_{~bcd}e_c\wedge e_d$, we obtain $R_{1212}=R_{2323}=R_{3131}=1$ (other non-zero components are determined by  the symmetry of $R_{abcd}$) to  construct the scalar curvature  
%%%%%%%%%%%%%%%%%%%
\be
R=R^{a}_{~bab}=6. 
\ee
%%%%%%%%%%%%%%%%%%%%  
}
%%%%%%%%%%%%%%%%%%%%%%%%%%%%
\begin{align}
&\omega^1_{~~2}(=-\omega^2_{~~1}) %=-\cot\chi ~e^2
= -\cos\chi ~d\theta, ~~\omega^3_{~~1}(=-\omega^1_{~~3}) %=\cot\chi ~e^3
= \cos\chi\sin\theta ~d\phi, \nn\\
&\omega^2_{~~3}(=-\omega^3_{~~2})%=-\frac{\cot\theta}{\sin\chi} ~e^3
= -\cos\theta ~d\phi .  \label{spinconnes3oneform}
\end{align}
%%%%%%%%%%%%%%%%%%%%%%%%%%%%%%%%%
The matrix form of the spin connection is then constructed as 
%%%%%%%%%%%%%%
\be
\omega=\sum_{a<b}\omega_{ab}\sigma^{ab} , 
\label{formspincon}
\ee
%%%%%%%%%%%%%
where $\sigma^{ab}$ $(a,b=1,2,3)$ are the $SO(3)$ generators in the spinor representation: 
%%%%%%%%%%%%%%%
\be
\sigma_{ab}=-i\frac{1}{4}[\gamma_a, \gamma_b]=\frac{1}{2}\epsilon_{abc}\gamma_c.  
\ee
%%%%%%%%%%%%%%%
$\gamma_a$ $(a=1,2,3)$ denote the $SO(3)$ gamma matrices, which throughout the paper we will take  
%%%%%%%%%%%%%%%%%%
\be
\gamma_1=\sigma_3, ~~\gamma_2=\sigma_1,~~\gamma_3=\sigma_2.  
\label{gammasigmacorr}
\ee
%%%%%%%%%%%%%%%%%%%%%
(\ref{formspincon}) is now represented as  
%%%%%%%%%%%%%%%%
\be
\omega_{\text{S}} 
=\frac{1}{2} 
\begin{pmatrix}
-\cos\theta  ~d\phi & i\cos\chi ~d\theta +\cos\chi\sin\theta ~d\phi \\
-i\cos\chi ~d\theta +\cos\chi\sin\theta ~d\phi  & \cos\theta  ~d\phi 
\end{pmatrix}.   
\label{schexpsu2diracsphefafie}
\ee
%%%%%%%%%%%%%%%%
Notice  that the holonomy of $S^3\simeq SO(4)/SO(3)$  is $SO(3)\simeq SU(2)$, and  the spin connection (\ref{schexpsu2diracsphefafie}) is formally equivalent to the $SU(2)$ monopole gauge field with minimal charge (see Sec.\ref{sec:nonrelaLandaumodel}). 
In the Dirac gauge (\ref{diracgaugedreibein}), the spin connection is  given by  
%%%%%%%%%%%%%%%%%%%%
\be
\omega_{\text{D}}=\frac{1}{2}(1-\cos\chi)\begin{pmatrix}
\sin^2\theta d\phi & -(id\theta +\sin\theta\cos\theta d\phi) e^{-i\phi} \\
(id\theta -\sin\theta\cos\theta d\phi) e^{i\phi} & -\sin^2\theta d\phi
\end{pmatrix}. \label{omegadirac}
\ee
%%%%%%%%%%%%%%%%%%%% 
(\ref{schexpsu2diracsphefafie}) and (\ref{omegadirac}) are related by the $SU(2)$ gauge transformation: 
%%%%%%%%%%%%%%%%%%%%%%%%%
\be
\omega_{\text{S}} ={g^{(1/2)}}^{\dagger}~\omega_{\text{D}} ~g^{(1/2)} -i{g^{(1/2)}}^{\dagger}~dg^{(1/2)}, 
\label{omegagaugetrans}
\ee
%%%%%%%%%%%%%%%%%%%%%%
where 
%%%%%%%%%%%%%%%
\be
g^{(1/2)}(\theta, \phi) = e^{-i\frac{1}{2}\phi\sigma_3}e^{-i\frac{1}{2}\theta\sigma_2} =
\begin{pmatrix}
e^{-i\frac{1}{2}\phi} \cos\frac{\theta}{2} & -e^{-i\frac{1}{2}\phi} \sin\frac{\theta}{2} \\
e^{i\frac{1}{2}\phi}\sin\frac{\theta}{2} & e^{i\frac{1}{2}\phi} \cos\frac{\theta}{2}
\end{pmatrix}. 
\ee
%%%%%%%%%%%%%%%%
$g(\theta, \phi)$ is the $SU(2)$ group element corresponding to the $SO(3)$ element (\ref{so3rotation}),\footnote{ In other words, 
%%%%%%%%%%%%%%%%%%
\be
g^{(1/2)}(\theta, \phi)~\gamma_a~ {g^{(1/2)}}^{\dagger}(\theta, \phi)=\gamma_b ~O(\theta,\phi)_{ba}, 
\ee
%%%%%%%%%%%%%%%%%%%%%%%%
or 
%%%%%%%%%%%%%%%%%%%%%%%
\be
O(\theta, \phi)_{ab}=\frac{1}{2}\tr ({g^{(1/2)}}^{\dagger}(\theta, \phi)~ \gamma_a~ g^{(1/2)}(\theta, \phi)~\gamma_b)=\frac{1}{2}\tr (g^{(1/2)}(\theta, \phi)~ \gamma_b~ {g^{(1/2)}}^{\dagger}(\theta, \phi)~\gamma_a).
\ee
%%%%%%%%%%%%%%%%%%%%%%%
} 
%%%%%%%%%%%%%%%%%%%%%%%%
\be
{g^{(1/2)}}^{\dagger}(\theta, \phi)~\gamma_a~ g^{(1/2)}(\theta, \phi)=~O(\theta,\phi)_{ab}\gamma_b. 
\label{sigmatranssu2}
\ee
%%%%%%%%%%%%%%%%%%%%%%%%%%%%%
For the local polar coordinates on $S^3$, the gauge field takes  a simple form in the Schwinger gauge (\ref{schexpsu2diracsphefafie}), while in the Dirac gauge the representation is  rather clumsy (\ref{omegadirac}).    
On the other hand, in the target space  Cartesian  coordinates,  the Schwinger gauge representation  (\ref{schexpsu2diracsphefafie}) becomes lengthy  
%%%%%%%%%%%%%%%%%
\begin{align}
&\omega_{\text{S}}= -\frac{1}{2{(1-{x_4}^2)\sqrt{{x_1}^2+{x_2}^2}}} \nn\\
& \!\!\!\!\!\!\!\!\!\!\!\!\!\!\!\!\!\!\!\!\!\times \biggl( x_4\sqrt{1-{x_4}^2}(x_2dx_1-x_1dx_2)\sigma_1   +x_4(x_1x_3dx_1+x_2x_3dx_2+({x_1}^2+{x_2}^2)dx_3) \sigma_2 -\frac{x_3}{\sqrt{{x_1}^2+{x_2}^2}} (x_2dx_1-x_1dx_2)\sigma_3\biggr), 
\label{schwingers3gaugen} 
\end{align}
%%%%%%%%%%%%%%%%%%%%
while the Dirac gauge representation  (\ref{omegadirac}) is much concise   
%%%%%%%%%%%%%%%%%%%%%%%
\be
\omega_{\text{D}}=-\frac{1}{2(1+x_4)}\epsilon_{abc}x_b \sigma_c dx_a. 
\label{su2diracfund}
\ee
%%%%%%%%%%%%%%%%%%%%%%%
Thus, the Schwinger gauge is an appropriate gauge in the usage of the local polar coordinates on $S^3$, and the Dirac gauge in the target space Cartesian coordinates.  
The spin connection  (\ref{schwingers3gaugen}) has the singularity  both at the  north pole $x_4=1$ and the south pole $x_4=-1$, and hence the name, the Schwinger gauge \cite{Felsager1998, Hasebe-2015}.  Meanwhile,  the singularity of  (\ref{su2diracfund}) is only at the south pole $x_4=-1$, and the name the Dirac gauge.

%%%%%%%%%%%%%%%%%%%%%%%%%%%%%%%%%%%%%%%%%%%%%%%%%
\section{Non-relativistic Landau Model}\label{sec:nonrelaLandaumodel}
%%%%%%%%%%%%%%%%%%%%%%%%%%%%%%%%%%%%%%%%%%%%%%%%%%

In this section, we perform a through investigation of the eigenvalue problem of the $SO(4)$ Landau Hamiltonian. 
The obtained results are utilized throughout the paper, and  we provide a detail explanation for  readers not to stumble against any logical gap or technical difficulty.  We  first present an expanded discussions of    
 \cite{Nair-Daemi-2004,Hasebe-2014-2} about  the mathematical background of the 
$SO(4)$ Landau model (Sec.\ref{subsec:chiralhopfsu2mono}), the  $SO(4)$ operators and their algebraic relations (Sec.\ref{subsec:so4operators}) and   the  $SO(4)$ Landau problem (Sec.\ref{sec:so4landauproblem}).     
Next in Sec.\ref{subsec:gaugefixanal}, we fix the gauge and provide new results about the   basic properties of the $SO(4)$ monopole harmonics (Sec.\ref{subsec:so4monoprop}) and the $SO(4)$   covariance  (Sec.\ref{subsec:so4covari}) in which we give a verification of the $SO(4)$ monopole harmonics to be the eigenstates of the $SO(4)$ Landau Hamiltonian.  In Sec.\ref{subsec:redso4harmo}, we check that the derived $SO(4)$ monopole harmonics indeed reduce to the known $SO(4)$ spherical harmonics in the free background limit.

%The readers may hopefully grasp the important structure of the $SO(4)$ Landau model without any 

%The Dirac gauge is adopted to provide explicit coordinate representations to operators and wave functions. 

%and address  relations to the formulae of the Nair and Randjbar-Daemi's work.  
%In \cite{Nair-Daemi-2004}, Nair and Randjbar-Daemi solved the  eigenvalue problem of the $SO(4)$ Landau Hamiltonian based on the coset space method without refering to any particular gauge, and their formalism is rather abstract.   

 %In \cite{Hasebe-2014-2}, we fixed the Dirac gauge   and derived  the lowest Landau level eigenstates  by taking a fully symmetric product of the chiral Hopf spin ors. 
 
%By fixing the gauge, we can explicitly confirm that the present result is reduced to the  

For notational brevity, we adopt the following abbreviation of the angular coordinates in (\ref{fs3coordinatesspherical}):   
%%%%%%%%%%%%%%
\be
\bs{\chi}=(\chi,\theta, \phi)
\ee
%%%%%%%%%%%%%
and 
%%%%%%%%%%%%
\be
-\bs{\chi}=(-\chi,\theta, \phi).
\ee
%%%%%%%%%%%%
The sign flip of $\bs{\chi}$ represents the parity transformation on $S^3$: 
%%%%%%%%%%%%%%%%%%
\be
(x_1, x_2, x_3, x_4) ~~\rightarrow ~~(-x_1,-x_2,-x_3,x_4), 
\label{partity3D}
\ee
%%%%%%%%%%%%%%%%
which interchanges  the left-handed and right-handed coordinate systems, and so we call it the LR transformation.     

%%%%%%%%%%%%%%%%%%%%%%%%%%%%%%%%%%%%%%%%%%%%%%%
\subsection{The chiral Hopf map and the $SU(2)$ monopole gauge field}\label{subsec:chiralhopfsu2mono}
%%%%%%%%%%%%%%%%%%%%%%%%%%%%%%%%%%%%%%%%%%%%%%%%%

The underlying geometry of the $SO(4)$ Landau model is  the  chiral Hopf map \cite{Hasebe-2014-2}, 
%%%%%%%%%%%%%%%%%%%%%%%
\be
S^3_{{L}}\otimes S^3_{{R}} ~\overset{S^3_{\text{diag}}}{\longrightarrow}~S^3.  
\label{chiralhopfabst}
\ee
%%%%%%%%%%%%%%%%%%%%%
Here, the projected space $S^3$ denotes the base-manifold, and  $S_D^3\simeq SU(2)$ corresponds to the $SU(2)$ fiber, and the map gives a set-up of the $SO(4)$ Landau model on $S^3$ with $SU(2)$ monopole at the center.             We represent the coordinates on $S_L$ and $S_R$ as two two-component chiral Hopf spinors, $\psi_L=(\psi_{L1}~\psi_{L2})$ and $\psi_R=(\psi_{R1}~\psi_{R2})$, subject to 
%%%%%%%%%%%%%%%%%%%%%%%%%%%%%%
\be
{\psi_L}{\psi_L}^{\dagger}={\psi_R}{\psi_R}^{\dagger}=\frac{1}{2}. 
\ee
%%%%%%%%%%%%%%%%%%%%%%%%%%%%
(\ref{chiralhopfabst}) can  be expressed as 
%%%%%%%%%%%%%%%%%%%%%%%%%%%
\be
\psi_L, ~\psi_R ~\longrightarrow ~{x_{\mu}}={\psi_L}{\bar{q}_{\mu}}{\psi_R}^{\dagger} + {\psi_R}
{{{q}_{\mu}}}{\psi_L}^{\dagger}  ~~~(\mu=1,2,3,4),
\label{1.5Hopfmap}
\ee
%%%%%%%%%%%%%%%%%%%%%%
where  $q_{\mu}$ and $\bar{q}_{\mu}$ are the  quaternions and conjugate-quaternions,    
%%%%%%%%%%%%%%%%%%%%%%%%%%%%%%
\be
q_{\mu}=(q_i, 1)=(-i\sigma_i, 1),~~~~\bar{q}_{\mu}=(-q_i, 1 )=(i\sigma_i, 1). 
\ee
%%%%%%%%%%%%%%%%%%%%%%%%%%%%%%
From (\ref{1.5Hopfmap}), we have 
%%%%%%%%%%%%%%%%%%%%
\be
\sum_{\mu=1}^4 x_{\mu}x_{\mu}=4 ({\psi_L}{\psi_L}^{\dagger})({\psi_R}{\psi_R}^{\dagger})=1. 
\ee
%%%%%%%%%%%%%%%%%%%
$x_{\mu}$ are invariant under the simultaneous $SU(2)$ transformation of $\psi_L$ and $\psi_R$,    
$\psi_{L/R} \rightarrow  e^{\alpha_i q_i} \psi_{L/R}$, and we denote such diagonal $SU(2)$ rotation  as $SU(2)_D$.  
The chiral Hopf spinors, $\psi_{L}$ and $\psi_{R}$, are represented as 
%%%%%%%%%%%%%%%%%%%%%%
\be
\psi_{L}(x)=\phi~\Psi_{L}^{(1/2)}(x), ~~~~~\psi_{R}(x)=\phi~\Psi_{R}^{(1/2)}(x), 
\label{su2su2bispinors}
\ee
%%%%%%%%%%%%%%%%%%%%%%
where $\phi=(\phi_1 ~\phi_2)$ is a normalized  two-component spinor $\phi\phi^{\dagger}=1/2$ representing the $S^3_D$-fibre.  $\Psi_{L}^{(1/2)}$ and $\Psi_R^{(1/2)}$ are given by 
%%%%%%%%%%%%%%%%%%%%%%%%%%%
\be
\Psi_{L}^{(1/2)}(x)=\frac{1}{\sqrt{2(1+x_4)}}(1+ x_{\mu}{q}_{\mu}), ~~~~~~~~~
\Psi_{R}^{(1/2)}(x)=\frac{1}{\sqrt{2(1+x_4)}}(1+ x_{\mu}\bar{q}_{\mu})
={{\Psi_L}^{(1/2)}(x)}^{\dagger}, 
\label{defofmatrixM}
\ee
%%%%%%%%%%%%%%%%%%%%%%%%%%%%%
each of which is an $SU(2)$ group element and their squares  yield 
${\Psi_L^{(1/2)}}^2=({\Psi_R^{(1/2)}}^{\dagger})^2=    x_{\mu}q_{\mu}$. 
Using $\Psi_L^{(1/2)}$ and $\Psi_R^{(1/2)}$,  we can derive the $SU(2)$ monopole connection  as 
%%%%%%%%%%%%%%%%%%%
\be
A^{(1/2)}=-i\frac{1}{2}({\Psi_L^{(1/2)}}^{\dagger}d\Psi^{(1/2)}_L+{\Psi_R^{(1/2)}}^{\dagger}d\Psi^{(1/2)}_R) = -\frac{1}{2(1+x_4)}\epsilon_{ijk}x_j\sigma_k dx_i. 
\label{derigaugefield}
\ee
%%%%%%%%%%%%%%%%%%%%
Note that the gauge connection (\ref{derigaugefield}) is formally identical to the spin connection  (\ref{su2diracfund}).  With the angular coordinates, $\Psi_{L}$ and $\Psi_R$ can be expressed as\footnote{Indeed, 
%%%%%%%%%%%%%%%%%%%%
\begin{align}
&\Psi_L^{(1/2)}=\frac{1}{\sqrt{2(1+x_4)}}\begin{pmatrix}
1+x_4-ix_3 & -x_2-ix_1 \\
x_2+ix_1 & 1+x_4+ix_3
\end{pmatrix} =
\begin{pmatrix}
\cos\frac{\chi}{2}-i\sin\frac{\chi}{2}\cos\theta & -i\sin\frac{\chi}{2}\sin\theta e^{-i\phi} \\
-i\sin\frac{\chi}{2}\sin\theta e^{i\phi}  & \cos\frac{\chi}{2}+i\sin\frac{\chi}{2}\cos\theta
\end{pmatrix} =e^{-i\frac{1}{2}\chi\hat{\bs{x}}\cdot \bs{\sigma}} , \nn\\
&\Psi_R^{(1/2)}=
\frac{1}{\sqrt{2(1+x_4)}}\begin{pmatrix}
1+x_4+ix_3 & x_2+ix_1 \\
-x_2-ix_1 & 1+x_4-ix_3
\end{pmatrix}= 
\begin{pmatrix}
\cos\frac{\chi}{2}+i\sin\frac{\chi}{2}\cos\theta & i\sin\frac{\chi}{2}\sin\theta e^{-i\phi} \\
i\sin\frac{\chi}{2}\sin\theta e^{i\phi}  & \cos\frac{\chi}{2}i\sin\frac{\chi}{2}\cos\theta
\end{pmatrix}=e^{i\frac{1}{2}\chi\hat{\bs{x}}\cdot \bs{\sigma}} . \label{largepsis}
\end{align}
%%%%%%%%%%%%%%%%%%%%
    }  
%%%%%%%%%%%%%%%%%%%%%%%%
\be
\Psi_L^{(1/2)}(x)= e^{-i\frac{1}{2}\chi\hat{\bs{x}}\cdot \bs{\sigma}},~~~~~~~~\Psi_R^{(1/2)}(x)= e^{i\frac{1}{2}\chi\hat{\bs{x}}\cdot \bs{\sigma}},  \label{su2elemfundmm} 
\ee
%%%%%%%%%%%%%%%%%%%%%%%
where $\hat{\bs{x}}$ signifies a position  on the ($S^2$-)equator of $S^3$:  
%%%%%%%%%%%%%%%%%
\be
\hat{\bs{x}}=(\sin\theta\cos\phi, \sin\theta\sin\phi, \cos\theta).  
\ee
%%%%%%%%%%%%%%%%%

Promoting the Pauli matrices in (\ref{su2elemfundmm}) to $SU(2)$ arbitrary matrices with  spin magnitude $I/2$, we introduce 
%%%%%%%%%%%%%%%%%%%%
\be
\Psi^{(I/2)}_{\text{D}}(\bs{\chi}) \equiv e^{-i{\chi}\hat{\bs{x}} \cdot \bs{S}^{(I/2)}}, 
\label{diracgaugedfunc}
\ee
%%%%%%%%%%%%%%%%%%%%
and 
%%%%%%%%%%%%%%%%%%%%%%%%%%%
\be
\Psi_{L}^{(I/2)}(x)=\Psi_{\text{D}}^{(I/2)}(\bs{\chi}),~~~ \Psi^{(I/2)}_{R}(x)=\Psi_{\text{D}}^{(I/2)}(-\bs{\chi}) . 
\label{defofmatrixMtwunitaryo}
\ee
%%%%%%%%%%%%%%%%%%%%%%%%%%%%% 
% $\bs{S}^{(I/2)}$ is the $SU(2)$ spin matrix of spin index $I/2$. 
$\Psi_L$ and $\Psi_R$ are interchanged by the LR transformation (\ref{partity3D}). 
They satisfy  
%%%%%%%%%% 
\be
{\Psi^{(I/2)}_{\text{D}}(\bs{\chi})}^{-1} ={\Psi^{(I/2)}_{\text{D}}(\bs{\chi})}^{\dagger} =\Psi^{(I/2)}_{\text{D}}(-\bs{\chi}) .
\ee
%%%%%%%%%%%
In a same manner to (\ref{derigaugefield}),  we obtain  the $SU(2)$ monopole gauge field  in the Dirac gauge 
%%%%%%%%%%%%%
\be
A^{(I/2)}=-i\frac{1}{2}({\Psi_L^{(I/2)}}^{\dagger}d\Psi^{(I/2)}_L+{\Psi_R^{(I/2)}}^{\dagger}d\Psi^{(I/2)}_R)=-\frac{1}{1+x_4}\epsilon_{ijk}x_j S^{(I/2)}_k dx_i,
\label{gaugefieldasu2}
\ee
%%%%%%%%%%%%%%%
or 
%%%%%%%%%%%%%%%%%%%%%
\be
A^{(I/2)}_{i}=-\frac{1}{1+x_4}\epsilon_{ijk}x_{j}S_k^{(I/2)}~~~(i=1,2,3), ~~~~A^{(I/2)}_4=0.  
\label{composu2diracmonosu2gau}
\ee
%%%%%%%%%%%%%%%%%%%%%%
The winding number associated with the $SU(2)$ gauge field (\ref{gaugefieldasu2}) is 
%%%%%%%%%%%%%%%%%%%
\be
\nu=i\frac{1}{24\pi^2}\int_{S^3}\tr (-ig^{\dagger}dg)^3, 
\label{defnurhs}
\ee
%%%%%%%%%%%%%%%%%%%%%%%%
where 
%%%%%%%%%%%%%%%%%%%%%
\be
g={\Psi^{(I/2)}_{\text{D}}(\bs{\chi})}^2 =e^{-2i\chi\hat{\bs{x}}\cdot \bs{S}^{(I/2)}}, 
\ee
%%%%%%%%%%%%%%%%%%%%%%%%
and (\ref{defnurhs}) is evaluated as 
%%%%%%%%%%%%%%%%%%%%%%%%
\be
\nu=\frac{1}{6}I(I+1)(I+2). 
\ee
%%%%%%%%%%%%%%%%%%%%%%%%%%%
It should be mentioned that  the chiral Hopf map and the present $SU(2)$ monopole are naturally understood by embedding  $S^3$  in one-dimension higher  $S^4$ \cite{Hasebe-2014-2, Hasebe-2017}.

%%%%%%%%%%%%%%%%%%%%%%%%%%%%%%%%%%%%%%%%%%%%%%%
\subsection{$SO(4)$ operators}\label{subsec:so4operators}
%%%%%%%%%%%%%%%%%%%%%%%%%%%%%%%%%%%%%%%%%%%%%%%%%

%As discussed in Sec.\ref{subsec:chiralhopfsu2mono}, our set-up is $S^3$ with the $SU(2)$ monopole at the origin. 
The $SU(2)$ magnetic field is perpendicular to  $S^3$ surface, and so the present system respects the $SO(4)$ rotational symmetry. 
 We construct total angular momentum operators that generate the simultaneous $SO(4)$ rotations  of the base-manifold and the gauge space. We clarify  analogies and differences to  the $U(1)$ monopole system on $S^2$ \cite{Hasebe-2015}. 

%%%%%%%%%%%%%%%%%%%%%%%%%%%%%%%
\subsubsection{$SO(4)$ angular momentum operators}
%%%%%%%%%%%%%%%%%%%%%%%%%%%%%%

 For $A_{\mu}$ $(\mu=1,2,3,4)$ (\ref{composu2diracmonosu2gau}), let us introduce  the covariant derivative  
%%%%%%%%%%%%%%%%%%%%%
\be 
D_{\mu} =\partial_{\mu}+iA_{\mu}, 
\ee 
%%%%%%%%%%%%%%%%%%%%%
 and the field strength is given by 
%%%%%%%%%%%%%%%%%%%%%%%
\be
F_{\mu\nu}=-i[D_{\mu}, D_{\nu}] =\partial_{\mu} A_{\nu} -\partial_{\nu} A_{\mu} +i[A_{\mu}, A_{\nu}]  
\ee
%%%%%%%%%%%%%%%%%%%%%%
with 
%%%%%%%%%%%%%%%%%%%%
\be
F_{ij}=-x_{i}A_{j}+x_{j}A_{i} +\epsilon_{ijk}S_k^{(I/2)} , ~~~~F_{i 4}=(1+x_4)A_{j} =-\epsilon_{ijk}x_{j} S_k^{(I/2)} . 
\label{compfieldstreng}
\ee
%%%%%%%%%%%%%%%%%%%
With the dreibein (\ref{diracgaugedreibein}),   (\ref{compfieldstreng}) is  concisely represented  as  
%%%%%%%%%%%%%%%
\be
F=\frac{1}{2}F_{\mu\nu}dx^{\mu}\wedge dx^{\nu}=\frac{1}{2}\epsilon_{ijk}e_{\text{D}}^i\wedge e_{\text{D}}^j ~S_k^{(I/2)}.
\ee
%%%%%%%%%%%%%%% 
The $SO(4)$ covariant angular momentum is introduced as 
%%%%%%%%%%%%%%%%%%%%%%%%
\be
\Lambda_{\mu\nu}=-ix_{\mu}D_{\nu} +ix_{\nu} D_{\mu} ~~~  (\mu, \nu=1,2,3, 4), 
\ee
%%%%%%%%%%%%%%%%%%%%%%%% 
and  the  conserved $SO(4)$ angular momentum operator consists of particle angular momentum  and the field angular momentum of the monopole: 
%%%%%%%%%%%%%%%%%%%%%%%%%%%%%%
\be 
 L_{\mu\nu}=\Lambda_{\mu\nu}+ F_{\mu\nu}. 
 \label{totalangularsum}
\ee 
%%%%%%%%%%%%%%%%%%%%%%%%%%%%%%
In the Dirac gauge, $L_{\mu\nu}$ (\ref{totalangularsum}) are expressed as 
%%%%%%%%%%%%%%%%%%%%%%%%%%%
\be
L_{ij} =l_{ij} +\epsilon_{ijk}S_k^{(I/2)}, ~~~L_{ i 4} =l_{i 4} -\frac{1}{1+x_4}\epsilon_{ijk}x_{j}S_k^{(I/2)}, 
\label{SO(4)totalangulaexp}
\ee
%%%%%%%%%%%%%%%%%%%%%%%%%% 
where  $l_{\mu\nu}$ denote the free $SO(4)$ angular momentum operators 
%%%%%%%%%%%%%%%%%%%%%%%%%%%%%%%%%
\be
l_{\mu\nu}=-ix_{\mu}\partial_{\nu} +ix_{\nu}\partial_{\mu}. 
\ee
%%%%%%%%%%%%%%%%%%%%%%%%%%%%%%%%%%
With the explicit coordinate representations, it is straightforward to check that $T_{\mu\nu}=L_{\mu\nu}$, $\Lambda_{\mu\nu}$ and $F_{\mu\nu}$  transform as two-rank tensors under the $SO(4)$  transformations generated by $L_{\mu\nu}$: 
%%%%%%%%%%%%%%%%%
\be
[L_{\mu\nu}, T_{\rho\sigma}] =i\delta_{\mu\rho}T_{\nu\sigma}-i\delta_{\mu\sigma}T_{\nu\rho}+i\delta_{\nu\sigma}T_{\mu\rho} - i\delta_{\nu\rho}T_{\mu\sigma}. 
\label{so4algebralmunu}
\ee
%%%%%%%%%%%%%%%%
From the orthogonality between $\Lambda_{\mu\nu}$ and $F_{\mu\nu}$    
%%%%%%%%%%%%%%%%%%%%%%%%%%%%%%%
\be
\sum_{\mu < \nu =1}^4 F_{\mu\nu}{\Lambda_{\mu\nu}} =\sum_{\mu < \nu =1}^4 {\Lambda_{\mu\nu}}F_{\mu\nu}=0, 
\ee
%%%%%%%%%%%%%%%%%%%%%%%%%%%%%%%%%%%
the $SO(4)$ Casimir is given by 
%%%%%%%%%%%%%%%%%%%%
\be
\sum_{\mu < \nu =1}^4 {L_{\mu\nu}}^2=\sum_{\mu < \nu =1}^4 {\Lambda_{\mu\nu}}^2+\sum_{\mu < \nu =1}^4 {F_{\mu\nu}}^2, 
\label{so4casimirllamf}
\ee
%%%%%%%%%%%%%%%%%%%%%
which  can be rewritten as
%%%%%%%%%%%%%%%%%
\be
\sum_{\mu<\nu}{L_{\mu\nu}}^2
=\sum_{\mu<\nu}{l_{\mu\nu}}^2+\frac{2}{1+x_4}\bs{l}\cdot \bs{S}^{(I/2)} +\frac{1}{2(1+x_4)} I(I+2) -\frac{1}{(1+x_4)^2} (\bs{x}\cdot \bs{S}^{(I/2)})^2, 
\label{so4casnonrel}
\ee
%%%%%%%%%%%%%%%%%
with $\bs{l}$  the free $SO(3)$ angular momentum operator: 
%%%%%%%%%%%%%%%%%
\be
l_i=\frac{1}{2}\epsilon_{ijk}l_{jk}=-i\epsilon_{ijk}x_j \partial_k.
\ee
%%%%%%%%%%%%%%%%%%%%
The first term is the $SO(4)$ free angular momentum Casimir, and the second term is formally equivalent to the spin-orbit coupling in three-dimension.  
Meanwhile, the $SO(3)$ Casimir of the $SO(3)$ Landau model on $S^2$  is represented as \cite{Hasebe-2015}    
%%%%%%%%%%%%%%%
\be
\sum_{i=1}^3 {{\hat{L}}_i}^2 =\sum_{i=1}^3{{l}_i}^2 +\frac{I}{1+x_3}l_3+\frac{1}{2(1+x_3)}I^2,   
\label{so3casnonrel}
\ee
%%%%%%%%%%%%%%%%%%
where 
%%%%%%%%%%%%%%%%
\be
{\hat{L}}_i\equiv -i\epsilon_{ijk}x_j(\partial_k+i\hat{A}_k)+\hat{F}_i.  
\label{so3genepre}
\ee
%%%%%%%%%%%%%%%
$\hat{A}_i=-I\frac{1}{2(1+x_3)}\epsilon_{ij3}x_j$ is the $U(1)$ monopole gauge field  and $\hat{F}_i=\epsilon_{ijk}\partial_j \hat{A}_k=\frac{I}{2}x_i$.  
Comparison between (\ref{so4casnonrel}) and (\ref{so3casnonrel}) shows the the last term of (\ref{so4casnonrel}), $-\frac{1}{(1+x_4)^2}(\bs{x}\cdot \bs{S}^{(I/2)})^2$, is specific  to the $SO(4)$ Landau model with the non-Abelian gauge field.\footnote{When $I=1$, this term is reduced to  
%%%%%%%%%%%%
$-\frac{1}{4}\frac{1-x_4}{1+x_4}.$ 
%%%%%%%%%%%
} 

%%%%%%%%%%%%%%%%%%%%%%%%%%%%%%%
\subsubsection{$SU(2)_L\otimes SU(2)_R$ group generators}
%%%%%%%%%%%%%%%%%%%%%%%%%%%%%%%%%

Since $SO(4)\simeq SU(2)_L\otimes SU(2)_R$, 
we can construct $su(2)_L\oplus su(2)_R$ generators from the $so(4)$ generators: 
%%%%%%%%%%%%%%%%%%%%%%%%%%%
\be
L_i=\frac{1}{4}\eta_{\mu\nu}^i L_{\mu\nu}, ~~~~\bar{L}_i=\frac{1}{4}\bar{\eta}^i_{\mu\nu}L_{\mu\nu} , 
\label{su2operatorstwo}
\ee
%%%%%%%%%%%%%%%%%%%%%%%%%%%%
which satisfy 
%%%%%%%%%%%%%%%%%%%%%
\be
[L_i, L_j]=i\epsilon_{ijk}L_k, ~~~~[\bar{L}_i, \bar{L}_j]=i\epsilon_{ijk}\bar{L}_k, ~~~[L_i, \bar{L}_j]=0. 
\label{su2algebras}
\ee
%%%%%%%%%%%%%%%%%%%%%%
Here, $\eta_{\mu\nu}^i$ and $\bar{\eta}_{\mu\nu}^i$ are the 't Hooft symbols: 
%%%%%%%%%%%%%%%%%%%
\be
\eta_{\mu\nu}^i=\epsilon_{\mu\nu i 4}+\delta_{\mu i}\delta_{\nu 4} -\delta_{\mu 4}\delta_{\nu i},~~~~~\bar{\eta}_{\mu\nu}^i=\epsilon_{\mu\nu i 4}-\delta_{\mu i}\delta_{\nu 4} +\delta_{\mu 4}\delta_{\nu i}. 
\label{thooftsymbolsdef}
\ee
%%%%%%%%%%%%%%%%%%%
The two independent $su(2)$ algebras (\ref{su2algebras}) can be verified by the $so(4)$ algebra and properties of the 't Hooft symbols (\ref{thooftsymbolsdef}).\footnote{ The following properties will be useful: 
%%%%%%%%%%%%%%%%
\be
\eta_{\mu\nu}^{i}\eta_{\mu\rho}^{j}=\delta_{ij}\delta_{\nu\rho}+\epsilon_{ijk}\eta_{\nu\rho}^{k} , ~~~\bar{\eta}_{\mu\nu}^{i}\bar{\eta}_{\mu\rho}^{j}=\delta_{ij}\delta_{\nu\rho}+\epsilon_{ijk}\bar{\eta}_{\nu\rho}^{k}. 
\ee
%%%%%%%%%%%%%%%%
}   
The $SO(4)$ Casimir is also given by a simple sum of the two  $SU(2)$ Casimirs: 
%%%%%%%%%%%%%%%%%%%%
\be
\sum_{\mu < \nu =1}^4 {L_{\mu\nu}}^2=2({L_i}^2+{\bar{L}_i}^2) 
\ee
%%%%%%%%%%%%%%%%%%%%%
whose eigenvalues are readily obtained as  
%%%%%%%%%%%%
\be
\sum_{\mu<\nu}{L_{\mu\nu}}^2 =2l_L(l_L+1) +2l_R(l_R+1).  
\label{so4eigenvalulllr}
\ee
%%%%%%%%%%%%%
Here, 
$l_L$ and $l_R$ denote the $SU(2)_L$ and $SU(2)_R$ Casimir indices,  
%%%%%%%%%%%%%%%%
\be
\bs{L}^2=l_L(l_L+1),~~~~\bar{\bs{L}}^2=l_R(l_R+1), 
\ee
%%%%%%%%%%%%%%%
and their sum is bounded below by the monopole charge
%%%%%%%%%%%%%%%%%%%%%
\be
l_L+l_R=n+\frac{I}{2}.~~~(n=0,1,2,\cdots)
\ee
%%%%%%%%%%%%%%%%%%%%%%
$n$ comes from the particle angular momentum, while $I/2$ comes from the field angular momentum of the monopole  
(recall (\ref{totalangularsum})). 
In the case $I=0$, the $SO(4)$ irreducible representation is reduced to the $SO(4)$ spherical harmonics with the $SO(4)$ indices $(l_L, l_R)=(n/2, n/2)$ (Appendix \ref{sphericalharmoso4}), and so each of $l_L$ and $l_R$ is bounded below by $n/2$:  
%%%%%%%%%%%%%%%
\be
l_L, l_R \ge \frac{n}{2}. 
\ee
%%%%%%%%%%%%%%
Consequently, for given $n$, $(l_L, l_R)$ can take the following $(I+1)$ distinct values 
%%%%%%%%%%%%%%%%%%%%%%%
\be
(l_L, l_R) =(\frac{n}{2}+\frac{I}{2}, \frac{n}{2}), ~(\frac{n}{2}+\frac{I}{2}-\frac{1}{2}, \frac{n}{2}+\frac{1}{2}), ~(\frac{n}{2}+\frac{I}{2}-1, \frac{n}{2}+1),~  \cdots~,(\frac{n}{2}, \frac{n}{2}+\frac{I}{2}). 
\label{llandns}
\ee
%%%%%%%%%%%%%%%%%%%%%%%%
Introducing the chirality parameter $s$ \cite{Hasebe-2014-2} 
%%%%%%%%%%%%%
\be
s\equiv l_L-l_R= \frac{I}{2}, \frac{I}{2}-1, \frac{I}{2}-2, \cdots, -\frac{I}{2},
\ee
%%%%%%%%%%%%%%%%
we specify  $(l_L, l_R)$  as 
%%%%%%%%%%%%%%%%%%
\be
(l_L, l_R) =(\frac{n}{2}+\frac{I}{4} +\frac{s}{2}, \frac{n}{2}+\frac{I}{4}-\frac{s}{2}) .  
\label{preciselllrdef}
\ee
%%%%%%%%%%%%%%%%%%%
Essentially, the Landau level $n$ corresponds to the sum of the two $SU(2)$ indices, while the chirality parameter $s$  their difference.  The $SO(4)$ Casimir eigenvalue (\ref{so4eigenvalulllr}) is now represented as 
%%%%%%%%%%%%%%%%%%%%
\be
\sum_{\mu < \nu =1}^4 {L_{\mu\nu}}^2 =(n+\frac{I}{2})(n+\frac{I}{2}+2)+s^2. 
\label{so4casimireigenns}
\ee
%%%%%%%%%%%%%%%%%%%
 
%%%%%%%%%%%%%%%%%%%%%%%%%%%%%%%
\subsubsection{$SU(2)_{\text{diag}}$ generators and ``boost'' generators}
%%%%%%%%%%%%%%%%%%%%%%%%%%%%%%%%%

In the Dirac gauge, (\ref{su2operatorstwo}) are explicitly represented as  
%%%%%%%%%%%%%%%%%%%%%%%%
\begin{align}
&L_i=-i\frac{1}{2}\epsilon_{ijk}x_j\partial_k -i\frac{1}{2}x_i\partial_4 +i\frac{1}{2}x_4\partial_i +\frac{1}{2}S_i^{(I/2)}-\frac{1}{2(1+x_4)}\epsilon_{ijk}x_jS_k^{(I/2)},\nn\\
&\bar{L}_i=-i\frac{1}{2}\epsilon_{ijk}x_j\partial_k +i\frac{1}{2}x_i\partial_4 -i\frac{1}{2}x_4\partial_i +\frac{1}{2}S_i^{(I/2)}+\frac{1}{2(1+x_4)}\epsilon_{ijk}x_jS_k^{(I/2)}. \label{lisu2diracex}
\end{align}
%%%%%%%%%%%%%%%%%%%%%%%
%The $SU(2)_L$ and $SU(2)_R$ angular momentum operators,   
$L_i$ and $\bar{L}_i$ are interchanged by the LR transformation (\ref{partity3D}). 
For the comparison to the $SO(3)$ Landau model \cite{Hasebe-2015}, it is useful to introduce the $SU(2)$ diagonal group generators and ``boost'' generators. 
From the two sets of $SU(2)$ operators, we define  the $SU(2)$ diagonal group generators,  
%%%%%%%%%%%%%%%%%%%%%%
\be
L_i^{\text{diag}}\equiv L_i +\bar{L}_i=-i\epsilon_{ijk}x_j\partial_k+S_i^{(I/2)}  
\label{defsu2diagop}
\ee
%%%%%%%%%%%%%%%%%%%%%%%%
where 
%%%%%%%%%%%%%%%%%%%%%%%
\be
l_i=-i\epsilon_{ijk}x_j\partial_k.
\ee
%%%%%%%%%%%%%%%%%%%%%%%%
$L_i^D$ do not depend on $x_4$ and are formally identical to  the angular momentum operators of a free particle with higher spin $I/2$ in three dimension\footnote{Recall that in the beginning $S_i^{(I/2)}$ were the $SU(2)$ gauge group generators and the particle was a spinless particle, but here we reinterpret $S_i^{(I/2)}$ as the intrinsic spin of  particle. This interpretation is similar to that of Wilczek \cite{Wilczek1982-1}. } that obviously satisfy the $su(2)$ algebra.  
In the $SO(3)$ Landau model on $S^2$ \cite{Hasebe-2015}, the $SO(2)$ operator $\hat{L}_z$  (\ref{so3genepre}) consists  of the free angular momentum  and the $U(1)$ gauge generator, $\hat{L}_z =-i\frac{\partial}{\partial \phi}-\frac{I}{2}$, and then (\ref{defsu2diagop}) may be regarded as  its  $SU(2)$ generalization. From  analogy to the Lorentz group, we refer to the remaining $SO(4)$ operators as the ``boost'' operators:  
%%%%%%%%%%%%%%%%%%%%
\be
K_i=L_i-\bar{L}_i=-ix_i\partial_4 +ix_4\partial_i -\frac{1}{1+x_4}\epsilon_{ijk}x_jS_k^{(I/2)}. 
\label{defsu2k}
\ee
%%%%%%%%%%%%%%%%%%%%%%
It is easy to see that $\bs{L}^{\text{diag}}$ and $\bs{K}$ satisfy \cite{Nair-Daemi-2004} 
%%%%%%%%%%%%%%%%%%%%%%%%%
\be
[L_i^{\text{diag}}, L_j^{\text{diag}}]=i\epsilon_{ijk}L_k^{\text{diag}}, ~~~[L_i^{\text{diag}}, K_j]=i\epsilon_{ijk}K_k, ~~~[K_i, K_j]=i\epsilon_{ijk}L_k^{\text{diag}}. 
\ee
%%%%%%%%%%%%%%%%%%%%%%% 
$K_i$ thus transform as the $SU(2)_{\text{diag}}$ vector. 
Though $\bs{K}^2$ (and ${\bs{L}^{\text{diag}}}^2$) is not invariant under general $SO(4)$ transformations   
%%%%%%%%%%%%%%%%%%%%%%%%%%%
\be
[K_i, \bs{K}^2]=-[K_i, {\bs{L}^{\text{diag}}}^2]=i\epsilon_{ijk}(K_jL_k^{\text{diag}}-L_j^{\text{diag}}K_k) \neq 0, 
\ee
%%%%%%%%%%%%%%%%%%%%%%%%%%%%%%
 ${\bs{K}}^2$ is invariant under the $SU(2)_D$ transformations
%%%%%%%%%%%%%%%%%%%%%
\be
[L_i^{\text{diag}}, \bs{K}^2]=0.  
\ee
%%%%%%%%%%%%%%%%%%%%
While ${\bs{L}^{\text{diag}}}^2$ represents  $SU(2)$ Casimir,  the square of the boost operators 
yields  \cite{Nair-Daemi-2004} 
%%%%%%%%%%%%%%%%%%%%%%
\be
{\bs{K}}^2=(\bs{L}-\bar{\bs{L}})^2=2({\bs{L}}^2+{\bar{\bs{L}}}^2)-{\bs{L}^{\text{diag}}}^2=\sum_{\mu<\nu}{L_{\mu\nu}}^2-{\bs{L}^{\text{diag}}}^2. 
\label{Ksqeigen}
\ee
%%%%%%%%%%%%%%%%%%%%%%%
In the $SO(3)$ Landau model \cite{Hasebe-2015},  the boost generators $K_i$ correspond to $\hat{L}_x$ and $\hat{L}_y$  (\ref{so3genepre}), and the sum of their squares gives 
%%%%%%%%%%%%%%%%%%%%
\be
{\hat{L}_x}^2+{\hat{L}_y}^2={\hat{\bs{L}}}^2-{\hat{L}_z}^2  . 
\label{lxysquares}
\ee
%%%%%%%%%%%%%%%%%%
The $SO(3)$ Landau Hamiltonian can be obtained from (\ref{lxysquares}) by replacing the eigenvalues of $\hat{L}_z$ with the $U(1)$ gauge generators ${I}/{2}$   %\cite{Hasebe-2015} 
%%%%%%%%%%%%%%%%%%%%%%%%%
\be
{\hat{L}_x}^2+{\hat{L}_y}^2~~\rightarrow~~{\hat{\bs{L}}}^2-(I/2)^2 .  
\ee
%%%%%%%%%%%%%%%%%%%%%%%
We heuristically apply this procedure to the present case and replace  $\bs{L}^D$ in (\ref{Ksqeigen}) with the gauge group generator $\bs{S}^{(I/2)}$ to obtain 
%%%%%%%%%%%%%%%%%%%%%%%
\be
\bs{K}^2~\rightarrow~\sum_{\mu<\nu}{L_{\mu\nu}}^2 - {\bs{S}^{(I/2)}}^2, 
\label{ksquareeq}
\ee
%%%%%%%%%%%%%%%%%%%%%%%
which may give  $SO(4)$ Landau Hamiltonian. We shall confirm it in Sec.\ref{sec:so4landauproblem}. 

%%%%%%%%%%%%%%%%%%%%%%%%
\subsection{$SO(4)$ Landau problem}\label{sec:so4landauproblem}
%%%%%%%%%%%%%%%%%%%%%%%%

%%%%%%%%%%%%%%%%%%%%%%%%%%%%%%%%%
\subsubsection{$SO(4)$ Landau levels and subbands}
%%%%%%%%%%%%%%%%%%%%%%%%%%%%%%%%%%%

From the  Landau Hamiltonian on $\mathbb{R}^4$ 
%%%%%%%%%%%%%%%%%%%%%%%%
\be
H=-\frac{1}{2M}\sum_{\mu=1}^4{D_{\mu}}^2=-\frac{1}{2M}  \frac{\partial^2}{\partial r^2}-\frac{3}{2M r}\frac{\partial}{\partial r} +    \frac{1}{2Mr^2}\sum_{\mu<\nu =1}^4{\Lambda_{\mu\nu}}^2, 
\ee
%%%%%%%%%%%%%%%%%%%%%%%%%%
 the $SO(4)$ Landau Hamiltonian on $S^3$ is obtained as  
%%%%%%%%%%%%%%%%%%%%%
\be
H=-\frac{1}{2M}\sum_{\mu}{D_{\mu}}^2|_{r=1}=\frac{1}{2M}\sum_{\mu<\nu}{\Lambda_{\mu\nu}}^2 ,  
\label{so4llhamil}
\ee
%%%%%%%%%%%%%%%%%%%%%%
which is  invariant under the $SO(4)$ global rotations generated by $L_{\mu\nu}$.    
Using the relation (\ref{so4casimirllamf}), we can express  (\ref{so4llhamil})  as 
%%%%%%%%%%%%%%%%%%
\be
H=\frac{1}{2M}\sum_{\mu<\nu}{L_{\mu\nu}}^2- \frac{1}{2M}\sum_{\mu<\nu}{F_{\mu\nu}}^2,   
\label{hamso4mod}
\ee
%%%%%%%%%%%%%%%%%%
where  the second term on the right-hand side is equal to the $SO(3)$ Casimir 
%%%%%%%%%%%%%%%%%%%%%
\be
\sum_{\mu<\nu} {F_{\mu\nu}}^2=\sum_{i=1}^3{S_i^{(I/2)}}^2=\frac{1}{4}I(I+2), 
\ee
%%%%%%%%%%%%%%%%%%%% 
and then 
%%%%%%%%%%%%%%%%%%%%%%
\be
H=\frac{1}{2M}(\sum_{\mu<\nu}{L_{\mu\nu}}^2- {\bs{S}^{(I/2)}}^2)=\frac{1}{M}\sum_{i=1}^3 ({L_i}^2+{\bar{L}_i}^2) - \frac{1}{2M}{\bs{S}^{(I/2)}}^2.  \label{so4hamso4}
\ee
%%%%%%%%%%%%%%%%%%%%%%
We thus verified that  (\ref{ksquareeq}) is equal to the $SO(4)$ Landau Hamiltonian up to the unimportant proportional factor. 
$H$ (\ref{so4hamso4}) is  obviously invariant under the LR transformation (\ref{partity3D}), $i.e.$, $L_i ~\leftrightarrow ~\bar{L}_i$, and    respects the LR symmetry.   
From  (\ref{so4eigenvalulllr}) or (\ref{so4casimireigenns}), we can readily derive  
the energy of the $SO(4)$ Landau Hamiltonian (\ref{so4hamso4}):   
%%%%%%%%%%%%%%%
\begin{align}
E_n(s)&=\frac{1}{M}(l_L(l_L+1)+l_R(l_R+1)) -\frac{1}{8M}I(I+2) \nn\\
&= \frac{1}{2M}(n(n+2)+\frac{I}{2}(2n+1)+s^2). 
\label{so4energynonrel}
\end{align}
%%%%%%%%%%%%%%%
Notice that the energy eigenvalues depend  both on $n$ and $s$.  While $n$ denotes the  Landau level index,  the chirality parameter $s$ corresponds to subband  of each Landau level.\footnote{The energy interval between adjacent subbands,   $\Delta E_n(s)={s}/(MR^2)$, %($s$ and $s+1$) 
   collapses  in the thermodynamic limit $I, R~\rightarrow~\infty$ with $I/(R^2)$ fixed. %(for subbands  $s <\!< I$).  
The subbands (for $s <\!< I$) thus become degenerate in the thermodynamic limit, and hence $n$ is referred to as the Landau level and $s$ subband.   The origin of subbands of Landau levels may be  accounted for by  the curvature of the sphere.  %(Also notice that the above argument cannot be applied  for subbands with $|s|~\sim ~I$.)  
} 
%%%%%%%%%%%%%%%%%%%%%%%%%%%%%%%%%%%%%%%%%%%%%%%%%%%%%%%%%%%%
\begin{figure}[tbph]
\includegraphics*[width=165mm]{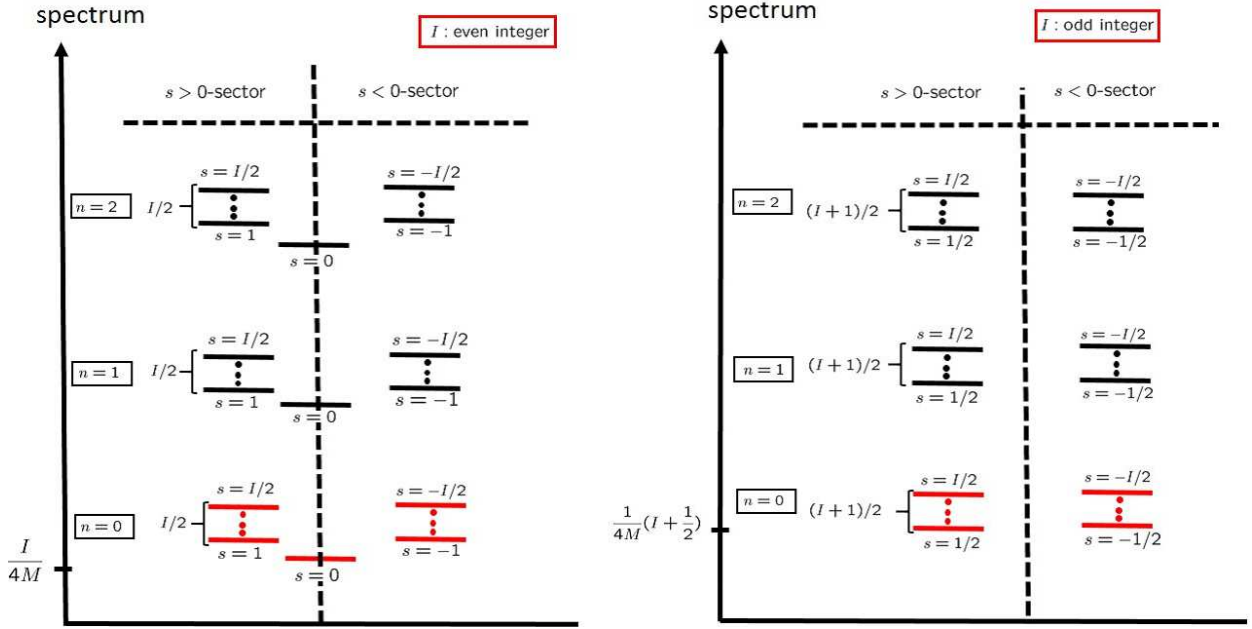}
\caption{Energy spectra of the $SO(4)$ Landau model. Each Landau level $n$ is split to subbands indexed by $s$.}
\label{sch-SO4spect.fig}
\end{figure}
%%%%%%%%%%%%%%%%%%%%%%%%%%%%%%%%%%%%%%%%%%%%%%%%%%%%%%%%%%%%%% 
Notice that the energy eigenvalue (\ref{so4energynonrel}) is invariant under the sign flip of the chirality parameter: 
%%%%%%%%%%%%%%%%%
\be
s ~~\leftrightarrow~~-s, 
\ee
%%%%%%%%%%%%%%%
which is a direct consequence of the LR symmetry, since  the $SU(2)_L$ and $SU(2)_R$ quantum numbers (\ref{preciselllrdef}) are interchanged by the LR transformation. 

The dimension of the $SO(4)$ irreducible representation  specified by $n$ and $s$ is   
%%%%%%%%%%%%%%%%%%%
\be
d_n(s)=(2l_L+1)(2l_R+1) =(n+\frac{I}{2}+1+s)(n+\frac{I}{2}+1-s). 
\label{dimso4irrpsns}
\ee
%%%%%%%%%%%%%%%%%%%%%%
Since $E_n(s)$ depends on $s^2$ (\ref{so4energynonrel}),  the $SO(4)$ eigenstates with $(n, s)$ and $(n, -s)$ are degenerate.     
The degeneracy  of the subband of the Landau level $E_n(|s|)$  for $s\neq 0$ is given by 
%%%%%%%%%%%%%%%%%%%%%%%%%%%%%%%
\be
2d_n(|s|)=2((n+\frac{I}{2}+1)^2-|s|^2). 
\ee
%%%%%%%%%%%%%%%%%%%%%%%%%%%%%%%%%%%
When $I$ is odd, $s$ can take $s=0$ and the degeneracy is  
%%%%%%%%%%%%%%%%%%%%%%%%%%
\be
d_n(s=0)=(n+\frac{I}{2}+1)^2. 
\ee
%%%%%%%%%%%%%%%%%%%%%%%%%%
A schematic picture  is given by Fig.\ref{sch-SO4spect.fig}.

%%%%%%%%%%%%%%%%%%%%%%%%%%%
\subsubsection{Landau level eigenstates}\label{sec:LLLeigenstates}
%%%%%%%%%%%%%%%%%%%%%%%%%%%%%

In \cite{Hasebe-2014-2}, we constructed the  lowest Landau level ($n=0$) basis states by taking the symmetric product of the chiral Hopf spinors. Here, we provide a precise meaning of  the construction.   
For $n=0$, the $SO(4)\simeq SU(2)_L\otimes SU(2)_R$ indices are given by  
%%%%%%%%%%%%%%%%%%%%%%%
\be
l_L=\frac{1}{2}(\frac{I}{2}+s),~~~l_R=\frac{1}{2}(\frac{I}{2}-s). 
\label{LLLrelationlandns}
\ee
%%%%%%%%%%%%%%%%%%%%%%%%%
For each of $SU(2)$, we take the fully symmetric product of the chiral Hopf spinors (\ref{su2su2bispinors}): 
%%%%%%%%%%%%%%%%%%%%%%%%
\begin{align}
&\Psi_{l_L, m_L}^{L}=\frac{1}{\sqrt{(l_L+m_L)!(l_L-m_L)!}}{\psi_{L1}}^{l_L+m_L} {\psi_{L2}}^{l_L-m_L}, \nn\\
&\Psi_{l_R, m_R}^{R}=\frac{1}{\sqrt{(l_L+m_L)!(l_R-m_R)!}}{\psi_{R1}}^{l_R+m_R} {\psi_{R2}}^{l_R-m_R},  
\end{align}
%%%%%%%%%%%%%%%%%%%%%%%% 
with $m_L=l_L, l_L-1, \cdots, -l_L$ and $m_R=l_R, l_R-1, \cdots, -l_R$, and  
 the lowest Landau level basis states are constructed as 
%%%%%%%%%%%%%%%%%
\be
\Phi_{ m_L,m_R}^{(0,s,I/2)}=\frac{1}{\sqrt{I!}}~\Psi_{l_L, m_L}^{L}\otimes \Psi_{l_R, m_R}^{R}. 
\ee
%%%%%%%%%%%%%%%

Recall that the $S^3_{D}$-fibre $\phi$ is common to $\psi_L$ and $\psi_R$ (\ref{su2su2bispinors}). As the expansion basis of $\Phi_{ m_L,m_R}^{(0,s,I/2)}$, 
we adopt the $SU(2)_D$ higher spin representation  made of $\phi$ :  
%%%%%%%%%%%%%%%%%
\be
\bs{e}^{(I/2)}_{A}=\frac{1}{\sqrt{(\frac{I}{2}+A)!(\frac{I}{2}-A)!}} {\phi_1}^{\frac{I}{2}+A} {\phi_2}^{\frac{I}{2}-A}. ~~~(A=I/2, I/2-1, \cdots, -I/2)
\label{absbasisconc}
\ee
%%%%%%%%%%%%%%%%%%
Using the basis (\ref{absbasisconc}), we   expand $\Phi_{m_L,m_R}^{(0, s,I/2)}$ as 
%%%%%%%%%%%%%%%%%%
\be
\Phi_{m_L,m_R}^{(0, s,I/2)}(x) =\sum_{A=-I/2}^{I/2}{\Phi}_{m_L,m_R}^{(0,s,I/2)}(x)_A ~ \bs{e}^{(I/2)}_{A}     
\label{defcoeffic}
\ee
%%%%%%%%%%%%%%%%
to define the expansion coefficients carrying the internal $SU(2)$ gauge index  $A$. 
In particular for $s=I/2$ and $s=-I/2$, the coefficients are obtained as 
%%%%%%%%%%%%%%%%%%%%%%
\begin{align}
&{\Phi}_{m_L,0}^{(0,I/2,I/2)}(x)_A = \Psi_L(x)_{A, m_L}=\Psi(\bs{\chi})_{A, m_L} , ~~~~~~~(m_L=\frac{I}{2}, \frac{I}{2}-1, \cdots, -\frac{I}{2})\nn\\
&{\Phi}_{0, m_R}^{(0,-I/2,I/2)}(x)_A = \Psi_R(x)_{A, m_R} =\Psi(-\bs{\chi})_{A, m_R}, ~~~(m_R=\frac{I}{2}, \frac{I}{2}-1, \cdots, -\frac{I}{2})
\end{align}
%%%%%%%%%%%%%%%%%%%%
where $\Psi(\bs{\chi})$ is the $SU(2)$ group element in the Dirac gauge (\ref{diracgaugedfunc}).  
Also from other exercises, such as 
%%%%%%%%%%%%%%%%%%%%
$\Phi^{(0, 0, 1)}_{1/2, -1/2}(x)_0= \frac{1}{\sqrt{2}} 
%\begin{pmatrix}
%\sqrt{2}~\psi_{L11} \psi_{R21} \\
(\Psi_{L 11} \Psi_{R22} + \Psi_{L12} \Psi_{R21}), %  \\
%\sqrt{2}~\psi_{L12}\psi_{R22}
%\end{pmatrix}, 
$
%%%%%%%%%%%%%%%
we can deduce a general formula for the lowest Landau level eigenstate: 
%%%%%%%%%%%%%%%%%%
\be
\Phi^{(n, s, I/2)}_{m_L, m_R}(x)_{A} =
\sum_{m'_L=-l_L}^{l_L}\sum_{m'_R=-l_R}^{l_R} \langle {I}/{2}, A|  l_L, m'_L; l_R, m'_R\rangle     \Psi^{(l_L)}(\bs{\chi})_{m'_L , m_L}  \Psi^{(l_R)}(-\bs{\chi})_{m'_R , m_R}   , 
\label{basiswithnormalcrude}
\ee
%%%%%%%%%%%%%%%% 
where $l_L$ and $l_R$ are given by (\ref{LLLrelationlandns}) and   $\langle {I}/{2}, A|  l_L, m'_L; l_R, m'_R\rangle$ denotes the  Clebsch-Gordan coefficient. 
%where we added the normalization condition. 
Replacing the relation (\ref{LLLrelationlandns}) of the lowest Landau level with that of the higher Landau level (\ref{preciselllrdef}),  we may expect that (\ref{basiswithnormalcrude})  realizes the higher Landau level basis states. (This expectation turns out to be true as we shall see below.)    
We will refer to (\ref{basiswithnormalcrude})  as the $SO(4)$ monopole harmonics.\footnote{In the  terminology, the usual monopole harmonics \cite{Wu-Yang-1976} are called the $SO(3)$ monopole harmonics, and the Yang's $SU(2)$ monopole harmonics \cite{Yang-1978-II}  will be the $SO(5)$ monopole harmonics.}

 Nair and Randjbar-Daemi gave the first derivation of the Landau level eigenstates  \cite{Nair-Daemi-2004}, in which harmonic expansion on  coset space \cite{Salam-Strathdee-1982} was applied (see \cite{Ishiki-Takayama-Tsuchiya-2006, Ishiki-Shimasaki-Takayama-Tsuchiya-2006} also).   
On the coset
%%%%%%%%%%%%%%%%%%%%
\be
S^3 \simeq SU(2)_L\otimes SU(2)_R/ SU(2)_{\text{diag}},  
\ee
%%%%%%%%%%%%%%%%%%%%
the following matrix element was considered 
%%%%%%%%%%%%%%%%%%%%%
\be
\Phi^{[l_L, l_R, I/2]}_{m_L, m_R}(x)_{A}= \langle {I}/{2}, A| g_L \otimes g_R  |l_L, m_L; l_R, m_R\rangle.     
\label{groupeleso4}
\ee
%%%%%%%%%%%%%%%%%%%%%%
Inserting the complete basis relation 
%%%%%%%%%%%%%%%%%
\be
1=\sum_{m_L=-l_L}^{l_L}\sum_{m_R=-l_R}^{l_R} |l_L, m_L; l_R, m_R\rangle  \langle l_L, m_L; l_R, m_R| 
\ee
%%%%%%%%%%%%%%%%%
to (\ref{groupeleso4}), we have\footnote{In \cite{Ishiki-Takayama-Tsuchiya-2006, Ishiki-Shimasaki-Takayama-Tsuchiya-2006}, (\ref{higherLLeigenstates}) is  referred to as the spin(-weighted) spherical harmonics on $S^3$.} 
%%%%%%%%%%%%%%%%%%%%%%%%%%%%%%%%
\begin{align}
\Phi^{(n, s, I/2)}_{m_L, m_R}(x)_{A}&=\sum_{m'_L=-l_L}^{l_L}\sum_{m'_R=-l_R}^{l_R} \langle {I}/{2}, A|  l_L, m'_L; l_R, m'_R\rangle     \langle l_L, m'_L; l_R, m'_R| g_L \otimes g_R|l_L, m_L; l_R, m_R\rangle\nn\\
&=\sum_{m'_L=-l_L}^{l_L}\sum_{m'_R=-l_R}^{l_R} \langle {I}/{2}, A|  l_L, m'_L; l_R, m'_R\rangle     D^{(l_L)}(\bs{\chi})_{m'_L , m_L}  D^{(l_R)}(-\bs{\chi})_{m'_R , m_R}, 
\label{higherLLeigenstates}
\end{align}
%%%%%%%%%%%%%%%%%%%%%%%%%%%%%%%%%%
where $D$ is the Wigner's $D$ function 
%%%%%%%%%%%%%%%%%%%%%%%%
\be
D^{(l_L)}(\bs{\chi})_{m n} =\langle l, m|g_L|l, n\rangle,~~~~~~D^{(l_R)}(-\bs{\chi})_{m n} =\langle l, m|g_R|l, n\rangle. 
\label{dfunctionformula}
\ee
%%%%%%%%%%%%%%%%%%%%%%%%%
$\Psi^{(l)}(\bs{\chi})$ corresponds to $D^{(l)}(\bs{\chi})$ in the Dirac gauge (Appendix \ref{append:gaugediracschwingerform}), and so we find (\ref{basiswithnormalcrude}) is equivalent to  (\ref{higherLLeigenstates}).   
By binding the left magnetic quantum numbers ($m'_L$ and  $m'_R$) of  two $D$ functions  
 with the Clebsch-Gordan coefficients,  we can construct the $SO(4)$ Landau level basis states with  internal magnetic quantum number ($A$) as in  (\ref{higherLLeigenstates}).  The  condition 
%%%%%%%%%%%%%%%%%%%%%%%%%%%%
\be
\int_{S^3} d\Omega_3 ~{{\Phi}^{(n, s, \frac{I}{2})}_{ m_L, m_R }(\bs{\chi})_{A}}^*~{\Phi}^{(n', s', \frac{I'}{2})}_{ m_L', m'_R }(\bs{\chi})_{A'} =\delta_{m_L m'_L}\delta_{m_R m'_R}\delta_{A A'}\delta_{ nn'}\delta_{ss'}\delta_{II'}, 
\ee
%%%%%%%%%%%%%%%%%%%%%%%%%%%%%%
determines the normalization constant  as 
%%%%%%%%%%%%%%%%%%%%%%%%
\begin{align}
&\Phi^{(n, s, I/2)}_{m_L, m_R}(x)_{A} \nn\\
&~~=\sqrt{\frac{(2l_L+1)(2l_R+1)}{2\pi^2}}\sum_{m'_L=-l_L}^{l_L}\sum_{m'_R=-l_R}^{l_R} \langle {I}/{2}, A|  l_L, m'_L; l_R, m'_R\rangle     D^{(l_L)}(\bs{\chi})_{m'_L , m_L}  D^{(l_R)}(-\bs{\chi})_{m'_R , m_R}.  
\label{basiswithnormal}
\end{align}
%%%%%%%%%%%%%%%%%%%%%%%%%%%%
Using (\ref{higherLLeigenstates}), we can construct a vector-like notation of the $SO(4)$ monopole harmonics: 
%%%%%%%%%%%%%%%%%%%%%%%%%%
\be
\bs{\Phi}^{(n, s, \frac{I}{2})}_{ m_L, m_R }(x)  =\frac{1}{\sqrt{I+1 }}
\begin{pmatrix}
\Phi^{(n, s, \frac{I}{2})}_{ m_L, m_R }(x)_{I/2} \\
\Phi^{(n, s, \frac{I}{2})}_{ m_L, m_R }(x)_{I/2 -1}\\
\vdots \\
\Phi^{(n, s, \frac{I}{2})}_{ m_L, m_R }(x)_{-{I}/2}
\end{pmatrix} , 
\label{vectorrepsorep}
\ee
%%%%%%%%%%%%%%%%%%%%%%%%%%
which satisfies 
%%%%%%%%%%%%%%%%%%%%%%%%%%%%%%%%%
\be
\int_{S^3} d\Omega_3 ~{\bs{\Phi}^{(n, s, \frac{I}{2})}_{ m_L, m_R }(x)}^{\dagger}~\bs{\Phi}^{(n', s', \frac{I'}{2})}_{ m_L', m'_R }(x) =\delta_{m_L m'_L}\delta_{m_R m'_R}\delta_{ nn'}\delta_{ss'}\delta_{II'}. 
\label{normalllso4}
\ee
%%%%%%%%%%%%%%%%%%%%%%%%%%%%% 
Since the $D$-functions depend on the $SO(4)$ Casimir indices determined only through $n+\frac{I}{2}$ (and $s$),  different $n$ and $I/2$ can give rise to same $D$ functions if  $n+\frac{I}{2}$ is fixed.  
 The Clebsch-Gordan coefficients account for their difference.

%%%%%%%%%%%%%%%%%%%%%%%%%%%%%%%%%%%%%%%%%%
\subsection{Gauge fixing analysis}\label{subsec:gaugefixanal}
%%%%%%%%%%%%%%%%%%%%%%%%%%%%%%%%%%%%%%%%%%%% 
 
 With gauge fixing, we further pursue the properties of the $SO(4)$ Landau Hamiltonian eigenstates. 
 
%%%%%%%%%%%%%%%%%%%%%%%%%%%%%%%%%%%%%%%%%%%%%%
\subsubsection{Dirac gauge and Schwinger gauge}\label{subsec:diracschgauges}
%%%%%%%%%%%%%%%%%%%%%%%%%%%%%%%%%%%%%%%%%%%%%

We first establish relations between the Dirac and the Schwinger gauges.  
In the Schwinger gauge, the $D$-function is given by 
%%%%%%%%%%%%%%%%%%
\be
\Psi_{\text{S}}^{(I/2)}(\bs{\chi})  = D^{(I/2)}(\chi, -\theta, -\phi) \equiv  e^{-i\chi S_z^{(I/2)}}e^{i\theta S_y^{(I/2)}}e^{i\phi S_z^{(I/2)}},  
\ee
%%%%%%%%%%%%%%%%%%%% 
which is related to  $\Psi_{\text{D}}^{(l)}(\bs{\chi})$ (\ref{diracgaugedfunc})   
by the $SU(2)$ gauge transformation (see Appendix \ref{append:gaugediracschwingerform} for details)
%%%%%%%%%%%%%%%%%%%%
\be
\Psi^{(l)}_{\text{D}}(\bs{\chi}) =g^{(l)}(\theta, \phi) ~\Psi^{(l)}_{\text{S}} (\bs{\chi}),  
\label{transdfuncschwin}
\ee
%%%%%%%%%%%%%%%%%%%%%%
where 
%%%%%%%%%%%%%%%%%%%%%%
\be
g^{(l)}(\theta,\phi) =e^{-i\phi S_z^{(l)}} e^{-i\theta S_y^{(l)}}.   
\label{gaugefunctionspinl}
\ee
%%%%%%%%%%%%%%%%%%%%%%
As in the case of the Dirac gauge (\ref{gaugefieldasu2}),   the $SU(2)$ gauge field in the Schwinger gauge can be expressed as  
%%%%%%%%%%%%%%%%%%%%%%%%%%%%%%%
\begin{align}
A_{\text{S}}&=-i \frac{1}{2}(\Psi_Ld{\Psi_L}^{\dagger}+\Psi_Rd{\Psi_R}^{\dagger}) & = S_x^{(I/2)} \cos\chi\sin\theta d\phi -S_y^{(I/2)} \cos\chi d\theta -S_z^{(I/2)}\cos\theta d\phi   ,  
\label{schwingergenegauge}
\end{align}
%%%%%%%%%%%%%%%%%%%
where 
%%%%%%%%%%%%%%%%%%%%%%%
\be
\Psi_L(x)=\Psi_{\text{S}}^{(I/2)}(\bs{\chi}),~~~~\Psi_R(x)\equiv \Psi_{\text{S}}^{(I/2)}(-\bs{\chi}). 
\label{mmandddschwin}
\ee
%%%%%%%%%%%%%%%%%%%%%%%
%%%%%%%%%%%%%%%%%%%%%%%%
We can read off the components of the gauge field from (\ref{schwingergenegauge}) as 
%%%%%%%%%%%%%%%%%%%%%%
\be
A_{\chi}=0,~~~~~~~A_{\theta}=-\cos\chi ~S_y^{(I/2)}, ~~~~A_{\phi}=\cos\chi \sin\theta ~S_x^{(I/2)} -\cos\theta ~S_{z}^{(I/2)}. 
\label{compsu2schw}
\ee
%%%%%%%%%%%%%%%%%%%%%%%%
For $I=1$, (\ref{compsu2schw}) is reduced to the $SU(2)$ spin connection (\ref{schexpsu2diracsphefafie}). 
From (\ref{transdfuncschwin}), one may find   
%%%%%%%%%%%%%%%%%%%%%%%%%%%%%%%
\be
A_{\text{S}} = {g^{(I/2)}}^{\dagger}A_{\text{D}} g^{(I/2)} -i{g^{(I/2)}}^{\dagger}dg^{(I/2)}. 
\label{gaugefieldtranssu2}
\ee
%%%%%%%%%%%%%%%%%%%%%%%%%%%%%%%%%
Actually this is a generalization of (\ref{omegagaugetrans}) for arbitrary spin magnitude.   
With (\ref{mmandddschwin}), in the Schwinger gauge the $SO(4)$ monopole harmonics are constructed as\footnote{Unlike the Dirac gauge  $\Psi_{\text{D}}^{(l)}(-\bs{\chi})=\Psi_{\text{D}}^{(l)}(\bs{\chi})^{-1}$, in the Schwinger gauge  $\Psi_{\text{S}}^{(l)}(-\bs{\chi})\neq \Psi_{\text{S}}^{(l)}(\bs{\chi})^{-1}$.}  
%%%%%%%%%%%%%%%%%%%%
\begin{align}
&\Phi^{(n, s, I/2)}_{m_L, m_R}(\bs{\chi})_{A} =\sqrt{\frac{(2l_L+1)(2l_R+1)}{2\pi^2}}\nn\\
&~~~~~~~~~~~~~~~~~\times\sum_{m'_L=-l_L}^{l_L}\sum_{m'_R=-l_R}^{l_R} \langle {I}/{2}, A|  l_L, m'_L; l_R, m'_R\rangle     \Psi^{(l_L)}_{\text{S}}(\bs{\chi})_{m'_L , m_L}  \Psi_{\text{S}}^{(l_R)}(-\bs{\chi})_{m'_R , m_R}. 
\label{basiswithnormalsch}
\end{align}
%%%%%%%%%%%%%%%%%%%% 
For instance, for $(n, I/2, s)=(1, 1/2, 1/2)$, the $SU(2)_L\otimes SI(2)_R$ indices are given by $(l_L, l_R)=(1, 1/2)$ and the dimension of the multiplet is $(2l_L+1)(2l_R+1)|_{(l_L, l_R)=(1, 1/2)}=6$.\footnote{In this case, the $SO(4)$ monopole harmonics (\ref{basiswithnormalsch}) are explicitly given by 
%%%%%%%%%%%%%%%%%%%%%%%%%%%%%%
\begin{align}
&\bs{\Phi}^{(1,1/2,1/2)}_{1,1/2} =i\frac{1}{\pi} e^{i\frac{3}{2}\phi}\sin\chi \sin\theta   
\begin{pmatrix}
e^{-i\frac{1}{2}\chi }\cos\frac{\theta}{2} \\
-e^{i\frac{1}{2}\chi}\sin\frac{\theta}{2}
\end{pmatrix}
,~~~\bs{\Phi}^{(1,1/2,1/2)}_{1,-1/2} =\frac{1}{\pi} e^{i\frac{1}{2}\phi}  (\cos\chi-i\sin\chi \cos\theta)
\begin{pmatrix}
e^{-i\frac{1}{2}\chi} \cos\frac{\theta}{2}  \\
-e^{i\frac{1}{2}\chi} \sin\frac{\theta}{2} 
\end{pmatrix}, 
\nn\\ 
&\bs{\Phi}^{(1,1/2,1/2)}_{0,1/2} =\frac{1}{\sqrt{2}\pi} e^{i\frac{1}{2}\phi} \begin{pmatrix}
-e^{-i\frac{1}{2}\chi} \cos\frac{\theta}{2} (e^{-i\chi} +2i\sin\chi\cos\theta) \\ 
e^{i\frac{1}{2}\chi} \sin\frac{\theta}{2} (e^{i\chi} +2i\sin\chi\cos\theta)
\end{pmatrix}, ~~\bs{\Phi}^{(1,1/2,1/2)}_{0,-1/2} =\frac{1}{\sqrt{2}\pi} e^{-i\frac{1}{2}\phi} \begin{pmatrix}
e^{-i\frac{1}{2}\chi} \sin\frac{\theta}{2} (e^{-i\chi} -2i\sin\chi\cos\theta) \\ 
e^{i\frac{1}{2}\chi} \cos\frac{\theta}{2} (e^{i\chi} -2i\sin\chi\cos\theta)
\end{pmatrix},\nn\\
&\bs{\Phi}^{(1,1/2,1/2)}_{-1,1/2} =-\frac{1}{\pi} e^{-i\frac{1}{2}\phi} (\cos\chi+i\sin\chi\cos\theta)
\begin{pmatrix}
e^{-i\frac{1}{2}\chi} \sin\frac{\theta}{2} \\
e^{i\frac{1}{2}\chi} \cos\frac{\theta}{2}
\end{pmatrix}, ~ \bs{\Phi}^{(1,1/2,1/2)}_{-1,-1/2} = -i\frac{1}{\pi} e^{-i\frac{3}{2}\phi} \sin\chi\sin\theta
\begin{pmatrix}
e^{-i\frac{1}{2}\chi}\sin\frac{\theta}{2} \\
e^{i\frac{1}{2}\chi} \cos\frac{\theta}{2}
\end{pmatrix}. 
\end{align}
%%%%%%%%%%%%%%%%%%%%%%%%%%%%%%%
}
As the gauge field undergoes the gauge transformation (\ref{gaugefieldtranssu2}), the $SO(4)$ monopole harmonics in the Dirac and the Schwinger gauges should be related as 
%%%%%%%%%%%%%%%%%%%%%
\be
\bs{\Phi}_{m_L, m_R}^{(n,s, I/2)}(\bs{\chi})_{\text{D}}=g^{({I}/{2})}(\theta, \phi) ~\bs{\Phi}_{m_L, m_R}^{(n, s, I/2)}(\bs{\chi})_{\text{S}}.   
\label{dirschphigau} 
\ee
%%%%%%%%%%%%%%%%%%%%%%
It is easy to verify (\ref{dirschphigau}) from (\ref{transdfuncschwin}) and the  property of the Clebsch-Gordan coefficient 
%%%%%%%%%%%%%%%%%%%%%%%
\be
\langle {I}/{2}, A|  l_L, m'_L; l_R, m'_R\rangle  ~ g^{(l_L)}(\theta, \phi)_{m'_L m_L}~ g^{(l_R)}(\theta, \phi)_{m'_R m_R} = g^{(I/2)}(\theta, \phi)_{A B} ~\langle {I}/{2}, B|  l_L, m_L; l_R, m_R\rangle .  
\ee
%%%%%%%%%%%%%%%%%%%%%%%

%%%%%%%%%%%%%%%%%%%%%%%%%%%%%%%
\subsubsection{Properties of the $SO(4)$ monopole harmonics }\label{subsec:so4monoprop}
%%%%%%%%%%%%%%%%%%%%%%%%%%%%%%%%%

%%%%%%%%%%%%%%%%%%%%%%%%%%%%%%%
%\subsubsection{Complex conjugation and integration formulae}\label{subsec:compintmono}
%%%%%%%%%%%%%%%%%%%%%%%%%%%%%%%%%

%Several properties of the $SO(4)$ monopole harmonics are as follows.   
From the properties of the $D$ function and the Clebsch-Gordan coefficients 
%%%%%%%%%%%%%%%%%%%%%%%%%%%%%%%%%%%%
\be
{D^{(l)}({\chi}, \theta, \phi)_{m  ,n}}^* = (-1)^{m-n} D^{(l)}({\chi}, \theta, \phi)_{-m, -n}, ~~~~~C^{J, -M}_{l, -m; ~l', -m'} = (-1)^{l+l'+J+2M} C^{J, M}_{l, m; ~l', m'}, 
\ee
%%%%%%%%%%%%%%%%%%%%%%%%%%%%%%%%%%%
the complex conjugate of the $SO(4)$ monopole harmonics is given by 
%%%%%%%%%%%%%%%%%%%%%%%%%%%%%%%%
\be
{\Phi^{(n, s, I/2)}_{m_L, m_R}(\bs{\chi})_{A}}^*=(-1)^{l_L+l_R+\frac{I}{2}-m_L-m_R-A} ~\Phi^{(n, s, I/2)}_{-m_L, -m_R}(\bs{\chi})_{-A}|_{l_L=\frac{1}{2}(n+\frac{I}{2}+s), l_R=\frac{1}{2}(n+\frac{I}{2}-s) }.  
\label{compconjso4mo}
\ee
%%%%%%%%%%%%%%%%%%%%%%%%%%%%%%
Notice that the complex conjugate flips both of the $SO(4)$ magnetic quantum numbers, $m_L$ and $m_R$. 
Integration of the product of three $SO(4)$ monopole harmonics is given by  (see Appendix \ref{append:integral})  
%%%%%%%%%%%%%%%%%%%%%%%%%%%%%%%%
\begin{align}
&\frac{1}{I+1}\int_{S^3} d\Omega_3~ (\sum_{A=-\frac{I}{2}}^{\frac{I}{2}}~\Phi_{m_L,m_R}^{[l_L,l_R,\frac{I}{2}]}(\bs{\chi})_A^*\cdot \Phi^{[\frac{p}{2}, \frac{p}{2}, 0]}_{m'_L, m'_R}(\bs{\chi})\cdot {\Phi_{m''_L,m''_R}^{[l_L,l_R,\frac{I}{2}]}(\bs{\chi})}_A)\nn\\
&=\sqrt{\frac{(p+1)(2l_L+1)(2l_R+1)}{2\pi^2}}~(-1)^{-(l_L+l_R+\frac{I}{2}+\frac{p}{2})}~\begin{Bmatrix}
l_L & l_R & \frac{I}{2} \\
l_R & l_L & \frac{p}{2}
\end{Bmatrix}~C_{\frac{p}{2}, m'_L~;~l_L, m''_L}^{l_L, m_L}~C_{\frac{p}{2}, m'_R~;~l_R, m''_R}^{l_R, m_R}, 
\label{threeintegralso4lleigen}
\end{align}
%%%%%%%%%%%%%%%%%%%%%%%%%%%%%%%
and 
%%%%%%%%%%%%%%%%%%%%%%%%%%%%%%%%%%%%
\begin{align}
&\frac{1}{I+1}\int_{S^3} d\Omega_3~ (\sum_{A=-\frac{I}{2}}^{\frac{I}{2}}~\Phi_{m_L,m_R}^{[l_L,l_R,\frac{I}{2}]}(\bs{\chi})_A^*\cdot \Phi^{[\frac{p}{2}, \frac{p}{2}, 0]}_{m'_L, m'_R}(\bs{\chi})\cdot {\Phi_{m''_R,m''_L}^{[l_R,l_L,\frac{I}{2}]}(\bs{\chi})}_A)\nn\\
&=\sqrt{\frac{(p+1)(2l_L+1)(2l_R+1)}{2\pi^2}}~(-1)^{2l_L+\frac{3}{2}I+\frac{3}{2}p }~\begin{Bmatrix}
l_L & l_R & \frac{I}{2} \\
l_L & l_R & \frac{p}{2}
\end{Bmatrix}~C_{\frac{p}{2}, m'_L~;~l_R, m''_R}^{l_L, m_L}~C_{\frac{p}{2}, m'_R~;~l_L, m''_L}^{l_R, m_R}.  \label{2threeintegralso4lleigen}
\end{align}
%%%%%%%%%%%%%%%%%%%%%%%%%%%%%%%%%%%%%%
where 
%%%%%%%%%%%%%%%%%%%%%%%%%%
\be
l_L=\frac{1}{2}(n+\frac{I}{2})+\frac{1}{2}s,~~~~~~~l_R=\frac{1}{2}(n+\frac{I}{2})-\frac{1}{2}s,  
\label{lllrnsres}
\ee
%%%%%%%%%%%%%%%%%%%%%%%%%%%%%%
and $\{\cdots\}$ on the right-hand side is the 6-j symbol : 
%%%%%%%%%%%%%%%%%%%%%%%%%
\begin{align}
&\begin{Bmatrix}  
j_1 & j_2 & j_3 \\
j_4 & j_5 & j_6
\end{Bmatrix} \nn\\
&\equiv \sum_{\text{all} ~m} (-1)^{\sum_j (j_i-m_i)} \begin{pmatrix}
j_1 & j_2 & j_3 \\
-m_1 & -m_2 & -m_3
\end{pmatrix}
\begin{pmatrix}
j_1 & j_5 & j_6 \\
-m_1 & -m_5 & m_6
\end{pmatrix}
\begin{pmatrix}
j_4 & j_2 & j_6 \\
m_4 & m_2 & -m_6
\end{pmatrix}
\begin{pmatrix}
j_4 & j_5 & j_3 \\
-m_4 & -m_5 & m_3
\end{pmatrix}.  
\end{align}
%%%%%%%%%%%%%%%%%%%%%%%%
%(\ref{threeintegralso4lleigen})  is a special case of the general formula Eq.(3.11) of \cite{Ishiki-Takayama-Tsuchiya-2006}
The integration formula (\ref{threeintegralso4lleigen}) will be crucial to derive the matrix geometry of the $SO(4)$ Landau level in Sec.\ref{sec:matrixgeo}.

%%%%%%%%%%%%%%%%%%%%%%%%%%%%%%%
\subsubsection{$SO(4)$ covariance}\label{subsec:so4covari}
%%%%%%%%%%%%%%%%%%%%%%%%%%%%%%%%%
 
 The $SO(4)$ covariance of the $SO(4)$  monopole harmonics is essential for the monopole harmonics to be the eigenstates of the $SO(4)$ Landau Hamiltonian.  The gauge fixing indeed allows us to demonstrate the covariance of the  $SO(4)$ monopole harmonics. 

 In the Dirac gauge,   the angular momentum operators  are represented as (\ref{lisu2diracex})  and the $SO(4)$ monopole harmonics are  (\ref{vectorrepsorep}) with (\ref{basiswithnormalcrude}). Using these, we can explicitly show\footnote{We checked the validity of (\ref{covsu2su2}) for low dimensional representations by using formula manipulation system,   Mathematica.    For the chiral Hopf spinors  $\bs{\psi}_{L\alpha}=\bs{\Phi}^{(0,1/2,1/2)}_{\alpha, 0}$ and $\bs{\psi}_{R\alpha}=\bs{\Phi}^{(0,-1/2,1/2)}_{0, \alpha}$ $(\alpha=1,2)$, we can check it by hand:  
%%%%%%%%%%%%%%%%%%%%%%%
\begin{align}
&(l_L, l_R)=(1/2, 0)~:~L_i\bs{\psi}_{L\alpha}=\bs{\psi}_{L\beta}\frac{1}{2}(\sigma_i)_{\beta \alpha},~~~ \bar{L}_i\bs{\psi}_{L\alpha}=0 , \nn\\
&(l_L, l_R)=(0, 1/2)~:~{L}_i\bs{\psi}_{R\alpha}=0 ,~~~~~~~~~~~~~~~~~~ \bar{L}_i\bs{\psi}_{R\alpha}=\bs{\psi}_{R\beta}\frac{1}{2}(\sigma_i)_{\beta \alpha}. 
\label{SO(4)coveigen}
\end{align}
%%%%%%%%%%%%%%%%%%%%%%%%%
$\bs{\psi}_{L\alpha}$ and $\bs{\psi}_{R\alpha}$ are related to (\ref{largepsis}) as 
%%%%%%%%%%%%%%%%%%%
\be
\Psi_L^{(1/2)}=(\bs{\psi}_{L1}~~\bs{\psi}_{L2}),~~~~\Psi_R^{(1/2)}=(\bs{\psi}_{R1}~~\bs{\psi}_{R2}).
\ee
%%%%%%%%%%%%%%%%%%%
Generally for $\Psi_L$ and $\Psi_R$, 
%%%%%%%%%%%%%%%%%%%%%%
\begin{align}
&(l_L, l_R)=(I/2, 0)~:~L_i\Psi_{L}=\Psi_{L}S_i^{(I/2)},~~~ \bar{L}_i\Psi_{L}=0 , \nn\\
&(l_L, l_R)=(0, I/2)~:~{L}_i\Psi_{R}=0 ,~~~~~~~~~~~~~ \bar{L}_i \Psi_{R}=\Psi_{R} S_i^{(I/2)}. 
\end{align}
%%%%%%%%%%%%%%%%%%%%%%%
} 
%%%%%%%%%%%%%%%%%%%%%%%%%%
\be
L_i \bs{\Phi}_{m_L, m_R}^{(n, s, \frac{I}{2})} = \sum_{m'_L=-l_L}^{l_L}\bs{\Phi}_{m'_L, m_R}^{(n, s, \frac{I}{2})} (S_i^{(l_L)})_{m'_L m_L}, ~~~\bar{L}_i \bs{\Phi}_{m_L, m_R}^{(n, s, \frac{I}{2})} = \sum_{m'_R=-l_R}^{l_R}\bs{\Phi}_{m_L, m'_R}^{(n, s, \frac{I}{2})} (S_i^{(l_R)})_{m'_R m_R}, 
\label{covsu2su2}
\ee
%%%%%%%%%%%%%%%%%%%%%%%% 
or more concisely,
%%%%%%%%%%%%%%%%%%%%%%%%%%%%%%%
\be 
L_{i}\bs{\Phi}^{(n, s, \frac{I}{2})} =  {(S_i^{(l_L)})}^t\bs{\Phi}^{(n, s, \frac{I}{2})},~~~~~ \bar{L}_{i}\bs{\Phi}^{(n, s, \frac{I}{2})} =\bs{\Phi}^{(n, s, \frac{I}{2})} S_i^{(l_R)},  
\label{llbarphisact}
\ee
%%%%%%%%%%%%%%%%%%%%%%%%%%%%%%
which manifests that the $SO(4)$ monopole harmonics are the irreducible representation of the $SU(2)_L$ and $SU(2)_R$. 
From 
%%%%%%%%%%%%%%
\be
L_{\mu\nu}=\eta_{\mu\nu}^iL_i+\bar{\eta}_{\mu\nu}^i \bar{L}_i, 
\ee
%%%%%%%%%%%%%%%
we have\footnote{For $M=(m_L, m_R)$ or 
%%%%%%%%%%%%%%%%%%
\be
\bs{\Phi}_{M}^{(n, s, \frac{I}{2})}\equiv  \bs{\Phi}_{m_L, m_R}^{(n, s, \frac{I}{2})}, 
\ee
%%%%%%%%%%%%%%%%% 
the $SO(4)$ transformation (\ref{so4transgenephi}) can be concisely expressed as 
%%%%%%%%%%%%%%%%%%%%%%
\be
L_{\mu\nu} \bs{\Phi}_{M}^{(n, s, \frac{I}{2})}=  \sum_{N =1}^{d(n, s)}\bs{\Phi}_{N}^{(n, s, \frac{I}{2})}(\Sigma_{\mu\nu})_{N M} , 
\ee
%%%%%%%%%%%%%%%%%%%%%
where $\Sigma_{\mu\nu}$ are the  $SO(4)$  generators of the $(l_L, l_R)$ representation: 
%%%%%%%%%%%%%%%%%%%
\be
\Sigma_{\mu\nu} =\eta_{\mu\nu}^i S_i^{(l_L)}\otimes \bs{1} +\bar{\eta}_{\mu\nu}^i \bs{1}\otimes S^{(l_R)}_{i}. 
\ee
%%%%%%%%%%%%%%%%%%%%%
} 
%%%%%%%%%%%%%%%%%%%%%%%%%%%
\be
L_{\mu\nu}\bs{\Phi}^{(n, s, \frac{I}{2})} = \eta_{\mu\nu}^i {(S_i^{(l_L)})}^t\bs{\Phi}^{(n, s, \frac{I}{2})}  + \bar{\eta}_{\mu\nu}^i\bs{\Phi}^{(n, s, \frac{I}{2})} S_i^{(l_R)}, \label{so4transgenephi}
\ee
%%%%%%%%%%%%%%%%%%%%%%%%%%% 
and then  
%%%%%%%%%%%%%%%%%%%%%%%%%%%%
\begin{align}
\sum_{\mu<\nu}{L_{\mu\nu}}^2 \bs{\Phi}_{m_L, m_R}^{(n, s, \frac{I}{2})}& = 
2\bs{\Phi}_{m'_L, m_R}^{(n, s, \frac{I}{2})} {(S_i^{(l_L)})^2}_{m'_L m_L} + 2\bs{\Phi}_{m_L, m'_R}^{(n, s, \frac{I}{2})} {(S_i^{(l_R)})^2}_{m'_R m_R} \nn\\
&=2(l_L(l_L+1) +l_R(l_R+1))  \bs{\Phi}_{m'_L, m_R}^{(n, s, \frac{I}{2})}. 
\end{align}
%%%%%%%%%%%%%%%%%%%%%%%%%%%%
%where we used 
%%%%%%%%%%%%%%%%%%%%%%%%%%%
%\begin{align}
%&\eta_{\mu\nu}^i \eta_{\mu\nu}^j =\bar{\eta}_{\mu\nu}^i \bar{\eta}_{\mu\nu}^j =4\delta_{ij},~~~~~~~\eta_{\mu\nu}^i \bar{\eta}_{\mu\nu}^j =0, \nn\\
%&({S_i^{(l)}}^2)_{mm'}=l(l+1)\delta_{mm'}.
%\end{align}
%%%%%%%%%%%%%%%%%%%%%%%%%%%%
Obviously,  $ \bs{\Phi}_{m'_L, m_R}^{(n, s, \frac{I}{2})}$ are the Landau level eigenstates with the $SO(4)$ index (\ref{lllrnsres}).

%%%%%%%%%%%%%%%%%%%%%%%%%%%%%%%%%%%%%%
\subsection{Reduction to the $SO(4)$ spherical harmonics}\label{subsec:redso4harmo}
%%%%%%%%%%%%%%%%%%%%%%%%%%%%%%%%%%%%

We can check that the $SO(4)$ monopole harmonics are reduced to the $SO(4)$ spherical harmonics in the free $SU(2)$ background  limit. % We check this in the following. 
In literature \cite{Biedenharn1961,Domokos-1967,Hochstadt-book}, the $SO(4)$ spherical harmonics is usually given by (see Appendix \ref{sphericalharmoso4})
%%%%%%%%%%%%%%%%%%%
\begin{align}
Y_{n l m}(\bs{\chi}) &=2^l l!\sqrt{\frac{2(n+1)(n-l)!}{\pi(n+l+1)!}}~\sin^l(\chi)~C_{n-l}^{l+1}(\cos\chi) \cdot Y_{lm}(\theta,\phi), \label{so4litspheharmo}\\
&~~~~~~~~~~~~~(l=0,1,2,\cdots, n~\text{and}~m=-l, -l+1, \cdots, l) \nn
\end{align}
%%%%%%%%%%%%%%%%%%%%
where $Y_{lm}$ are the $SO(3)$ spherical harmonics and $C_{n-l}^{l+1}$ are the Gegenbauer polynomials: 
%%%%%%%%%%%%%%%%%%%%%%%%
\be
C_{n}^{\alpha} (x) \equiv 
 \frac{(-2)^n}{n!} \frac{\Gamma(n+\alpha) \Gamma(n+2\alpha)}{\Gamma(\alpha)\Gamma(2n+2\alpha)} (1-x^2)^{-\alpha+\frac{1}{2}}\frac{d^n}{dx^n} [(1-x^2)^{n+\alpha-\frac{1}{2}}].
\ee
%%%%%%%%%%%%%%%%%%%%%%%% 
The $SO(4)$ spherical harmonics 
(\ref{so4litspheharmo}) carry the $SU(2)_L\otimes SU(2)_R$ index   
%%%%%%%%%%%%%%%%%%%%%%%
\be
(l_L, l_R) =(\frac{n}{2}, \frac{n}{2}). 
\ee
%%%%%%%%%%%%%%%%%%%%
Meanwhile, the $SO(4)$ monopole harmonics in the Dirac gauge (\ref{basiswithnormal}) yield  $(n/2, n/2)$ representation  as\footnote{ 
In the  Schwinger gauge (\ref{basiswithnormalsch}), we have 
%%%%%%%%%%%%%%%%%%%%%%%%%%%%%
\be
\Phi_{(\text{S})m_L, m_R}^{(n,0,0)}(x) 
=(-i)^n\sqrt{\frac{n+1}{4\pi^2}}\sum_{m=-{n}/{2}}^{n/2} i^{2m} \Psi_{(\text{S})}^{(n/2)} (\bs{\chi} )_{m, m_L}
\Psi^{(n/2)} _{(\text{S})}(-\bs{\chi})_{-m, m_R}, 
\label{schgaugesphericalharmo} 
\ee
%%%%%%%%%%%%%%%%%%%%%%%%%%%%%%%
which is equal to (\ref{diragaugesphericalharmo}): 
%%%%%%%%%%%%%%%%%%%%%%%
\be
\Phi_{(\text{S})m_L, m_R}^{(n,0,0)}(x) =\Phi_{m_L, m_R}^{(n,0,0)}(x).
\ee
%%%%%%%%%%%%%%%%%%%%%%%%
The $SO(4)$ free angular momentum operator acts to $\Phi_{m_L, m_R}^{(n,0,0)} (\bs{\chi})$ as 
%%%%%%%%%%%%%%%%%%%%%%%%
\be
l_{\mu\nu} \Phi_{m_L, m_R}^{(n,0,0)} (x) = \eta_{\mu\nu}^i \sum_{m'_L=-n/2}^{n/2}\Phi_{m'_L, m_R}^{(n,0,0)} (x) (S_i^{(n/2)})_{m'_L m_L}+\bar{\eta}_{\mu\nu}^i \sum_{m'_R=-n/2}^{n/2}\Phi_{m_L, m'_R}^{(n,0,0)} (x) (S_i^{(n/2)})_{m'_R m_R}.
\ee
%%%%%%%%%%%%%%%%%%%%%%%%%% 
} 
%%%%%%%%%%%%%%%%%%%%%%%%%%%%
\begin{align}
\Phi_{m_L, m_R}^{(n,0,0)} (x) &=\frac{n+1}{\sqrt{2\pi^2}}\sum_{m'_L, m'_R =-{n}/{2}}^{n/2} 
\langle 0,0 |\frac{n}{2}, m'_L;  \frac{n}{2}, m'_R\rangle \Psi^{(n/2)}  (\bs{\chi})_{m'_L m_L} \Psi^{(n/2)}  (-\bs{\chi})_{m'_R m_R}\nn\\
&=(-i)^n\sqrt{\frac{n+1}{2\pi^2}}\sum_{m=-{n}/{2}}^{n/2} i^{2m} \Psi^{(n/2)} (\bs{\chi} )_{m, m_L}
\Psi^{(n/2)} (-\bs{\chi})_{-m, m_R} , \label{diragaugesphericalharmo}
\end{align}
%%%%%%%%%%%%%%%%%%%%%%%%%%%%%%
where we used 
%%%%%%%%%%%%%%%%%%%%%%%%%%%%%%%%%%%%%%%%%%%
\be
\langle 0,0 |\frac{n}{2}, m'_L;  \frac{n}{2}, m'_R\rangle =(-i)^n \frac{1}{\sqrt{n+1}} i^{2m'_L}\delta_{m'_L+m'_R, 0}.
\ee
%%%%%%%%%%%%%%%%%%%%%%%%%%%%%%%%%%%%%%%%
For instance, 
%%%%%%%%%%%%%%%%%%%%%%%%%%%
\begin{align}
n=1~:~\Phi_{m_L, m_R}^{(n=1,0,0)} (\bs{\chi})&=\frac{1}{\pi}\begin{pmatrix}
   i\sin\chi \sin\theta e^{i\phi} &                \cos\chi -i\sin\chi\cos\theta  \\
- \cos\chi -i\sin\chi\cos\theta   &     -i\sin\chi \sin\theta e^{-i\phi}              
\end{pmatrix}_{m_L, m_R}\nn\\
&=\frac{1}{\pi}
\begin{pmatrix}
ix_1-x_2 & x_4 -ix_3 \\
-x_4 -ix_3 & -ix_1-x_2
\end{pmatrix}_{m_L, m_R}.  \label{xsandphin1}
\end{align}
%%%%%%%%%%%%%%%%%%%%%%%%%%%%%
The Clebsch-Gordan coefficients simply relate the two superficially different expressions of the $SO(4)$ spherical harmonics, %of $(n/2, n/2)$ representations %
 (\ref{so4litspheharmo}) and (\ref{diragaugesphericalharmo}),  
as   
%%%%%%%%%%%%%%%%%%%%%%%%
\be
Y_{nlm}(x)=i^l\sum_{m_L, m_R=-n/2}^{n/2} \langle l,m| \frac{n}{2}, m_L; \frac{n}{2}, m_R\rangle  ~\Phi_{m_L, m_R}^{(n,0,0)} (x). 
\ee
%%%%%%%%%%%%%%%%%%%%%%%%%%%
The $SO(4)$ monopole harmonics are thus reduced to  the $SO(4)$ spherical harmonics in the free background limit.

%%%%%%%%%%%%%%%%%%%%%%%%%%%%%%%%
\section{Spinor Landau Model}\label{sec:spinorlandau}
%%%%%%%%%%%%%%%%%%%%%%%%%%%%%%%%

Before going to the analysis of relativistic Landau models,  we investigate the spinor $SO(4)$ Landau model whose  Hamiltonian is the square of a relativistic Landau operator. 
The analysis of the spinor Landau model is a preliminary step to more complicated  relativistic Landau models, but the spinor Landau model has importance  in its own right. 
%The eigenstates of the relativistic Landau operator eigenstates are automatically those of the spinor Landau operator, and inversely 
%Appropriate linear combinations of the eigenstates of the spinor $SO(4)$ Landau model constitute the relativistic eigenstates, and we first discuss the eigenvalue problem of the spinor Landau model in this section.  
The spinor Landau model includes a synthesized connection of the spin and  the gauge connections just like the relativistic Landau models. %We systematically analyze such Landau models with synthetic connection. 
In Sec.\ref{subsec:syntheconn} we discuss  special properties of the synthesized connection to present a basic idea to solve the eigenvalue problem.  Based on the observation, we introduce the $SO(4)$ angular momentum operators in Sec.\ref{subsec:synthesizedang} and explicitly solve the eigenvalue problem in Sec.\ref{subsec:spinorlandaumodeleigen}.  We discuss the physics described by the spinor $SO(4)$ Landau model in Sec.\ref{subsec:intspinorham}.  

%The spinor Landau model contains a synthesized connection of the spin and  the gauge connections just as in the relativistic Landau models. We demonstrate how we can solve the eigenvalue problem  of the Landau models with synthesized connection.   The physical meaning of the spinor Landau model is also clarified. 

%%%%%%%%%%%%%%%%%%%%%%%%%%%%%%%%%%%
\subsection{Synthesized connection}\label{subsec:syntheconn}
%%%%%%%%%%%%%%%%%%%%%%%%%%%%%%%%%%%%%

As mentioned in Sec.\ref{sec:nonrelaLandaumodel}, the components of the spin connection  and the $SU(2)$ gauge field are exactly equivalent and their difference is just the representation of the $SU(2)$ generators: 
%%%%%%%%%%%%%%%%%%%%%%%%%
\be
\omega_{\mu}=\omega_{\mu}^i~\frac{1}{2}\sigma_i, ~~~~~A_{\mu} =\omega_{\mu}^i ~S_i^{(I/2)}. 
\ee
%%%%%%%%%%%%%%%%%%%%%%%%%%%
%where, in the Dirac gauge, 
%%%%%%%%%%%%%%%%
%\be
%\omega_{\text{D}}^i=-\frac{1}{1+x_4}\epsilon_{ijk}x_j dx_k. 
%\ee
%%%%%%%%%%%%%%%%%%
The synthesized connection of the spin connection and the gauge field is:\footnote{
In the local coordinates on $S^3$, the synthesized connection is given by   
%%%%%%%%%%%%%%%%%%%%%%%%%%%%%%
\be
\mathcal{A}_{\alpha}=\omega_{\alpha} \otimes \bs{1}_{I+1} +\bs{1}_{2}\otimes A_{\alpha}^{(I/2)}=\omega_{\alpha}^i(\frac{1}{2}\sigma_i\otimes \bs{1}_{I+1} +\bs{1}_2\otimes S_i^{(I/2)}).    ~~~~~(\alpha=\chi, 
\theta, \phi)
\ee
%%%%%%%%%%%%%%%%%%%%%%%%%%%%%%
}    
%%%%%%%%%%%%%%%%%%%%%%%%
\be
\mathcal{A}_{\mu}=\omega_{\mu}\otimes \bs{1}_{I+1} +\bs{1}_{2}\otimes A_{\mu}^{(I/2)}=\omega_{\mu}^i(\frac{1}{2}\sigma_i\otimes \bs{1}_{I+1} +\bs{1}_2\otimes S_i^{(I/2)}).   ~~~~~~(\mu=1,2,3,4) 
\label{synthegaufiel}
\ee
%%%%%%%%%%%%%%%%%%%%%%%%%%%
Notice that in the present model, the  common factor  $\omega_{\mu}^i$  can be extracted in front of the synthesized $SU(2)$ generators.  We can then decompose the  $SU(2)$ representations to two direct sum of the two irreducible representations.    
The  generators of the synthesized $SU(2)$ gauge group are 
%%%%%%%%%%%%%%%%%%%%%%%%%%
\be
\frac{1}{2}\sigma_i \otimes \bs{1}_{I+1} +\bs{1}_2\otimes S_i^{(I/2)},  
\label{synthegene}
\ee
%%%%%%%%%%%%%%%%%%%%%%%%
which are irreducibly  decomposed as 
%%%%%%%%%%%%%%%%%%%%%
\be
\frac{1}{2}~\otimes ~\frac{I}{2} =   J^+  (\equiv  \frac{I}{2}+\frac{1}{2}) ~\oplus ~  J^-  (\equiv  \frac{I}{2}-\frac{1}{2}) 
\ee
%%%%%%%%%%%%%%%%%%%%%%%
or more explicitly 
%%%%%%%%%%%%%%%%%%%%%%%%
\be
\frac{1}{2}\sigma_i \otimes \bs{1}_{I+1} + \bs{1}_2\otimes S_i^{(I/2)}~~{\Longrightarrow}~~U^{\dagger} (\frac{1}{2}\sigma_i \otimes \bs{1}_{I+1} + \bs{1}_2\otimes S_i^{(I/2)})
   U=\begin{pmatrix}
S_i^{(J^+)} & 0 \\
0 & S_i^{(J^-) }
\end{pmatrix}.  
\ee
%%%%%%%%%%%%%%%%%%%%%%
$U$ denotes  $2(I+1)\times 2(I+1)$ unitary matrix constructed by the Clebsch-Gordan coefficients: 
%%%%%%%%%%%%%%%%%%%%%%%%%%%
\be
U=\begin{pmatrix}
 {C^+} & {C^-} 
\end{pmatrix} 
\equiv \begin{pmatrix}
C^{J^+, A^+}_{\frac{1}{2}, \frac{1}{2}\sigma~;~\frac{I}{2}, m}        &    C^{J^-, A^-}_{\frac{1}{2}, \frac{1}{2}\sigma~;~\frac{I}{2}, m} 
\end{pmatrix}. 
 \label{cgcoefficients}
\ee
%%%%%%%%%%%%%%%%%%%%%%%%%
The column is specified by the index $(\sigma, m)$ $(\sigma=+, -,~m={I}/{2}, {I}/{2}-1, \cdots, -{I}/{2})$ and the row by $A^+(=J^+, J^+-1, \cdots, -J^+)$  ,  $A^-(=J^-, J^--1, \cdots, -J^{-})$.\footnote{For instance $I/2=1/2$, we have 
%%%%%%%%%%%%%%%%%%%%%%
\be
U=\frac{1}{\sqrt{2}}
\left(
\begin{array}{ccc:c}
\sqrt{2} & 0 & 0 & 0 \\
0 & 1 & 0 & 1 \\
0 & 1 & 0 & -1 \\
0 & 0 & \sqrt{2} & 0 
\end{array}
\right).
\ee
%%%%%%%%%%%%%%%%%%%%%%%%|
}
 $C^+$ and $C^-$ represent  $2(I+1)\times (I+2)$ and $2(I+1)\times I$ rectangular matrices, respectively. 
The unitary transformation (\ref{cgcoefficients}) decomposes the synthesized $SU(2)$ connection space to the direct sum of the two $SU(2)$ spaces of spin  $J^+$ and $J^-$, such as  
%%%%%%%%%%%%%%%%%%%%%
\be
U^{\dagger}\mathcal{A}_{\mu}U=\begin{pmatrix}
A_{\mu}^{(J^+)} & 0 \\
0 & A_{\mu}^{(J^-)}
\end{pmatrix}. 
%, ~~~~~~U^{\dagger}\mathcal{D}_{\mu} U =\begin{pmatrix}
%D_{\mu}^{(J^+)} & 0 \\
%0 & D_{\mu}^{(J^-)}
%\end{pmatrix}. 
\ee
%%%%%%%%%%%%%%%%%%%%%%  

As usual, we construct the covariant derivatives for the synthesized connection:  
%%%%%%%%%%%%%%%%%
\be
\mathcal{D}_{\mu}=\partial_{\mu}+i\mathcal{A}_{\mu} \equiv \partial_{\mu}+i\omega_{\mu} \otimes \bs{1} +i\bs{1}\otimes A_{\mu},  
\ee
%%%%%%%%%%%%%%%%%
and the field strength:\footnote{
More concisely, 
%%%%%%%%%%%%%%%%%%%
\be
\mathcal{F}=\frac{1}{2}\mathcal{F}_{\mu\nu} dx^{\mu}\wedge dx^{\nu} =\frac{1}{2}\epsilon_{ijk}e^i\wedge e^j (\frac{1}{2}\sigma_k\otimes \bs{1}_{I+1} +\bs{1}_2\otimes S_k^{(I/2)}). 
\ee
%%%%%%%%%%%%%%%%%%%
} 
%%%%%%%%%%%%%%%%%%%%%%%%
\be
\mathcal{F}_{\mu\nu}=-i[\mathcal{D}_{\mu}, \mathcal{D}_{\nu}]=f_{\mu\nu}\otimes \bs{1}+\bs{1}\otimes F_{\mu\nu}=f_{\mu\nu}^i (\frac{1}{2}\sigma_i\otimes \bs{1}_{I+1} +\bs{1}_2\otimes S_i^{(I/2)}), 
\ee
%%%%%%%%%%%%%%%%%%%%%%%%%%%
where 
%%%%%%%%%%%%%%%%%%
\be
f_{\mu\nu}=\partial_{\mu}\omega_{\nu}-\partial_{\nu}\omega_{\mu}+i[\omega_{\mu}, \omega_{\nu}].
\ee
%%%%%%%%%%%%%%%  
They are also block diagonalized as 
%%%%%%%%%%%%%%%%%%%%%
\be
U^{\dagger}\mathcal{D}_{\mu} U =\begin{pmatrix}
D_{\mu}^{(J^+)} & 0 \\
0 & D_{\mu}^{(J^-)}
\end{pmatrix}, ~~~~~~
U^{\dagger}\mathcal{F} U=\begin{pmatrix}
F^{(J^+)} & 0 \\
0 & F_{\mu}^{(J^-)}
\end{pmatrix}. 
\ee
%%%%%%%%%%%%%%%%%%%%%%
The spinor Landau model is thus decomposed to the direct sum of two $SO(4)$ Landau models of  $SU(2)$ gauge fields with $SU(2)$ index  $J^+$ and $J^-$, and the eigenvalue problem is boiled down to those  of the $SO(4)$ Landau models with $J^+$ and $J^-$ sectors. Therefore, to solve the eigenvalue problem, what we need to do  is  just to apply the method of Sec.\ref{sec:nonrelaLandaumodel} to each of the sectors.   

%%%%%%%%%%%%%%%%%%%%%%%%%%%%%%%%%%%
\subsection{$SO(4)$ synthesized angular momentum}\label{subsec:synthesizedang}
%%%%%%%%%%%%%%%%%%%%%%%%%%%%%%%%%%%%%

As the gauge connection is simply replaced with the synthesized connection, we introduce the synthesized $SO(4)$  angular momentum  operators as   
%%%%%%%%%%%%%%%%%%%%%
\be
\mathit{\Lambda}_{\mu\nu} = -ix_{\mu}\mathcal{D}_{\nu} + ix_{\nu}\mathcal{D}_{\mu},~~~~
\mathcal{L}_{\mu\nu}=\mathit{\Lambda}_{\mu\nu} +\mathcal{F}_{\mu\nu} . \label{totalso4ang}
\ee
%%%%%%%%%%%%%%%%%%%%%%%
By the unitary transformation (\ref{cgcoefficients}),  they are transformed to the block-diagonalized forms:  
%%%%%%%%%%%%%%%%%%%%
\be
U^{\dagger} \mathit{\Lambda}_{\mu\nu} U =\begin{pmatrix}
\Lambda_{\mu\nu}^{(J^+)} & 0 \\
0 & \Lambda_{\mu\nu}^{(J^-)} 
\end{pmatrix},~~~~~U^{\dagger} \mathcal{L}_{\mu\nu} U =\begin{pmatrix}
L_{\mu\nu}^{(J^+)} & 0 \\
0 & L^{(J^-)}_{\mu\nu} 
\end{pmatrix}. 
\ee
%%%%%%%%%%%%%%%%%%%%%%
It is obvious that they satisfy the similar relations in Sec.\ref{subsec:so4operators}:
%%%%%%%%%%%%%%%%%%%%
\be
\sum_{\mu<\nu}\mathit{\Lambda}_{\mu\nu}\mathcal{F}_{\mu\nu}=\sum_{\mu<\nu}\mathcal{F}_{\mu\nu}\mathit{\Lambda}_{\mu\nu}=0, ~~~~~\sum_{\mu<\nu}{\mathcal{L}_{\mu\nu}}^2=\sum_{\mu<\nu}{\mathit{\Lambda}_{\mu\nu}}^2 +\sum_{\mu<\nu}{\mathcal{F}_{\mu\nu}}^2.   \label{squaresynlso4}
\ee
%%%%%%%%%%%%%%%%%%
For 
%%%%%%%%%%%%%%%%%%%%%%%%%%%%
\be
\mathcal{T}_{\mu\nu}=\mathcal{L}_{\mu\nu},~\mathit{\Lambda}_{\mu\nu},~\mathcal{F}_{\mu\nu}, 
\ee
%%%%%%%%%%%%%%%%%%%%%%%%%%%%%%%%%%%%%%%%%%%%%%%%%%
 the $SO(4)$ covariance is realized as 
%%%%%%%%%%%%%%%%%%%%%%%%%%%
\be
 [\mathcal{L}_{\mu\nu}, \mathcal{T}_{\rho\sigma}]=i\delta_{\mu\rho}\mathcal{T}_{\nu\sigma}-i\delta_{\mu\sigma}\mathcal{T}_{\nu\rho}+i\delta_{\nu\sigma}\mathcal{T}_{\mu\rho}-i\delta_{\nu\rho}\mathcal{T}_{\mu\sigma}. 
\ee
%%%%%%%%%%%%%%%%%%%%%%%%%%%%%
The $SO(4)$ Casimir 
%%%%%%%%%%%%%%%%%%%%%
\be
\sum_{\mu<\nu}{\mathcal{L}_{\mu\nu}}^2=U\begin{pmatrix}
\sum_{\mu<\nu}{L_{\mu\nu}^{(J^+)}}^2 & 0 \\
0 & \sum_{\mu<\nu}{L_{\mu\nu}^{(J^-)}}^2
\end{pmatrix}U^{\dagger}, 
\ee
%%%%%%%%%%%%%%%%%%%%%%
is diagonalized as 
%%%%%%%%%%%%%%%%%%
\be
\sum_{\mu<\nu}{{\mathcal{L}_{\mu\nu}}}^2~~\Longrightarrow ~~
U^{\dagger}~\sum_{\mu<\nu}{\mathcal{L}_{\mu\nu}}^2 ~U= \begin{pmatrix} 
\mathcal{C}^+(n^+, s^+)  \bs{1}_{2J^{+}+1}  & 0 \\
0 & \mathcal{C}^-(n^-, s^-)  \bs{1}_{2J^{-}+1} 
\end{pmatrix} , 
\ee
%%%%%%%%%%%%%%%%%
where 
%%%%%%%%%%%%%%%
\begin{subequations}
\begin{align}
&\mathcal{C}^+(n^+, s^+) = (n^++J^+)(n^++J^+ +2 )+{s^+}^2 , \label{casa}\\
&\mathcal{C}^-(n^-, s^-)=(n^-+J^-)(n^-+J^- +2 )+{s^-}^2, \label{casb}
\end{align}
\end{subequations}
%%%%%%%%%%%%%%%
with $n^+, n^-=0, 1, 2, 3, \cdots$, $s^{+}=-J^+, -J^++1, \cdots,  J^+$ and $s^{-}=-J^-, -J^-+1, \cdots,  J^-$. 
There exist  degeneracies, $\mathcal{C}^+(n, s)=\mathcal{C}^+(n, -s) =\mathcal{C}^-(n+1, s)=\mathcal{C}^-(n+1, s)$.   
The corresponding eigenstates are given by the following $2(I+1)$ component  states: 
%%%%%%%%%%%%%%%%%%%%%%%%%
\begin{subequations}
\begin{align}
&\bs{\Psi}^{(n^+, \pm s^+,  J^+)}_{M_L^+, M_R^+} \equiv U\begin{pmatrix}
\bs{\Phi}^{(n^+, \pm s^+, J^+)}_{M_L^+, M_R^+} \\ \bs{0}_I 
\end{pmatrix}=C^+  \bs{\Phi}^{(n^+, \pm s^+, J^+)}_{M_L^+, M_R^+}, \label{so4eigenstatesvect1} \\
&\bs{\Psi}^{(n^-, \pm s^-, J^-)}_{M_L^-, M_R^-}\equiv U\begin{pmatrix}
\bs{0}_{I+2} \\  \bs{\Phi}^{(n^-, \pm s^-, J^-)}_{M_L^-, M_R^-}
\end{pmatrix}=C^-   \bs{\Phi}^{(n^-, \pm s^-, J^-)}_{M_L^-, M_R^-} ,  
\label{so4eigenstatesvect2}
\end{align}\label{so4eigenstatesvect}
\end{subequations}
%%%%%%%%%%%%%%%%%%%%%%%
where  $\bs{\Phi}^{(n^+, \pm s^+, J^+)}_{M_L^+, M_R^+}$ and $\bs{\Phi}^{(n^-, \pm s^-, J^-)}_{M_L^-, M_R^-}$ denote the  $SO(4)$ monopole harmonics (\ref{vectorrepsorep}) with $(I+2)$ and $I$ components, respectively.\footnote{Using the explicit form of $C^+$ and $C^-$ (\ref{cgcoefficients}),  the components of  (\ref{so4eigenstatesvect}) can be expressed as 
%%%%%%%%%%%%%%%%%%%%%%%%%%%%%%%%%%%%%%%%%%
\begin{align}
&{\Psi}^{(n^+, \pm s^+, J^+)}_{M_L^+, M_R^+}(x)_{\frac{1}{2}, \frac{1}{2}\sigma; \frac{I}{2}, m} =\sum_{M'_L=-L^+}^{L^+}  \sum_{M'_R=-R^+}^{R^+} \langle \frac{1}{2}, \frac{1}{2}\sigma ; \frac{I}{2}, m|P_{J^+}|L^+, M'_L; ~  R^+, M'_R\rangle \cdot \Psi^{(L^+)}(\bs{\chi})_{M'_L, M_L} \Psi^{(R^+)}(-\bs{\chi})_{M'_R, M_R}, \nn\\
&{\Psi}^{(n^-, \pm s^-, J^-)}_{M_L^-, M_R^-}(x)_{\frac{1}{2}, \frac{1}{2}\sigma; \frac{I}{2}, m} =\sum_{M'_L=-L^-}^{L^-}  \sum_{M'_R=-R^-}^{R^-} \langle \frac{1}{2}, \frac{1}{2}\sigma ; \frac{I}{2}, m|P_{J^-}|L^-, M'_L; ~  R^-, M'_R\rangle \cdot \Psi^{(L^-)}(\bs{\chi})_{M'_L, M_L} \Psi^{(R^-)}(-\bs{\chi})_{M'_R, M_R}, 
\end{align}
%%%%%%%%%%%%%%%%%%%%%%%%%%%%%%%%%%%%%%%%%%
where $\sigma=+1, -1$, $m={I}/{2}, {I}/{2}-1, \cdots, -{I}/{2}$, and $P^{J^{\pm}}$ is the projection operator to the Hilbert space of the $SU(2)$  index $J^{\pm}$: 
%%%%%%%%%%%%%%%%%%%%%%%%%%%%%%%
\be
P_{J^{\pm}}\equiv \sum_{A=-J^{\pm}}^{J^{\pm}} |J^{\pm}, A\rangle \langle J^{\pm}, A|.
\ee
%%%%%%%%%%%%%%%%%%%%%%%%%%%%%%%%%%  
}

%%%%%%%%%%%%%%%%%%%%%%%%%%%%%%
\subsection{Eigenvalue Problem }\label{subsec:spinorlandaumodeleigen}
%%%%%%%%%%%%%%%%%%%%%%%%%%%%%%%%

\begin{figure}[tbph]
\center
\hspace{-2cm}
\includegraphics*[width=140mm]{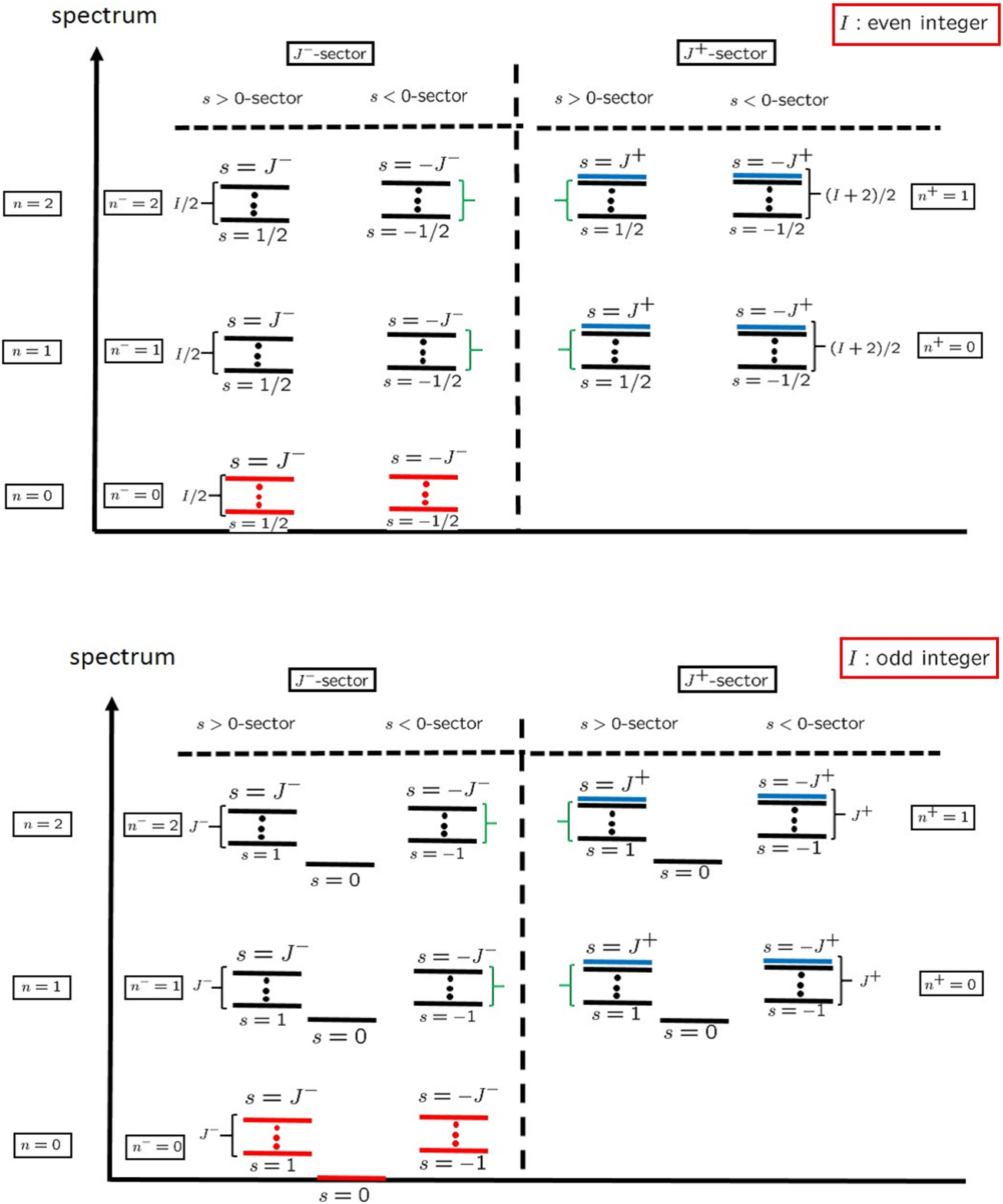}
\caption{Energy spectra of the spinor Landau model}
\label{Spinorspectra.fig}
%\vspace{-3mm}
\end{figure}

With the synthesized connection, %it is natural to consider the  spinor Landau Hamiltonian  given by $\frac{1}{2M}\sum_{\mu<\nu} {\mathit{\Lambda}_{\mu\nu}}^2$, and 
we adopt the following as the spinor Landau Hamiltonian
%%%%%%%%%%%%%%%%%%%%%%
\begin{align}
 H&=\sum_{\mu<\nu} {\mathit{\Lambda}_{\mu\nu}}^2 +\frac{1}{2}((I+3)C^+ {C^{+}}^{\dagger}  -(I-1)C^- {C^{-}}^{\dagger}) \nn\\  
 &= \sum_{\mu<\nu} {\mathit{\Lambda}_{\mu\nu}}^2 +\frac{1}{2} U 
\begin{pmatrix}
(I+3)\bs{1}_{I+2} & 0 \\
 0 & -(I-1) \bs{1}_{I}
\end{pmatrix} U . 
 \label{spinorlandaudef}
\end{align}
%%%%%%%%%%%%%%%%%%%%%%%%
The constant matrix term on the right-hand side was added for the spinor Landau Hamiltonian to be the precise square of the Weyl-Landau operator (see Sec.\ref{sec:relativisLandau}).  
From (\ref{squaresynlso4}) and 
%%%%%%%%%%
\be
\sum_{\mu<\nu}{\mathcal{F}_{\mu\nu}}^2 =U\begin{pmatrix} 
{\bs{S}^{(J^+)}}^2 & 0 \\
0 & {\bs{S}^{(J^+)}}^2 
\end{pmatrix} U^{\dagger} = U\begin{pmatrix} 
\frac{1}{4}(I+1)(I+3)\bs{1}_{I+2}  & 0 \\
0 & \frac{1}{4}(I-1)(I+1)\bs{1}_{I} 
\end{pmatrix} U^{\dagger}, 
\ee
%%%%%%%%%%%%
we find that (\ref{spinorlandaudef}) is essentially the $SO(4)$ Casimir made of $\mathcal{L}_{\mu\nu}$: 
%%%%%%%%%%%%%%%%%%%%%%%%%
\be
H=\sum_{\mu<\nu}{\mathcal{L}_{\mu\nu}}^2-\frac{1}{4}(I-1)(I+3).  
\label{spinorlandauham}
\ee
%%%%%%%%%%%%%%%%%%%%%%%%%%%%
The spinor Landau Hamiltonian is invariant under the $SO(4)$ rotations.   
%and exactly equal to the square of the Weyl-Landau operator (see  Sec.\ref{sec:relativisLandau}). 
Recalling the results in Sec.\ref{subsec:synthesizedang}, we can derive the energy eigenvalues, $E^+_n(s)$ for $J^+$-sector and  $E^-_n(s)$ for  $J^-$-sector. 
 For $E^{\pm}_n(s)$, the range of the indices are   $n=0, 1,2, \cdots, s=-J^{\pm}, -J^{\pm}+1, \cdots,  J^{\pm}$, and hence we have $E_n(s)=n(I+n+1)+s^2$ with $n=0,1,2,\cdots$ and $s=-J^+, -J^++1, \cdots, J^+$. In detail, 
%%%%%%%%%%%%%%%%%%%%%%
\begin{subequations}
\begin{align}
|s|=J^+~:~&E_n(J^{+})\equiv  E^+_{n-1}( \pm J^{+}) =(n+\frac{I-1}{2})^2, \label{so4spec1}\\
|s|\le J^-~:~&E_n(s)\equiv E^-_{n}(\pm s)  = E^+_{n-1} (\pm s)  = n(I+n+1)+s^2, \label{so4spec2}\\ 
n=0~~:~&E^-_0(s) =E^-_0(-s)=s^2~~~~( |s|\le J^-), \label{so4spec3}
\end{align}\label{so4spinorenergyspsum}
\end{subequations}
%%%%%%%%%%%%%%%%%%%%%
where 
%%%%%%%%%%%%%%%%%%%%%%%
\be
n=1,2,3, \cdots, ~~~~s=-\frac{I}{2}+\frac{1}{2}, -\frac{I}{2} +\frac{3}{2}, \cdots, \frac{I}{2}-\frac{1}{2}. 
\ee
%%%%%%%%%%%%%%%%%%%%%%%
The schematic picture of the spectrum is given by Fig.\ref{Spinorspectra.fig}.  
%%%%%%%%%%%%%%%%%%%%%%%%%%%%%%%%%%%%%%%%%%%%%%%%%%%%%%%%%%%%
%%%%%%%%%%%%%%%%%%%%%%%%%%%%%%%%%%%%%%%%%%%%%%%%%%%%%%%%%%%%%% 
Note that 
$J^-=\frac{I}{2}-\frac{1}{2} \ge 0$ is not well defined for $I=0$, and then the energy levels (\ref{so4spec2}) and (\ref{so4spec3}) vanish in the free background limit. (See Appendix \ref{subsec:freeweylsquare} also.)    
The degeneracies are 
%%%%%%%%%%%%%%%%%%%%%
\begin{subequations}
\begin{align}
&2(2L^+(n-1, J^+)+1)(2R^+(n-1, J^+)+1) = 2n(n+I+1)   ,\\
&4(2L^-(n, s)+1)(2R^-(n, s)+1)(1-\delta_{I, 0})
= (2n+I+2s+1)(2n+I-2s+1) (1-\delta_{I, 0})\\
&(2L^-(0, s)+1)(2R^-(0, s)+1)(2-\delta_{s, 0})(1-\delta_{I, 0})=(2-\delta_{s, 0})(\frac{I+1}{2}+s)(\frac{I+1}{2}-s)(1-\delta_{I, 0}). 
\label{LLLdegespinorLandau}
\end{align}
\end{subequations}
%%%%%%%%%%%%%%%%%%%%%%
In (\ref{LLLdegespinorLandau}),   $s=0$ is special:  
%%%%%%%%%%%%%%%%%%%%%%
\be
(\frac{I+1}{2})^2=1, ~2^2,~3^2, ~\cdots, 
\ee
%%%%%%%%%%%%%%%%%%%%%
which is equal to the degeneracy $(2j+1)^2$ $(j=0, 1/2, 1, 3/2, \cdots)$  of  two particles with  identical spin $j$.  
The explicit forms of the eigenstates are given by 
%%%%%%%%%%%%%%%%%%%%%%
\begin{subequations}
\begin{align}
&~~~~~~~~~~~~~~~~~\bs{\Psi}^{(n-1, \pm J^+ J^+)}_{M_L, M_R} ,~~~(n=1,2,\cdots) \\
&\bs{\Psi}^{(n, \pm s, J^-)}_{M_L, M_R} ,~~~\bs{\Psi}^{(n-1, \pm s,  J^+)}_{M_L, M_R} ,~~~(n=1,2,\cdots) \\
&\bs{\Psi}^{(0,\pm s,  J^-)}_{M_L, M_R}. \label{LLLdegespinorLandaustate}
\end{align}
\end{subequations}
%%%%%%%%%%%%%%%%%%%%%%
The lowest Landau level of the $SO(4)$ spinor Landau model 
(\ref{so4spec3}) comes only from that of the $J^-$-sector, and (\ref{so4spec2}) from   $n$th Landau level in the $J^+$-sector and  $(n-1)$th Landau level in the $J^-$-sector, because of  $n+J^+=(n+1)+J^-$.  These features are similar to those of the $SO(3)$  spinor Landau model on $S^2$  \cite{Hasebe-2015}.  
However, (\ref{so4spec1}) comes only from the $|s|=J^+$ subband of $n$th Landau level, which is a new feature not observed in the $SO(3)$ spinor Landau model.

%%%%%%%%%%%%%%%%%%%%%%%%%%%%%%%%%%%%%%%%%%%%%%%%%%%%%%%%%%%
\subsection{Interactions in the spinor Landau Hamiltonian}\label{subsec:intspinorham}
%%%%%%%%%%%%%%%%%%%%%%%%%%%%%%%%%%%%%%%%%%%%%%%%%%%%%%%%%%%

We discuss interactions that the spinor Landau model describes.  
In (\ref{so4casnonrel}), we saw that the non-relativistic $SO(4)$ Landau Hamiltonian includes the (gauge) spin-orbit interaction. 
 Since the synthesized connection consists of  two kinds of spins coming from the holonomy  and  gauge groups,  
the spinor Landau Hamiltonian is expected to represent their interactions also.    

We first decompose the $SO(4)$ synthesized angular momentum operator as  
%%%%%%%%%%%%%%%
\be
\mathcal{L}_{\mu\nu}= L_{\mu\nu}+s_{\mu\nu} 
\ee
%%%%%%%%%%%%%%%
where $L_{\mu\nu}$ contain the gauge spin and   the holonomy spin $s_{\mu\nu}$ : 
%%%%%%%%%%%%%%%%%%%%%%
\begin{subequations}
\begin{align}
&L_{\mu\nu}=-ix_{\mu}D_{\nu}+ix_{\nu}D_{\mu}+1\otimes F_{\mu\nu}, \\
&s_{\mu\nu}=(x_{\mu}\omega_{\nu}-x_{\nu}\omega_{\mu}+f_{\mu\nu})\otimes 1.  
\label{holonomypart}
\end{align}
\end{subequations}
%%%%%%%%%%%%%%%%%%%%%% 
The square of $SO(4)$ total angular momentum is derived as  
%%%%%%%%%%%%%%%%%%%%
\be
{\mathcal{L}_{\mu\nu}}^2= {L_{\mu\nu}}^2+2s_{\mu\nu}L_{\mu\nu}+{s_{\mu\nu}}^2, 
\label{squareoflmunusyn}
\ee
%%%%%%%%%%%%%%%%%%%
where we used 
%%%%%%%%%%%%%%%%%%%%%%%%%
\be
l_{\mu\nu} s_{\mu\nu}=-2ix_{\mu}\partial_{\nu}s_{\mu\nu}=0. 
\ee
%%%%%%%%%%%%%%%%%%%%%%%%%%
${L_{\mu\nu}}^2$ is given by (\ref{so4casnonrel}).  
The second term on the right-hand side of (\ref{squareoflmunusyn}) contains the holonomy spin-orbit interaction and the holonomy gauge spin-spin  interaction.  
Indeed in  the Dirac gauge 
%%%%%%%%%%%%%%%%%%%%%
\be 
\omega_i =-\frac{1}{2(1+x_4)}\epsilon_{ijk}x_j\sigma_k,~~~\omega_4=0, ~~~~
f_{ij}=-x_i\omega_j+x_j\omega_i+\frac{1}{2}\epsilon_{ijk}\sigma_k, ~~~f_{i4}=(1+x_4)\omega_i,  
\ee
%%%%%%%%%%%%%%%%%%%
the holonomy part (\ref{holonomypart}) is simply represented as 
%%%%%%%%%%%%%%%%%%%%%%%
\be
s_{ij}=\frac{1}{2}\epsilon_{ijk}\sigma_k, ~~~s_{i4}=\omega_i, 
\ee
%%%%%%%%%%%%%%%%%%%%%%%%%%%%%
and the second term of (\ref{squareoflmunusyn}) becomes 
%%%%%%%%%%%%%%%%%%%%%%%%%%%%%%%%%%%
\be
2\sum_{\mu<\nu}s_{\mu\nu}L_{\mu\nu}=\sigma_i \otimes L^D_i +2\omega_i \otimes K_i=\frac{1}{1+x_4} \bs{\sigma}\cdot \bs{l}+\frac{2}{1+x_4}\biggl(\bs{\sigma}\otimes \bs{S}^{(I/2)}-\frac{1}{2(1+x_4)} (\bs{x}\cdot \bs{\sigma})\otimes  (\bs{x}\cdot \bs{S}^{(I/2)})\biggr), 
\ee
%%%%%%%%%%%%%%%%%%%%%%%%%
where both $L^D_i$  (\ref{defsu2diagop}) and $K_i$ (\ref{defsu2k}) contain the gauge spin $S_i^{(I/2)}$.    
The last term on the right-hand side of (\ref{squareoflmunusyn}) can be expressed as     
%%%%%%%%%%%%%%%%%%%
\be
\sum_{\mu<\nu}{s_{\mu\nu}}^2={\omega_{\mu}}^2+\sum_{\mu<\nu}{f_{\mu\nu}}^2 
=\frac{1}{1+x_4}+\frac{1}{4} . 
\ee
%%%%%%%%%%%%%%%%%%%%%%%%%%
Consequently, the spinor Landau Hamiltonian (\ref{spinorlandauham}) is   represented as 
%%%%%%%%%%%%%%%%%%%%
\begin{align}
H&=\sum_{\mu<\nu}{l_{\mu\nu}}^2+\frac{2}{1+x_4}\bs{S}^{(I/2)} \cdot  \bs{l}+\frac{1}{1+x_4} \bs{\sigma}\cdot \bs{l}+\frac{2}{1+x_4}\biggl(\bs{\sigma}\otimes \bs{S}^{(I/2)}-\frac{1}{2(1+x_4)} (\bs{x}\cdot \bs{\sigma})\otimes  (\bs{x}\cdot \bs{S}^{(I/2)})\biggr)\nn\\
&~~  -\frac{1}{(1+x_4)^2} (\bs{x}\cdot \bs{S}^{(I/2)})^2+\frac{1-x_4}{4(1+x_4)}I(I+2) +\frac{2+x_4}{1+x_4}. 
\label{expandspinorhamso4}
\end{align}
%%%%%%%%%%%%%%%%%%%%%%%%% 
The second, third and fourth terms  respectively stand for the gauge spin-orbit,  holonomy spin-orbit, and  holonomy gauge spin-spin interactions.    With special combination of such interactions, the spinor Landau Hamiltonian (\ref{expandspinorhamso4}) respects the $SO(4)$ symmetry.

%%%%%%%%%%%%%%%%%%%%%%%%%%%%%%%%%%%%%%%%%%%%%%%%%%%%%%%
\section{Relativistic Landau Models}\label{sec:relativisLandau}
%%%%%%%%%%%%%%%%%%%%%%%%%%%%%%%%%%%%%%%%%%%%%%%%%%%%%%%

Behind the $SO(4)$ Landau models, we implicitly assumed $(3+1)$ space-time (whose spacial manifold is  $S^3$), and we refer to  $2\times 2$ Dirac operator as the Weyl operator and  $4\times 4$ Dirac operator as the Dirac operator simply.  
 The previous results of the spinor Landau models are applied to solve  the eigenvalue problems of the relativistic Landau models.   %We first analyze the Weyl-Landau model, and subsequently massive Dirac-Landau model and the supersymmetric Landau model.  %In the previous study \cite{Nair-Daemi-2004}, the analysis of the relativistic Landau model was sketched, and the relativistic model does not contain the spin connection. Meanwhile in the present work, 
 In Sec.\ref{subsec:so4weyllanmod}, we utilize the spin connection to construct the Weyl-Landau Hamiltonian on $S^3$ and analyze the eigenvalue problem, in which verification of  the $SO(4)$ invariance of the Dirac-Landau operator and an explicit form of the eigenstates are derived.   
 We also account for the existence of the zero-modes from the non-commutative geometry point of view.    
 It is shown that the obtained results are reduced to the known formulae of the free Weyl operator in the free background limit.  In Sec.\ref{subsec:diraclandauso4},   we make use of the results of the Weyl-Landau model to solve the eigenvalue problem of the massive Dirac-Landau model. The supersymmetric Landau operator made of the square of the Dirac-Landau operator is also analyzed.

 A convenient gauge to express the relativistic operators on $S^3$ is the Schwinger gauge, which we will adopt in this section.

%%%%%%%%%%%%%%%%%%%%%%%%%%%%%%%%%%%%%%%
\subsection{$SO(4)$ Weyl-Landau model}\label{subsec:so4weyllanmod}
%%%%%%%%%%%%%%%%%%%%%%%%%%%%%%%%%%%%%%%%%

 Let us begin with the construction of the  Weyl-Landau operator  
%%%%%%%%%%%%%%%%%%%%%%%%%%%
\be
-i\fsl{\mathcal{D}}=-ie_{a}^{~~\alpha}\gamma^{a}\mathcal{D}_{\alpha}=-ie_{a}^{~~\alpha}\gamma^{a}(\partial_{\alpha}+i\omega_{\alpha}\otimes \bs{1}_{I+1} +i\bs{1}_2\otimes A^{(I/2)}_{\alpha}).  
\label{formuwlop}
\ee
%%%%%%%%%%%%%%%%%%%%%%%%%%%
It is a $(2\cdot (I+1)) \times (2\cdot (I+1))$ matrix-valued differential operator.  
From (\ref{spinconnes3oneform}),  the covariant derivatives  
%%%%%%%%%%%%%%%%%%%%%%%%%%
\be
-i\mathcal{D}_{\alpha}=-i\partial_{\alpha}+\mathcal{A}_{\alpha}
\ee
%%%%%%%%%%%%%%%%%%%%%%%%%%%%%%%%%%
are given by 
%%%%%%%%%%%%%%%%%%%%%%%%%%%%%%%
\begin{align}
&-i\mathcal{D}_{\chi}=-i\partial_{\chi}, \nn\\
&-i\mathcal{D}_{\theta}=-i\partial_{\theta}  -\cos\chi (\frac{1}{2}\sigma_{y}\otimes 1 + 1\otimes S_y^{(I/2)} ),\nn\\
&-i\mathcal{D}_{\phi}=-i\partial_{\phi} +\sin\theta \cos\chi (\frac{1}{2}\sigma_x\otimes 1 + 1\otimes S_x^{(I/2)})-\cos\theta (\frac{1}{2} \sigma_{z}\otimes 1 + 1\otimes S_z^{(I/2)}),   
\end{align}
%%%%%%%%%%%%%%%%%%%%%%%%%%%%%%%
and the Weyl-Landau operator is expressed as 
%%%%%%%%%%%%%%%%%%%%%%%%%%%%%%%%%
\begin{align}
-i\fsl{\mathcal{D}}=&-i\gamma^1 \mathcal{D}_{\chi} -i\frac{1}{\sin\chi} \gamma^2 \mathcal{D}_{\theta}-i\frac{1}{\sin\chi\sin\theta}\gamma^3 \mathcal{D}_{\phi}\nn\\
=&-i\sigma_z \tilde{D}_{\chi} -i\frac{1}{\sin\chi} \sigma_x \tilde{D}_{\theta}-i\frac{1}{\sin\chi\sin\theta}\sigma_y   \tilde{D}_{\phi}, \label{weyldiracconcise}
\end{align}
%%%%%%%%%%%%%%%%%%%%%%%%%%%%%%%%%%%%%
or 
%%%%%%%%%%%%%%%%%%%%
\be
-i\fsl{\mathcal{D}}=
\begin{pmatrix} 
 -i\tilde{D}_{\chi}  & -i\frac{1}{\sin\chi} \tilde{D}_{\theta} -\frac{1}{\sin\chi\sin\theta}\tilde{D}_{\phi} \\
 -i\frac{1}{\sin\chi} \tilde{D}_{\theta} +\frac{1}{\sin\chi\sin\theta}\tilde{D}_{\phi} & i\tilde{D}_{\chi}
\end{pmatrix} , 
\label{matrixweyllandau}
\ee
%%%%%%%%%%%%%%%%%%%
where 
%%%%%%%%%%%%%%%%%%%%
\begin{align}
&\tilde{D}_{\chi} \equiv \partial_{\chi} +\cot\chi =D_{\chi}+\cot\chi, \nn\\
&\tilde{D}_{\theta} \equiv \partial_{\theta}-i\cos \chi S_y^{(I/2)}+\frac{1}{2}\cot\theta=D_{\theta}+\frac{1}{2}\cot\theta, \nn\\
&\tilde{D}_{\phi} \equiv \partial_{\phi} -i\cos\theta S_z^{(I/2)} +i\cos\chi\sin\theta S_x^{(I/2)}=D_{\phi}.
\end{align}
%%%%%%%%%%%%%%%%%%%%%
The last terms of $\tilde{D}_{\chi}$ and $\tilde{D}_{\theta}$ are non-hermitian terms that come from the spin connection (\ref{spinconnes3oneform}) (as in the case of the Dirac-Landau operator on $S^2$ \cite{Hasebe-2015}); $\cot\chi$ in $\tilde{D}_{\chi}$ is from $\omega^1_{~~2}$ and $\omega^3_{~~1}$, and  $\frac{1}{2}\cot\theta$ in $\tilde{D}_{\theta}$  from $\omega^2_{~~3}$\footnote{The spin connection $\fsl{\omega}=e_a^{~~\alpha}\gamma^a\omega_{\alpha}$ is given by 
%%%%%%%%%%%%%%%%%%%%%%%%%%%%%%%
\begin{align}
&\text{Schwinger gauge}~:~\fsl{\omega}_\text{S}=-i\cot\chi~\sigma_z -i\frac{1}{2\sin\chi}\cot\theta ~\sigma_x, \nn\\
&\text{Dirac~gauge}~:~\fsl{\omega}_\text{D}=i\tan(\frac{\chi}{2})~(\sin\theta\cos\phi ~\sigma_x +\sin\theta\sin\phi ~\sigma_y +  \cos\theta ~\sigma_z) =i \tan(\frac{\chi}{2})~  \hat{\bs{x}}\cdot \bs{\sigma}.
\end{align}
%%%%%%%%%%%%%%%%%%%%%%%%%%%%%%%
}.  
In the previous study of the $SO(3)$ Landau model \cite{Hasebe-2015}, we showed that the Dirac-Landau operator is invariant under the $SO(3)$ rotation. 
% The $SO(4)$ Landau model is a natural generalization of the $SO(3)$ Landau model, and   
In the present case, the Weyl-Landau operator (\ref{formuwlop}) is invariant under the transformations generated  by the $SO(4)$ synthesized angular momentum operators. 
 To see this, we use explicit coordinate representation of the  Weyl-Dirac operator and  $SO(4)$ angular momentum operators.  Having established the gauge transformation (Sec.\ref{subsec:diracschgauges}),  it is not difficult to  work  either in  the Dirac gauge or in the Schwinger gauge to demonstrate 
%%%%%%%%%%%%%%%%%%%%%%%%%%%
\be
[-i\fsl{\mathcal{D}}, \mathcal{L}_{\mu\nu}]=0.  
\label{invdiracso4}
\ee
%%%%%%%%%%%%%%%%%%%%%%%%%%%
%(The author checked the validity of (\ref{invdiracso4}) for several $I$s by Mathematica.)  
As the Weyl-Landau operator is  $SO(4)$ singlet,  
the Weyl-Landau operator eigenvalues should have degeneracies due to the existence of the $SO(4)$ symmetry. In other words, the Weyl-Landau operator eigenstates consist of the $SO(4)$ Landau level basis states of the spinor Landau model. 

\begin{figure}[tbph]
\center
\includegraphics*[width=130mm]{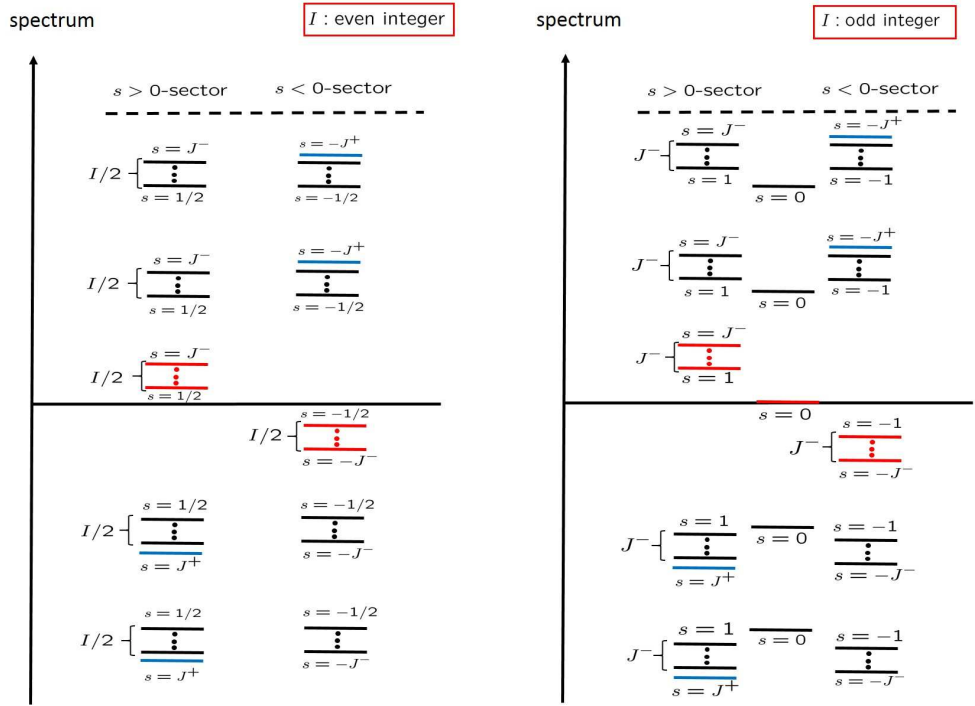}
\caption{The Weyl-Landau operator spectra. The spectral patterns  are different with respect to the parity of $I$. In particular  for odd $I$, there appear zero-modes ($s=0$). }
\label{schepDLspectra.fig}
\end{figure}

Let us first derive the eigenvalues of the Weyl-Landau operator. 
The square of the Weyl-Landau operator gives the spinor Landau Hamiltonian (\ref{spinorlandauham}): 
%%%%%%%%%%%%%%%%%%%%%%%%%%%%%%%%
\be
(-i\fsl{\mathcal{D}})^2 = \sum_{\mu<\nu} {\mathcal{L}_{\mu\nu}}^2 -{\bs{S}^{(I/2)}}^2+\frac{3}{4}= \sum_{\mu<\nu} {\mathcal{L}_{\mu\nu}}^2 - \frac{1}{4}(I+3)(I-1)  . \label{squareofweyllandau}
\ee
%%%%%%%%%%%%%%%%%%%%%%%%%%%%%%    
The first equation of (\ref{squareofweyllandau}) can be checked rigorously from the Weyl-Landau operator (\ref{matrixweyllandau})  and the $SO(4)$ angular momentum operators.\footnote{(\ref{squareofweyllandau}) can also be derived from a general formula \cite{Dolan-2003,Hasebe-2017, Coskun-Kurkcuoglu-Toga-2017}
%%%%%%%%%%%%%%%%%%%%%%
\be
(-i\fsl{\mathcal{D}})^2 =C_{SO(4)}-C_{SO(3)} +\frac{1}{8}\mathcal{R}_{S^3} 
\ee
%%%%%%%%%%%%%%%%%%%%%%%%
with 
%%%%%%%%%%%%%%%%%%%%%%
\be
\mathcal{R}_{S^3}=d(d-1)|_{d=3}=6. 
\ee
%%%%%%%%%%%%%%%%%%%%%%%
}   
With the results in Sec.\ref{subsec:spinorlandaumodeleigen},  we can readily obtain the Weyl-Landau operator eigenvalues as $\pm \lambda_n(s)=\pm \sqrt{n(n+I+1)+s^2}$ with $n=0,1,2,\cdots$ and $s=-J^+, -J^++1, \cdots, J^+$, or 
%%%%%%%%%%%%%%%%%%%%%%
\begin{subequations}
\begin{align}
|s|=J^+~&:~+{\lambda(n,\frac{I+1}{2} )} =+ (n+\frac{I+1}{2}),~~~~~~~-{\lambda(n,\frac{I+1}{2} )} =- (n+\frac{I+1}{2}) \label{firstdiraclandaueigenva}\\
|s|\le J^-~&:~+\lambda(n,s) =    +\sqrt{n(I+n+1)+s^2},~~~-\lambda(n,s) =    -\sqrt{n(I+n+1)+s^2}, \label{secondeigenvalues}\\ 
~n=0~&:~+\lambda(0,s)=s~~(s\ge 0),~~~~~~~~~~~~~~~~~~-\lambda(0,s)=s~~(s\le 0), \label{thirdeigenvalues}
\end{align}\label{weyllandauenerlev}
\end{subequations}
%%%%%%%%%%%%%%%%%%%%%
where 
%%%%%%%%%%%%%%%%%%%%%%%%%%%%%%%%%%%%%
\be 
n=1,2,3, \cdots,~~~~~~s=-\frac{I}{2}+\frac{1}{2}, -\frac{I}{2}+\frac{3}{2}, \cdots, \frac{I}{2}-\frac{1}{2}. 
\label{rangenands}
\ee
%%%%%%%%%%%%%%%%%%%%%%%%%%%%%%%%%%%%%
(In the free background limit $I=0$, there do not exist  
 %$J^-=\frac{I}{2}-\frac{1}{2} \ge 0$  is not well defined, meaning 
the eigenstates for (\ref{secondeigenvalues}) and (\ref{thirdeigenvalues}), as in the case of the spinor Landau model.)   
Notice that the zeroth Landau level (\ref{thirdeigenvalues}) does not explicitly depend on the monopole charge and  remains in low energy region even in  $I\rightarrow \infty$ limit. The schematic picture of the Weyl-Landau operator is given by Fig.\ref{schepDLspectra.fig}. 
%%%%%%%%%%%%%%%%%%%%%%%%%%%%%%%%%%%%%%%%%%%%%%%%%%%%%%%%%%%%
%%%%%%%%%%%%%%%%%%%%%%%%%%%%%%%%%%%%%%%%%%%%%%%%%%%%%%%%%%%%%% 
 The degeneracies for (\ref{weyllandauenerlev}) are respectively given by 
%%%%%%%%%%%%%%%%%%%%%
\begin{subequations}
\begin{align}
&(n+I+1)n   ,\label{nlandauschiralgenerelalandau1}\\
&2(n+\frac{I}{2}+\frac{1}{2}+s)(n+\frac{I}{2}+\frac{1}{2}-s)~\times (1-\delta_{I , 0}) ,\label{nlandauschiralgenerelalandau}\\ 
&(\frac{I+1}{2}+s)(\frac{I+1}{2}-s)~\times (1-\delta_{I , 0}).
\end{align}
\end{subequations}
%%%%%%%%%%%%%%%%%%%%%%
The explicit forms of the eigenstates are 
%%%%%%%%%%%%%%%%%%%%%%
\begin{subequations}
\begin{align}
\bs{\Psi}^{(n-1,  -J^+, J^+)}_{M_L, M_R} ,~~~~~~
&\bs{\Psi}^{(n-1, J^+, J^+)}_{M_L, M_R} , \label{diraclandauj+}\\
\alpha_{s} ~\bs{\Psi}^{(n-1,  s, J^+)}_{M_L, M_R}+\beta_{s} ~ \bs{\Psi}^{(n,  s, J^-)}_{M_L, M_R}
, ~~~~~&\beta_{s} ~\bs{\Psi}^{(n-1, - s, J^+)}_{M_L, M_R} -\alpha_{s}~\bs{\Psi}^{(n, - s, J^-)}_{M_L, M_R} ,  \label{secondeigenstate}\\%~~~(s=-J^-, -J^-+1,\cdots, J^-)\\
~~~~~~~~~~\bs{\Psi}^{(0, s, J^-)}_{M_L, M_R}~~~(s\ge 0),~~~
&\bs{\Psi}^{(0, s, J^-)}_{M_L, M_R}~~~(s\le 0) ,  \label{thirdeigenstate}
\end{align}\label{eigenstateweyllandauei}
\end{subequations}
%%%%%%%%%%%%%%%%%%%%%%
where $\alpha_{s}$ and $\beta_{s}$ are the coefficients (subject to ${\alpha_{s}}^2+{\beta_{s}}^2=1$) that are determined so that (\ref{secondeigenstate}) be the eigenstates of the Weyl-Landau operator with the eigenvalues (\ref{secondeigenvalues}). 
In Appendix \ref{append:examplesdirac}, we explicitly derive  $\alpha_s$ and $\beta_s$ and  construct the  Weyl-Landau operator eigenstates for several cases.     
As  mentioned above, the Weyl-Landau operator eigenstates (\ref{eigenstateweyllandauei}) are the $SO(4)\simeq SU(2)_L\otimes SU(2)_R$ irreducible representations with the indices $(L, R)$:  
%%%%%%%%%%%%%%%%%
\begin{subequations}
\begin{align}
(\frac{1}{2}(n-1), ~\frac{1}{2}(n+I)),~~~~&(\frac{1}{2}(n+I), ~\frac{1}{2}(n-1)), \\
(\frac{1}{2}(n+\frac{I}{2}-\frac{1}{2})+\frac{s}{2},~ \frac{1}{2}(n+\frac{I}{2}-\frac{1}{2})-\frac{s}{2})),~~~~&(\frac{1}{2}(n+\frac{I}{2}-\frac{1}{2})-\frac{s}{2}, ~\frac{1}{2}(n+\frac{I}{2}-\frac{1}{2})+\frac{s}{2})), \\
(\frac{1}{2}(\frac{I}{2}-\frac{1}{2})+\frac{s}{2}, ~\frac{1}{2}(\frac{I}{2}-\frac{1}{2})-\frac{s}{2})~(s\ge 0),~~& (\frac{1}{2}(\frac{I}{2}-\frac{1}{2})-\frac{s}{2}, ~\frac{1}{2}(\frac{I}{2}-\frac{1}{2})+\frac{s}{2})~(s\le 0). 
\label{n=0weyllandaurepso4}
\end{align}
\end{subequations}
%%%%%%%%%%%%%%%%%%%%
For $s=0$, two representations of (\ref{n=0weyllandaurepso4}) coincide to be  $(L, R)=(\frac{1}{4}(I-1), \frac{1}{4}(I-1))$.  
(\ref{secondeigenstate}) consists of both the $(n-1)$th Landau level basis states in  $J^+$-sector and the $n$th Landau level in $J^-$-sector, because   $n-1+J^+=n+J^-$.  This  feature is similar to that of the $SO(3)$ Dirac-Landau model on $S^2$  \cite{Hasebe-2015}.  
 Meanwhile, (\ref{diraclandauj+}) comes only from the non-relativistic $n$th Landau level with replacement of $I/2$ with $J^+$, and this is a new feature in the $SO(4)$ model.  

Notice that the zeroth Landau level 
(\ref{thirdeigenvalues}) comes only from the lowest Landau level of $J^-$-sector, and  the relativistic zeroth Landau level is exactly equal to that of the non-relativistic lowest Landau level with replacement of $I/2$ with $J^-=(I-1)/2$. This property is also observed in the $SO(3)$ model on $S^2$ \cite{Hasebe-2015}. 
For the zeroth Landau level, the spectrum of the subbands is exactly equal to the corresponding chirality parameter: 
%%%%%%%%%%%%%%%%%%%%%%%%
\be
-i\fsl{\mathcal{D}}_{n=0}=s. 
\ee
%%%%%%%%%%%%%%%%%%%%%%%%
Therefore the reflection symmetry between the positive and negative eigenvalues  in the zeroth Landau level is identical to the LR symmetry: 
%%%%%%%%%%%%%%%%%%%%%%%%%%%
\be
s~\leftrightarrow ~-s.
\ee
%%%%%%%%%%%%%%%%%%%%%%%%%%%%
Thus in the  relativistic Landau model, the left-right symmetry of the non-relativistic Landau model is realized as the chiral symmetry (of the zeroth Landau level). 

%%%%%%%%%%%%%%%%%%%%%%%%%%%%%%%%%%%%%%%%%%%%%%%%%%
\subsubsection{The lowest Landau level and the zero-modes }\label{subsub:lowestllzero}
%%%%%%%%%%%%%%%%%%%%%%%%%%%%%%%%%%%%%%%%%%%%%%%%%%%

In \cite{Hasebe-2017}, we showed that the total degeneracy of the $SO(4)$ zeroth Landau level is equal to the 2nd Chern number of the $SO(5)$ Landau model: 
%%%%%%%%%%%%%%%%
\be
\nu_{3\text{D}}^{\text{total}}=\sum_{s=-\frac{I-1}{2}}^{\frac{I-1}{2}} (\frac{I+1}{2}+s)(\frac{I+1}{2}-s) =\frac{1}{6}I(I+1)(I+2)=c_2, 
\ee
%%%%%%%%%%%%%%%
as a manifestation of the dimensional ladder of anomaly. 
The total degeneracy of the zeroth Landau level thus finds its topological origin in one dimension higher space. 

In even dimensions, the existence of Dirac-Landau operator zero-modes is accounted for by the Atiyah-Singer index theorem \cite{Dolan-2003, Hasebe-2014-1}. %, while the $SO(4)$  
Though  constructed on three-sphere,  the $SO(4)$ relativistic Landau model also accommodates the zero-modes %in the zeroth Landau level 
for odd $I$.  
Here we discuss the origin of such zero-modes.  
For odd $I=2q-1$ $(q=1,2,3,\cdots)$, the explicit form of the zero-modes  
is  given by 
%%%%%%%%%%%%%%%%%%%%%%%
\be
{\Psi}^{(n=0, s=0 , J^- )}_{M_L, M_R} ~\propto ~{\psi_{L1}}^{\frac{J^-}{2}+M_L}{\psi_{L2}}^{\frac{J^-}{2}-M_L}\otimes {\psi_{R1}}^{\frac{J^-}{2}+M_R}{\psi_{R2}}^{\frac{J^-}{2}-M_R}, ~~~~~(-\frac{J^-}{2}\le M_L,M_R \le \frac{J^-}{2})
\label{zeromodeexwave} 
\ee
%%%%%%%%%%%%%%%%%%%%%%%% 
with degeneracy 
%%%%%%%%%%%%%%%%%%
\be
\nu_{3\text{D}}^{\text{zero}}\equiv (J^-+1)^2 = q^2. 
\ee
%%%%%%%%%%%%%%%%%
Each of the $SU(2)$ representations, ${\psi_{L1}}^{\frac{J^-}{2}+M_L}{\psi_{L2}}^{\frac{J^-}{2}-M_L}$ and ${\psi_{R1}}^{\frac{J^-}{2}+M_R}{\psi_{R2}}^{\frac{J^-}{2}-M_R}$, is the zero-modes of the $SO(3)$ relativistic models with  $U(1)$ monopole charge $q/2=(I+1)/4$. Due to the Atiyah-Singer index theorem, the degeneracy of such zero-modes on two-sphere is equal to  the 1st Chern number, $c_1=\frac{1}{2\pi}\int_{S^2} \hat{F}=q$.   Therefore, the zero-mode degeneracy of the Weyl-Landau operator can be   expressed by the 1st Chern-number: 
%%%%%%%%%%%%%%%%%%%%%%%%%
\be
\nu_{3\text{D}}^{\text{zero}}={c_1}^2
\ee
%%%%%%%%%%%%%%%%%%%%%%%%
In this sense, the zero-modes on $S^3$ originates from the topological quantity in one dimension lower 2D space.   
From the viewpoint of the non-commutative geometry,  
the fuzzy three-sphere is represented as\cite{Ramgoolam2002, Nair-Daemi-2004}\footnote{ 
(\ref{sfclassicallimit}) is naturally induced from the chiral Hopf map (\ref{chiralhopfabst}): 
%%%%%%%%%%%%%%%%%%%%%%%
\be
S_L^3\otimes S_R^3~~~\overset{S^1_L\otimes S^1_R}{\longrightarrow}~~S^2_L\otimes S^2_R.
\ee
%%%%%%%%%%%%%%%%%%%%%% 
In arbitrary odd dimension, (\ref{sfclassicallimit}) is generalized as \cite{Ramgoolam2002}, 
%%%%%%%%%%%
 \be
S_F^{2k-1}~\simeq ~SO(2k)/(U(1)\times U(k-1)).
\ee 
%%%%%%%%%%%%%
} 
%%%%%%%%%%%%%%%
\be
S_F^3~\simeq ~(SU(2)_L\otimes SU(2)_R)/(U(1)_L\otimes U(1)_R)~\simeq ~S^2_L\otimes S^2_R, 
\label{sfclassicallimit}
\ee
%%%%%%%%%%%%%%%%
meaning that the fuzzy three-sphere is essentially the product of two independent fuzzy two-spheres. 
As the zero-mode degeneracy  is equal to the dimension of  fuzzy sphere \cite{Hasebe-2014-1}, it may be natural  that the zero-modes on $S^3$ is given by the product of the zero-modes on two $S^2$s.

%%%%%%%%%%%%%%%%%%%%%%%%%%%%%%%%%%%%%%%%%%%%%%%%%%%%%%%%%
\subsubsection{Reduction to the free Weyl model}\label{subsub:redufreeweyl}
%%%%%%%%%%%%%%%%%%%%%%%%%%%%%%%%%%%%%%%%%%%%%%%%%%%%%%

In  the free background limit $I\rightarrow 0$, the synthesized gauge field becomes the spin connection, % $SU(2)$ ``gauge'' field of spin $1/2$,
 and  $\mathcal{L}_{\mu\nu}|_{I/2=0}=-ix_{\mu}(\partial_{\nu}+i\omega_{\nu})+ix_{\nu}(\partial_{\mu}+i\omega_{\mu})$   is formally equivalent to the non-relativistic angular momentum,  $L_{\mu\nu}=-ix_{\mu}(\partial_{\nu}+iA^{(I/2)}_{\nu})+ix_{\nu}(\partial_{\mu}+iA^{(I/2)}_{\mu})$  with minimal monopole charge $I/2=1/2$ (see Appendices \ref{subsec:freeweyl} and \ref{subsec:freeweylsquare}). 
Therefore, the eigenstates of %the Casimir  
${\mathcal{L}_{\mu\nu}^{(I/2=0)}}^2$ are given by the $SO(4)$ monopole harmonics with  $I/2=1/2$. 

\begin{figure}[tbph]
\center
\includegraphics*[width=70mm]{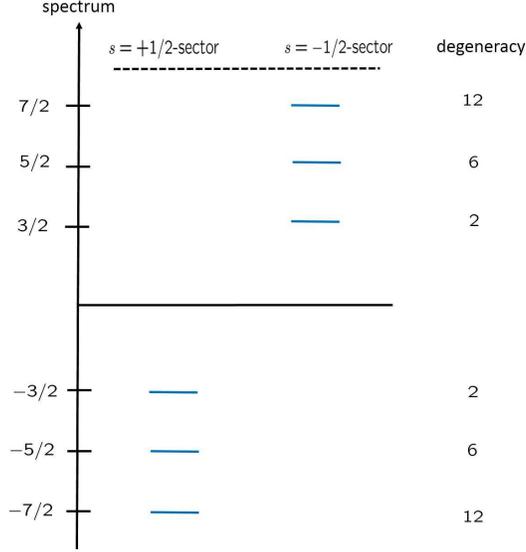}
\caption{The free Weyl operator spectrum. The special subbands with $s=\pm J^+$ (denoted by blue in Fig.\ref{schepDLspectra.fig}) survive in the free limit.} 
\label{free-dirac-spect.fig}
\end{figure}

Also in the relativistic Landau model  for $I=0$,  the Weyl-Landau operator is reduced to the free Weyl operator, $-i\fsl{\nabla}=-ie_a^{~~\alpha}\gamma^a(\partial_{\alpha}+i\omega_{\alpha})$, and the chirality parameter (\ref{rangenands}) becomes 
%%%%%%%%%%%%%%%%%
\be
s=+\frac{1}{2},~ -\frac{1}{2}, 
\ee
%%%%%%%%%%%%%%%% 
and  only (\ref{firstdiraclandaueigenva}) survives  in (\ref{weyllandauenerlev}) to give the eigenvalue 
%%%%%%%%%%%%%%%%%%%%%
\be
-i\fsl{\nabla}=\pm ({n+\frac{3}{2}}) ~~~~(n=0, 1,2,\cdots) , 
\label{spectfreeweyl}
\ee
%%%%%%%%%%%%%%%%%%%%%%%%
with degeneracy (\ref{nlandauschiralgenerelalandau1})
%%%%%%%%%%%%%%%%%%%%%%%%%%%%
\be
(n+2)(n+1). 
\ee
%%%%%%%%%%%%%%%%%%%%%%%%%%
(In the convention (\ref{spectfreeweyl}), $n$  starts from 0 not 1 unlike  (\ref{rangenands}).)  
These indeed coincide with the known results of the free Weyl operator  (Appendix \ref{subsec:freeweyl}). 
The free Weyl operator spectra are given by Fig.\ref{free-dirac-spect.fig}. 
%%%%%%%%%%%%%%%%%%%%%%%%%%%%%%%%%%%%%%%%%%%%%%%%%%%%%%%%%%%%
%%%%%%%%%%%%%%%%%%%%%%%%%%%%%%%%%%%%%%%%%%%%%%%%%%%%%%%%%%%%%% 
When $I=0$,  $J^-$ no longer exists, and only   (\ref{diraclandauj+}) with  $J^+=1/2$  becomes  the eigenstates: 
%%%%%%%%%%%%%%%%%%%%%%
\begin{align}
&-i\fsl{\nabla}=+(\frac{3}{2}+n)~:~\bs{\Psi}_{M_L^+, M_R^+}^{(n, -1/2, 1/2)}(x)=\bs{\Phi}_{M_L^+, M_R^+}^{(n, -1/2, 1/2)}(x), \nn\\
&-i\fsl{\nabla}=-(\frac{3}{2}+n)~:~\bs{\Psi}_{M_L^+, M_R^+}^{(n, +1/2, 1/2)}(x)=\bs{\Phi}_{M_L^+, M_R^+}^{(n, +1/2, 1/2)}(x), 
\label{freediracopeigredstate}
\end{align}
%%%%%%%%%%%%%%%%%%%%%%%
where $\bs{\Phi}^{(n, s, 1/2)}_{m_L, m_R}$ denote the $SO(4)$ monopole harmonics (\ref{vectorrepsorep}).   
We can also show that (\ref{freediracopeigredstate}) is transformed to the known expression of the free Weyl operator eigenstates  $\bs{\psi}^{(\pm)}_{n, l, m, \sigma}$ (\ref{eigenstatess3diracexplicitrepr}) as 
%%%%%%%%%%%%%%%%%%%%%
\begin{align}
&\bs{\psi}^{(+)}_{n, l, m, \sigma}(\chi, \theta, \phi)=i^l ~\sigma^m~ \sum_{M_L=-L}^L \sum_{M_R=-R}^R \langle l+\frac{1}{2}, -\sigma (m+\frac{1}{2}) | L, M_L; R, M_R\rangle|_{(L, R)=(\frac{n}{2}, \frac{n+1}{2})}  \bs{\Psi}_{M_L, M_R}^{(n, -1/2,1/2)}(x), \nn\\
&\bs{\psi}^{(-)}_{n, l, m, \sigma}(\chi, \theta, \phi)=i^l~ \sigma^m~ \sum_{M_L=-L}^L \sum_{M_R=-R}^R \langle l+\frac{1}{2}, -\sigma (m+\frac{1}{2}) | L, M_L; R, M_R\rangle|_{(L, R)=(\frac{n+1}{2}, \frac{n}{2})}    \bs{\Psi}_{M_L, M_R}^{(n, +1/2,1/2)}(x), 
\label{freediraceigenrelations}
\end{align}
%%%%%%%%%%%%%%%%%%%%%%%%%
with 
%%%%%%%%%%%%%%%
$\sigma=+, ~-.$ 
%%%%%%%%%%%%%%%%%
We thus established the relations between  the obtained results of the $SO(4)$ Weyl-Landau model and  the  free Weyl model \cite{CamporesiHiguchi1996, Tranutman1995,Tranutman1993}.

%%%%%%%%%%%%%%%%%%%%%%%%%%%%%%%%
\subsection{$SO(4)$ Dirac-Landau model}\label{subsec:diraclandauso4}
%%%%%%%%%%%%%%%%%%%%%%%%%%%%%%%%

Next we analyze the Weyl-Landau operator to the Dirac-Landau operator. 
From  $SO(4)$ gamma matrices  
%%%%%%%%%%%%%%%%%%%%%%
\be
\Gamma^{a}=\begin{pmatrix}
\gamma^a & 0 \\
0 & -\gamma^a 
\end{pmatrix}~(a=1,2,3),~~~\Gamma^{4}=\begin{pmatrix}
0 & 1_{2} \\
1_{2} & 0 
\end{pmatrix}, \label{so4diragamm}
\ee
%%%%%%%%%%%%%%%%%%%%%%%%%%%%%
 with $\gamma^a$ 
(\ref{gammasigmacorr}),  the massive Dirac-Landau Hamiltonian is constructed as 
%%%%%%%%%%%%%%%%%%%%
\be
-i\fsl{D}+M\Gamma^{4}=-ie_a^{~\alpha}\Gamma^a (\partial_{\alpha}+i\mathcal{A}_{\alpha})+M \Gamma^{4}=
\begin{pmatrix}
-i\fsl{\mathcal{D}} & M \\
M & i\fsl{\mathcal{D}} 
\end{pmatrix}.  \label{massivedirachamil1}
\ee
%%%%%%%%%%%%%%%%%%%%%
The mass term appears as the off-diagonal blocks. The $SO(4)$ angular momentum operators are similarly given by  
%%%%%%%%%%%%%%%%
\be
\mathcal{L}_{\mu\nu} =\begin{pmatrix}
\mathcal{L}_{\mu\nu} & 0 \\
0 & \mathcal{L}_{\mu\nu} 
\end{pmatrix}, 
\ee
%%%%%%%%%%%%%%%%%
with $\mathcal{L}_{\mu\nu}$ (\ref{totalso4ang}) on the right-hand side.  
The chirality matrix is 
%%%%%%%%%%%%%%%%%%%%%%%%%%%%%
\be
\Gamma^{5}=-\Gamma^1 \Gamma^2\Gamma^3\Gamma^4=\begin{pmatrix}
0 & -i1_{2}  \\
i 1_{2} & 0 
\end{pmatrix}. 
\ee
%%%%%%%%%%%%%%%%%%%%%%%%%%%%
% between the two different Dirac operators, and 
The massive Dirac Hamiltonian respects both of the $SO(4)$  symmetry
%%%%%%%%%%%%%%
\be
[-i\fsl{{D}}+M\Gamma^4, \mathcal{L}_{\mu\nu} ]=0, 
\ee
%%%%%%%%%%%%%%
and the chiral symmetry  
%%%%%%%%%%%%%%%%%%%%%%
\be
\{-i\fsl{D}+M\Gamma^{4}, \Gamma^{5}\}=0 
\label{chiralsymm1}
\ee
%%%%%%%%%%%%%%%%%%%%%
or 
%%%%%%%%%%%%%%%%%%%%
\be
\Gamma^{5} (-i\fsl{D}+M\Gamma^{4})  \Gamma^{5} = i\fsl{D}-M\Gamma^{4}.  
\label{g5g5flip}
\ee
%%%%%%%%%%%%%%%%%%%%%%
The chiral symmetry guarantees the reflection symmetry of the positive and negative energy levels with respect to the zero-energy. The Dirac-Landau operator (\ref{massivedirachamil1}) can then be regarded as a Hamiltonian of  chiral topological insulator.  
(\ref{g5g5flip}) suggests that the chiral transformation is equivalent to the sign flips  of the Dirac-Landau operator and the mass parameter: 
%%%%%%%%%%%%%%%%%%%%%
\begin{subequations}
\begin{align}
-i\fsl{\mathcal{D}}&~~\rightarrow~~+i\fsl{\mathcal{D}}, \label{changediracop}\\
M&~~\rightarrow~~-M.
\end{align}
\end{subequations}
%%%%%%%%%%%%%%%%%%%%
In particular for the zeroth Landau level, the chiral transformation corresponds to  
%%%%%%%%%%%%%%%%%%%%%%
\be
s~\rightarrow~-s. 
\ee
%%%%%%%%%%%%%%%%%%%%%
Thus, the left-right symmetry of the non-relativistic $SO(4)$ Landau model can be translated as  the chiral symmetry in the Dirac-Landau model.   

\begin{figure}[tbph]
\includegraphics*[width=155mm]{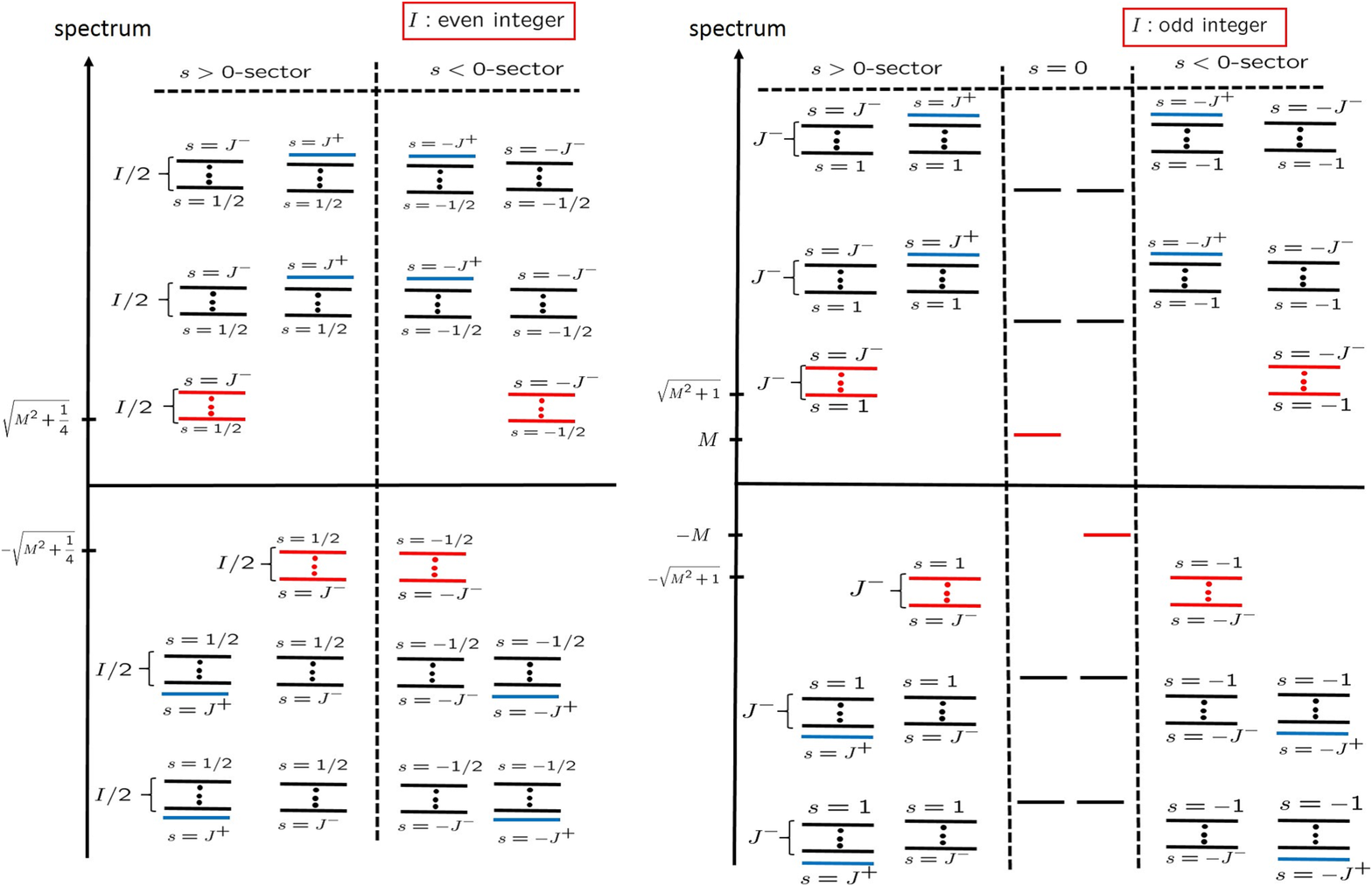}
\caption{
Energy spectra of the massive Dirac-Landau model.  
 }
\label{MassiveDL.fig}
\end{figure}

Having solved  the  eigenvalue problem of the Weyl-Landau operator, we can readily  analyze the massive  Dirac-Landau problem 
%%%%%%%%%%%%%%%%%%%%%%%%%%%%%
\be
(-i\fsl{D}+M\Gamma^{4})\bs{\Xi}_{\pm \Lambda_n(s)}=\pm \Lambda_n(s) \bs{\Xi}_{\pm \Lambda_n(s)}. 
\ee
%%%%%%%%%%%%%%%%%%%%%%%%%%%
Since $(-i\fsl{D}+M\Gamma^4)^2=(-i\fsl{D})^2+M^2=\lambda^2+M^2$, the eigenvalues are  derived as 
%%%%%%%%%%%%%%%%%%
\be
\pm \Lambda_n(s)=\pm \sqrt{{\lambda_n(s)}^2+M^2} , 
\label{Diracenergymass}
\ee
%%%%%%%%%%%%%%%%%%
or 
%%%%%%%%%%%%%%%%%%
\begin{subequations}
\begin{align}
&+\Lambda_n(\frac{I+1}{2})=+\sqrt{(\frac{I+1}{2}+n)^2+M^2},~~~~~~~  -\Lambda_n(\frac{I+1}{2})=-\sqrt{(\frac{I+1}{2}+n)^2+M^2},~~~  \label{masscase1}\\
&+\Lambda_n(s)=+\sqrt{n(I+1+n)+s^2 +M^2},~~~~~~-\Lambda_n(s)=-\sqrt{n(I+1+n)+s^2 +M^2}, \label{masscase2}
\\
&+\Lambda_0(s)=+\sqrt{s^2+M^2},~~~~~~~~~~~~~~~~~~~~~~~~~~~  -\Lambda_0(s)=-\sqrt{s^2+M^2}, \label{masscase3} 
\end{align}\label{eigenvaluesmassivediraclan}
\end{subequations}
%%%%%%%%%%%%%%%%%%
where $n=1,2,\cdots$ and $s=\frac{I-1}{2}, \frac{I-1}{2}-1, \cdots, -\frac{I-1}{2}$ (Fig.\ref{MassiveDL.fig}).  
%%%%%%%%%%%%%%%%%%%%%%%%%%%%%%%%%%%%%%%%%%%%%%%%%%%%%%%%%%%%
%%%%%%%%%%%%%%%%%%%%%%%%%%%%%%%%%%%%%%%%%%%%%%%%%%%%%%%%%%%%%%
Notice the gap between the positive and negative subbands of the zeroth Landau level is  $\sim M$, and the gap does not close even in the thermodynamic limit.  
Each of the eigenvalues (\ref{eigenvaluesmassivediraclan}) has the following degeneracy: 
%%%%%%%%%%%%%%%%%%%%
\begin{subequations}
\begin{align}
&2(n+I+1)n, \\
&(2n+I+2s+1)(2n+I-2s+1)\times (1-\delta_{I, 0}) , \\
&(\frac{I+1}{2}+s) ((\frac{I+1}{2}-s)) \times (2-\delta_{s,0} (1-\delta_{M, 0})) (1-\delta_{I, 0}) . 
\end{align}
\end{subequations}
%%%%%%%%%%%%%%%%%%%%%
For even $I$, zero-modes do not appear, while for odd $I$, the states withe energies $M$ and $-M$  $(s=0)$ coincide to become  degenerate zero-modes in the massless limit.  
For (\ref{masscase2}), the degenerate eigenstates are given by 
%%%%%%%%%%%%%%%%%%%%
\begin{subequations}
\begin{align}
&\bs{\Xi}^{(1)}_{+ \Lambda_n(s)}={\sqrt{\frac{\Lambda_n(s)+\lambda_n(s)}{2\Lambda_n(s)}}}
 \begin{pmatrix}
 ~\bs{\Psi}_{+\lambda_n(s)}\\
\frac{M}{\Lambda_n(s)+\lambda_n(s)}~\bs{\Psi}_{+\lambda_n(s)}
\end{pmatrix}, ~~~ \bs{\Xi}^{(2)}_{+ \Lambda_n(s)}={\sqrt{\frac{\Lambda_n(s)+\lambda_n(s)}{2\Lambda_n(s)}}} \begin{pmatrix}
\frac{M}{\Lambda_n(s)+\lambda_n(s)}~\bs{\Psi}_{-\lambda_n(s)} \\
  ~\bs{\Psi}_{-\lambda_n(s)}
\end{pmatrix}, \label{+lambdandiracmass} \\
&\bs{\Xi}^{(1)}_{- \Lambda_n(s)}={\sqrt{\frac{\Lambda_n(s)+\lambda_n(s)}{2\Lambda_n(s)}}} \begin{pmatrix}
-\frac{M}{\Lambda_n(s)+\lambda_n(s)} ~\bs{\Psi}_{+\lambda_n(s)} \\
\bs{\Psi}_{+\lambda_n(s)}
\end{pmatrix}, ~\bs{\Xi}^{(2)}_{- \Lambda_n(s)}={\sqrt{\frac{\Lambda_n(s)+\lambda_n(s)}{2\Lambda_n(s)}}}  \begin{pmatrix}
~\bs{\Psi}_{-\lambda_n(s)} \\
-\frac{ M}{\Lambda_n(s)+\lambda_n(s)} ~\bs{\Psi}_{-\lambda_n(s)} 
\end{pmatrix}, \label{eigenstatesfor-lambdabig}
\end{align}\label{diraclandaueigendetail}
\end{subequations}
%%%%%%%%%%%%%%%%%%%%%
where $\bs{\Psi}_{\pm \lambda_n(s)}$ represents the Weyl-Landau operator eigenstates with  eigenvalues $\pm\lambda_n(s)$ (\ref{eigenstateweyllandauei}).   
$\bs{\Xi}_{\pm \Lambda_n(s)}$ transforms as a representation of the $SO(4)$ group generated by $\mathcal{L}_{\mu\nu}$. 
The chiral symmetry relates the eigenstates (\ref{diraclandaueigendetail}) as  
%%%%%%%%%%%%%%%%%%
\be
\bs{\Xi}^{(1)}_{+ \Lambda_n(s)}=+i\Gamma^5 \bs{\Xi}^{(2)}_{- \Lambda_n(s)},~~~~~~ \bs{\Xi}^{(2)}_{+ \Lambda_n(s)}=-i\Gamma^5 \bs{\Xi}^{(2)}_{- \Lambda_n(s)}. 
\ee
%%%%%%%%%%%%%%%%%
Similarly  for (\ref{masscase1}) and (\ref{masscase3}), the degenerate eigenstates are respectively given by 
%%%%%%%%%%%%%%%%%%%%%%
\begin{subequations}
\begin{align}
&\bs{\Xi}^{(2)}_{\Lambda_n(J^+)},~\bs{\Xi}^{(1)}_{\Lambda_n(-J^+)},~~~~~~~~~~~~~~~~~~~~~~\bs{\Xi}^{(1)}_{-\Lambda_n(J^+)},~\bs{\Xi}^{(2)}_{-\Lambda_n(-J^+)}, \\
&\bs{\Xi}^{(1)}_{\Lambda_0(s)} ~(s\ge 0),~\bs{\Xi}^{(2)}_{\Lambda_0(s)} ~(s\le 0),~~~~~~~~\bs{\Xi}^{(2)}_{-\Lambda_0(s)}~(s\ge 0),~\bs{\Xi}^{(1)}_{-\Lambda_0(s)}~(s\le 0). \label{correszerothll}
\end{align}\label{otherdegenerateeigenstatemassive}
\end{subequations}
%%%%%%%%%%%%%%%%%%%%%%%
In the zeroth Landau level  of $s=0$, $\bs{\Xi}^{(1)}$ and $\bs{\Xi}^{(2)}$ (\ref{correszerothll}) are not independent: %we have the special relation
%%%%%%%%%%%%%%%%%%%%
\be
\bs{\Xi}^{(1)}_{\Lambda_0(0)=M}=\bs{\Xi}^{(2)}_{\Lambda_0(0)=M} =\frac{1}{\sqrt{2}} 
\begin{pmatrix}
\bs{\Psi}_{\lambda=0} \\
\bs{\Psi}_{\lambda=0}
\end{pmatrix},~~~~\bs{\Xi}^{(2)}_{-\Lambda_0(0)=-M}=-\bs{\Xi}^{(1)}_{\Lambda_0(0)=-M} =\frac{1}{\sqrt{2}} 
\begin{pmatrix}
\bs{\Psi}_{\lambda=0} \\
-\bs{\Psi}_{\lambda=0}
\end{pmatrix}.
\ee
%%%%%%%%%%%%%%%%%%%%%%%%

%%%%%%%%%%%%%%%%%%%%%%%%%%%%%%%%%%%%%%%%%
\subsubsection{Supersymmetric Landau model}\label{subsec:superlandau}
%%%%%%%%%%%%%%%%%%%%%%%%%%%%%%%%%%%%%%%%

The square of the Dirac-Landau operator yields a supersymmetric quantum mechanical Hamiltonian: 
%%%%%%%%%%%%%%%%%%%%%%%%
\be
H_{\text{SUSY}} =(-i\fsl{D})^2=\begin{pmatrix}
(-i\fsl{\mathcal{D}})^2 & 0 \\
0 & (+i\fsl{\mathcal{D}})^2
\end{pmatrix}=\{Q , Q^{\dagger}\}, 
\label{susyhamildef}
\ee
%%%%%%%%%%%%%%%%%%%%%%%%%
where  
%%%%%%%%%%%%%%%%%%%%%
\be
Q=\begin{pmatrix}
0 & -i\fsl{\mathcal{D}} \\
0 & 0 
\end{pmatrix}, ~~~Q^{\dagger}=\begin{pmatrix}
0 &  0 \\
-i\fsl{\mathcal{D}} & 0 
\end{pmatrix}.  
\ee
%%%%%%%%%%%%%%%%%%%
$Q$ and $Q^{\dagger}$ are supercharges that satisfy 
%%%%%%%%%%%%%%%%%%%%%%%%%%
\be
Q^2=0, ~~~{Q^{\dagger}}^2=0, ~~~
[H_{\text{SUSY}}, Q]=[H_{\text{SUSY}}, Q^{\dagger}]=0. 
\ee
%%%%%%%%%%%%%%%%%%%%%
The SUSY Hamiltonian (\ref{susyhamildef}) consists of two identical spinor Landau Hamiltonians, and its  eigenvalues are $\lambda_n(s)^2=n(n+I+1)+s^2$ with the degenerate eigenstates 
%%%%%%%%%%%%%%%%%%%%%%%
\be
\bs{\Xi}_{\lambda_n(s)}^{(1)} =
\begin{pmatrix}
\bs{\Psi}_{\lambda_n(s)} \\
0
\end{pmatrix},~~\bs{\Xi}_{\lambda_n(s)}^{(2)} =
\begin{pmatrix}
0 \\
\bs{\Psi}_{-\lambda_n(s)}
\end{pmatrix},~~\bs{\Xi}_{-\lambda_n(s)}^{(1)} =
\begin{pmatrix}
0 \\
\bs{\Psi}_{\lambda_n(s)} 
\end{pmatrix},~~\bs{\Xi}_{-\lambda_n(s)}^{(2)} =
\begin{pmatrix}
\bs{\Psi}_{-\lambda_n(s)} \\
0
\end{pmatrix}.
\ee
%%%%%%%%%%%%%%%%%%%%%%%  
The Witten parity 
%%%%%%%%%%%%%%%%%%%%%
\be
W%=\frac{2}{H_{\text{SUSY}}}QQ^{\dagger}-1 =\frac{1}{H_{\text{SUSY}}}[Q, Q^{\dagger}]
 =\frac{[Q, Q^{\dagger}]}{\{Q, Q^{\dagger}\}} 
\ee
%%%%%%%%%%%%%%%%%%%%%%%%
is given by 
%%%%%%%%%%%%%%%%%%%%%
\be
W= 
\frac{1}{\lambda^2}
\begin{pmatrix}
(-i\fsl{\mathcal{D}})^2 &  0  \\
0 & -(-i\fsl{\mathcal{D}})^2
\end{pmatrix}
=\begin{pmatrix}
1 & 0 \\
0 & -1
\end{pmatrix} . 
\ee
%%%%%%%%%%%%%%%%%%%%
$\begin{pmatrix}\bs{\Psi}_{\pm\lambda} \\0 \end{pmatrix}$ belongs to the Witten parity $+$ (``fermionic'') sector, while $\begin{pmatrix} 0 \\ \bs{\Psi}_{\pm\lambda}  \end{pmatrix}$  the Witten parity $-$ (``bosonic'') sector.  They are the  superpartner  related by the SUSY transformation: 
%%%%%%%%%%%%%%%%%%%
\be
\begin{pmatrix}
\bs{\Psi}_{\pm \lambda} \\
0 
\end{pmatrix} =\pm \frac{1}{\lambda} Q \begin{pmatrix}
0 \\
\bs{\Psi}_{\pm \lambda}
\end{pmatrix} , ~~~\begin{pmatrix}
0 \\
\bs{\Psi}_{\pm \lambda}
\end{pmatrix} =\pm \frac{1}{\lambda} Q^{\dagger} \begin{pmatrix}
\bs{\Psi}_{\pm \lambda} \\
0 
\end{pmatrix}. ~~~~~(\lambda\neq 0)
\ee
%%%%%%%%%%%%%%%%%%%%
For odd $I$, the ground-state of $H_{\text{SUSY}}$  is given by SUSY invariant zero-energy state (good SUSY),   while for even $I$, the ground-state  is  not  zero-energy state 
and  does not respect  the SUSY (broken SUSY). 
%;  
%for odd $I$, the SUSY is good, while for even $I$, the SUSY is broken. 
We can also consider the square of the massive Dirac-Landau Hamiltonian
%%%%%%%%%%%%%%%%%%%%
\be
H_M =(-i\fsl{D} +M\Gamma^4)^2 =(-i\fsl{D})^2+M^2=H_{\text{SUSY}}+M^2.   
\label{susyhammass}
\ee
%%%%%%%%%%%%%%%%%%%%%%
The mass term just shifts the zero-energy of the spectrum of $H_{\text{SUSY}}$.

%%%%%%%%%%%%%%%%%%%%%%%%%%%%%%%%%%%%%%%%%%%%%%%
\section{Matrix Geometry}\label{sec:matrixgeo}
%%%%%%%%%%%%%%%%%%%%%%%%%%%%%%%%%%%%%%%%%%%%%%%%%

In the above, we introduced the various $SO(4)$  Landau models and solved their eigenvalue problem. 
With the developed technologies and results, we are now ready to evaluate the matrix geometry of the $SO(4)$ Landau levels. %\footnote{  
%Though  we utilized the chiral Hopf map to derive an operator representation of the $S^3$ coordinates in the former works \cite{Hasebe-2017, Hasebe-2014-2}, the derivation is rather speculative and restricted to the lowest Landau level. }   
We concretely derive the matrix elements of  $S^3$ coordinates in each of the Landau levels based on  the level projection method  
% that is a most straightforward way 
to appreciate the emergent non-commutative geometry. 
%, and the odd dimensional matrix geometry in the context of the Landau model  is  the first observation. 
 We first derive the fuzzy three-sphere geometry of the $SO(4)$ non-relativistic Landau model in Sec.\ref{subsec:llprojection}, and next we explore fuzzy geometry  of the spinor Landau model and relativistic Landau models in Sec.\ref{subsec:llprojectionspinorll} and Sec.\ref{subsec:llprojectionrelll}. We point out that the massive Dirac-Landau model accommodates the two fuzzy three-spheres whose interaction is induced by the mass term (Sec.\ref{subsec:llprojectionrelll}). 

We switch the notation from  $\bs{\Phi}^{(n, s, I/2)}_{m_L, m_R}$ to   $\bs{\Phi}^{[l_L, l_R, I/2]}_{m_L, m_R}$ in this section.  

%%%%%%%%%%%%%%%%%%%%%%%%%%%%%%
\subsection{Landau level projection}\label{subsec:llprojection} % of the $SO(4)$ non-relativistic Landau model}
%%%%%%%%%%%%%%%%%%%%%%%%%%%%%%

We  apply the level projection method to the $SO(4)$ non-relativistic Landau model. 
The basis states of the Landau level subband $E_n(s)$ (\ref{so4energynonrel}) are the $SO(4)$ monopole harmonics 
%With a given monopole charge $I/2$, the energy of the  subband of the Landau level is determined by $|s|$ and $n$. 
%The degenerate Landau level subbands are spanned by the $SO(4)$ monopole harmonics 
carrying  the indices   
%%%%%%%%%%%%%%%%
\be
(n , I/2 , s), ~~~~(n, I/2, -s),  
\ee
%%%%%%%%%%%%%%%%%
or in the $SO(4)\simeq SU(2)\otimes SU(2)$ notation, 
%%%%%%%%%%%%%%%%%%%
\be
(l_L, l_R) =(\frac{1}{2}(n+\frac{I}{2}+{s}), \frac{1}{2}(n+\frac{I}{2}-{s})), ~~~~(l'_L, l'_R) =(l_R, l_L).
\label{defs12lllr}
\ee
%%%%%%%%%%%%%%%%%%%
The matrix elements of $x_{\mu}$ in the Landau level subband are then given by  
%We therefore consider the following matrix elements to  derive matrix geometry  
%%%%%%%%%%%%%%%%%%%%%%%
\be
X_{\mu}(n) =
\begin{pmatrix}
 \langle l_L, l_R|x_{\mu}|l_L, l_R\rangle    &    \langle l_L, l_R|x_{\mu}|l_R, l_L\rangle   \\
   \langle l_R, l_L|x_{\mu}|l_L, l_R\rangle    &    \langle l_R, l_L|x_{\mu}|l_R, l_L\rangle 
\end{pmatrix}. 
\label{toderivexmumatr}
\ee
%%%%%%%%%%%%%%%%%%%
Each of the blocks in (\ref{toderivexmumatr}) denotes a square matrix of $d(n, s)\times d(n, s)$ with 
%%%%%%%%%%%%%%%%%%%%%%%%%
\be
d(n, s)=d(n,-s)=(2l_L+1)(2l_R+1) =(n+\frac{I}{2}+s+1) (n+\frac{I}{2}-s+1). 
\ee
%%%%%%%%%%%%%%%%%%%%%%%%%%
To evaluate  (\ref{toderivexmumatr}), from (\ref{xsandphin1}) we represent $x_{\mu}$ as 
%%%%%%%%%%%%%%%%%%%%%%%%%%%%
\begin{align}
&x_1=-i\frac{\pi}{2}(\Phi^{[1/2, 1/2, 0]}_{1/2, 1/2} (x) - \Phi^{[1/2, 1/2, 0]}_{-1/2, -1/2} (x)), \nn\\
&x_2=-\frac{\pi}{2}(\Phi^{[1/2, 1/2, 0]}_{1/2, 1/2} (x) + \Phi^{[1/2, 1/2, 0]}_{-1/2, -1/2} (x)), \nn\\
&x_3=i\frac{\pi}{2}(\Phi^{[1/2, 1/2, 0]}_{1/2, -1/2} (x) + \Phi^{[1/2, 1/2, 0]}_{-1/2, 1/2} (x)), \nn\\
&x_4=\frac{\pi}{2}(\Phi^{[1/2, 1/2, 0]}_{1/2, -1/2} (x) - \Phi^{[1/2, 1/2, 0]}_{-1/2, 1/2} (x)), 
\label{s3coordinatessphericalrel}
\end{align}
%%%%%%%%%%%%%%%%%%%%%%%%%%%
 and derive the matrix elements, such as $\langle l_L, l_R|\Phi^{[1/2, 1/2, 0]}_{\sigma/2, \tau/2}|l_L, l_R\rangle$. %, to derive  $\langle l_L, l_R|x_{\mu}|l_L, l_R\rangle$. 
 The $SO(4)$ vector $x_{\mu}$ carries the $SU(2)_L\otimes SU(2)_R$ index, $(l_L, l_R)=(1/2, 1/2)$, and so the matrix elements of (\ref{toderivexmumatr}) take finite values only for the cases that each $SU(2)$ index  of  $SU(2)_L\otimes SU(2)_R$  between the ket and bra differs by $\pm 1/2$, $i.e.$ $|l_L-l_R|=1/2$ or  $|s|=1/2$ (recall (\ref{defs12lllr})).  
 %     =\int d\Omega_3~ {\bs{\Phi}^{(l_L, l_R, I/2)}_{m_L, m_R}}^{\dagger} ~\Phi^{(1/2, 1/2, %0)}_{\sigma/2, \tau/2} ~{\bs{\Phi}^{(l_L, l_R, I/2)}_{m_L, m_R}}$. 
%Using the formula (\ref{threeintegralso4lleigen}), we derive the matrix elements of $x_{\mu}$. 
Explicit evaluation of (\ref{toderivexmumatr}) is as follows.  
%Let us first consider the diagonal blocks in (\ref{toderivexmumatr}). 
%For the Landau level $(l_L, l_R) =(\frac{1}{2}(n+\frac{I}{2}+s), \frac{1}{2}(n+\frac{I}{2}-s) )$, we derive the $(2l_L+1)(2l_R+1) \times (2l_L+1)(2l_R+1)$ matrix. From 
From (\ref{threeintegralso4lleigen}), we have 
%, the intra Landau level matrix elements are evaluated by the following  formula: 
%%%%%%%%%%%%%%%%%%%%%%%%%%%%%
\begin{align}
&\langle l_L, l_R|\Phi^{[1/2, 1/2, 0]}_{\sigma/2, \tau/2}|l_L, l_R\rangle  \nn\\
&=\int d\Omega_3~ {\bs{\Phi}^{[l_L, l_R, I/2]}_{m_L, m_R}}^{\dagger} ~\Phi^{[1/2, 1/2, 0]}_{\sigma/2, \tau/2} ~{\bs{\Phi}^{[l_L, l_R, I/2]}_{m_L, m_R}}\nn\\
&=\sqrt{\frac{(2l_L+1)(2l_R+1)(I+1)}{\pi^2}}(-1)^{-(l_L+l_R+\frac{I}{2}+\frac{1}{2})}
\begin{Bmatrix}
l_L & l_R & {I}/2 \\
l_R & l_L & 1/2
\end{Bmatrix}
~C_{1/2, \sigma/2;~l_L, n_L}^{l_L, m_L}~C_{1/2, \tau/2;~l_R, n_R}^{l_R, m_R}\nn\\
&=0, ~~~~~~~~~~~(\sigma, \tau=+1, -1)
\end{align}
%%%%%%%%%%%%%%%%%%%%%%%%%%%%%%%%%
where we used   
%%%%%%%%%%%%%%%%%%%
%\be
%\begin{Bmatrix}
%l_L & l_R & {I}/2 \\
%l_R & l_L & 1/2
%\end{Bmatrix}=
$C_{1/2, \sigma/2;~l_L, n_L}^{l_L, m_L}=C_{1/2, \tau/2;~l_R, n_R}^{l_R, m_R}=0$.   
%\ee
%%%%%%%%%%%%%%%%%%%%
Therefore,  the diagonal blocks of (\ref{toderivexmumatr})  always vanish: 
%%%%%%%%%%%%%%%%%%%%%%%
\be
\langle l_L, l_R|x_{\mu}|l_L, l_R\rangle =\langle l_R, l_L|x_{\mu}|l_R, l_L\rangle=\bs{0}_{(2l_L+1)(2l_R+1)},  
\ee
%%%%%%%%%%%%%%%%%%%%%%%%%%%%%%%%%%%%%%%%%%%%%%%%%%%%% 
and then $X_{\mu}(n)$  take the form of 
%%%%%%%%%%%%%%%%%%%%%%
\be
X_{\mu}(n) =
\begin{pmatrix}
 \langle l_L, l_R|x_{\mu}|l_L, l_R\rangle    &    \langle l_L, l_R|x_{\mu}|l_R, l_L\rangle   \\
   \langle l_R, l_L|x_{\mu}|l_L, l_R\rangle    &    \langle l_R, l_L|x_{\mu}|l_R, l_L\rangle 
\end{pmatrix}
=\begin{pmatrix}
\bs{0}_{d(n, s)} &  Y^{(I)}_{\mu}(n) \\
(Y^{(I)}_{\mu}(n))^{\dagger} & \bs{0}_{d(n, s)}
\end{pmatrix}.  \label{s3matrixelexmu}
\ee
%%%%%%%%%%%%%%%%%%%%%%%
The  matrix elements of $Y_{\mu}^{(I)}(n)$ can be derived as follows. Also from (\ref{2threeintegralso4lleigen}), we have 
%%%%%%%%%%%%%%%%%%%%%%%%%%%%%%%%
\begin{align}
&\langle l_L, l_R|\Phi^{[1/2, 1/2, 0]}_{\sigma/2, \tau/2}|l_R, l_L\rangle\nn\\
&=\int d\Omega_3 ~{\bs{\Phi}_{m_L, m_R}^{[l_L, l_R, I/2]}}^{\dagger}~\Phi_{\sigma/2, \tau/2}^{[1/2, 1/2, 0]} ~\bs{\Phi}_{n_R, n_L}^{[l_R, l_L, I/2]}\nn\\
&=(-1)^{n+1}(-1)^{s+\frac{1}{2}} \frac{\sqrt{(2l_L+1)(2l_R+1)}}{\pi} \begin{Bmatrix}
l_L & l_R & I/2 \\
l_L & l_R & 1/2
\end{Bmatrix} ~C_{\frac{1}{2}, \frac{\sigma}{2};~l_R, n_R}^{l_L, m_L}~C_{\frac{1}{2}, \frac{\tau}{2};~l_L, n_L}^{l_R, m_R}\nn\\
&=-\tau \frac{1}{\pi}
%\frac{2}{\pi} \frac{I+1}{(2n+I+1)(2n+I+3)}
 ~\delta_{m_L, n_R+\frac{\sigma}{2}}~\delta_{m_R, n_L+\frac{\tau}{2}}\times \sqrt{(l_L-\tau n_L)(l_R+1+\sigma n_R)} \begin{Bmatrix}
l_L & l_R & I/2 \\
l_L & l_R & 1/2
\end{Bmatrix},  
\label{threeprodsecondv}
\end{align}
%%%%%%%%%%%%%%%%%%%%%%%%%%%%%%%%%
where we used (for $l_L=l_R+\frac{1}{2}$)
%%%%%%%%%%%%%%%%%%%%%%%%%
\be
C_{1/2, \sigma/2; l_R, n_R}^{l_L, m_L}=\sqrt{\frac{l_R+1+\sigma n_R}{2l_R+1}} ~\delta_{m_L, n_R+\frac{\sigma}{2}}, ~~~C_{1/2, \tau/2; l_R, n_R}^{l_L, m_L}=\tau \sqrt{\frac{l_L-\tau n_L}{2l_L+1}} ~\delta_{m_R, n_L+\frac{\tau}{2}}. 
\label{clebshexptwo}
\ee
%%%%%%%%%%%%%%%%%%%%%%%%%%
For even $I$,  
%%%%%%%%%%%%%%%%%%
\be
\begin{Bmatrix}
l_L & l_R & I/2 \\
l_L & l_R & 1/2
\end{Bmatrix}=0, ~~~~(I=0,2,4,\cdots)
\ee
%%%%%%%%%%%%%%%%%%
and then $Y_{\mu}=0$. 
Meanwhile for odd $I$, 
%%%%%%%%%%%%%%%%%%%%%%%%%%%%%%%%
\be
\begin{Bmatrix}
l_L & l_R & I/2 \\
l_L & l_R & 1/2
\end{Bmatrix} 
 =   (-1)^{n+1} \frac{2(I+1)}{(2n+I+1)(2n+I+3)}~\delta_{|s|, \frac{1}{2} }, 
 \label{explioddithreej}
\ee
%%%%%%%%%%%%%%%%%%%%%%%%%%%%%%%%%%
and, as mentioned above,  non-zero matrix elements appear only for   
%%%%%%%%%%%%%%%%%%%%
\be
|s|={1}/{2}, 
\ee
%%%%%%%%%%%%%%%%%%%%%%
%Substituting (\ref{explioddithreej}) to (\ref{threeprodsecondv}), 
and from (\ref{explioddithreej}) we obtain 
%%%%%%%%%%%%%%%%%%%%%%%%%%%
\be
Y_{\mu}^{(I)} (n) =\frac{I+1}{(2n+I)(2n+I+3)} \hat{Y}_{\mu}^{(2(l_L+l_R)=2n+I)}, 
\ee
%%%%%%%%%%%%%%%%%%%%%%%%%%
where 
%%%%%%%%%%%%%%%%%%%%%%%%%%%%%%%
\begin{align}
&\hat{Y}_1^{(I)}= \nn\\
&\!\!\!\!\!\!\!\!i\biggl(\delta_{m_L, n_R+\frac{1}{2}} \delta_{m_R, n_L+\frac{1}{2}} \sqrt{(l_L-n_L)(l_R+1+n_R)}+\delta_{m_L, n_R-\frac{1}{2}} \delta_{m_R, n_L-\frac{1}{2}} \sqrt{(l_L+n_L)(l_R+1-n_R)}\biggr)\biggr|_{(l_L,l_R)=(\frac{I}{4}+\frac{1}{4}, \frac{I}{4}-\frac{1}{4})}, \nn\\
&\hat{Y}_2^{(I)}=  \nn\\
&\!\!\!\!\!\!\!\!\biggl(\delta_{m_L, n_R+\frac{1}{2}} \delta_{m_R, n_L+\frac{1}{2}} \sqrt{(l_L-n_L)(l_R+1+n_R)}-\delta_{m_L, n_R-\frac{1}{2}} \delta_{m_R, n_L-\frac{1}{2}} \sqrt{(l_L+n_L)(l_R+1-n_R)}\biggr)\biggr|_{(l_L,l_R)=(\frac{I}{4}+\frac{1}{4}, \frac{I}{4}-\frac{1}{4})}, \nn\\
&\hat{Y}_3^{(I)}= \nn\\
&\!\!\!\!\!\!\!\!i\biggl(\delta_{m_L, n_R+\frac{1}{2}} \delta_{m_R, n_L-\frac{1}{2}} \sqrt{(l_L+n_L)(l_R+1+n_R)}-\delta_{m_L, n_R-\frac{1}{2}} \delta_{m_R, n_L+\frac{1}{2}} \sqrt{(l_L-n_L)(l_R+1-n_R)}\biggr)\biggr|_{(l_L,l_R)=(\frac{I}{4}+\frac{1}{4}, \frac{I}{4}-\frac{1}{4})}, \nn\\
&\hat{Y}_4^{(I)}= \nn\\
&\!\!\!\!\!\!\!\!\biggl(\delta_{m_L, n_R+\frac{1}{2}} \delta_{m_R, n_L-\frac{1}{2}} \sqrt{(l_L+n_L)(l_R+1+n_R)}+\delta_{m_L, n_R-\frac{1}{2}} \delta_{m_R, n_L+\frac{1}{2}} \sqrt{(l_L-n_L)(l_R+1-n_R)}\biggr)\biggr|_{(l_L,l_R)=(\frac{I}{4}+\frac{1}{4}, \frac{I}{4}-\frac{1}{4})}. 
\end{align}
%%%%%%%%%%%%%%%%%%%%%%%%%%%%%%%%%
$\hat{Y}_{\mu}^{(I)}$ are the off-diagonal blocks of the $SO(4)$ gamma matrices in the symmetric representation:\footnote{For instance,  
%%%%%%%%%%%%%%%
\be
\Gamma_{\mu}^{(I=1)} =\begin{pmatrix}
0 & i\sigma_i \\
-i\sigma_i & 0 
\end{pmatrix}, ~~\begin{pmatrix}
0 & 1_2 \\
1_2 & 0 
\end{pmatrix}. 
\ee
%%%%%%%%%%%%%%%%%
} 
%%%%%%%%%%%%%%%%%%%%%
\be
\Gamma_{\mu}^{(I)}=\begin{pmatrix}
\bs{0} & \hat{Y}_{\mu}^{(I)} \\
(\hat{Y}_{\mu}^{(I)})^{\dagger} & \bs{0}
\end{pmatrix}, ~~~~(I=1,3,5,\cdots) 
\label{so4gammasymmge}
\ee
%%%%%%%%%%%%%%%%%%%%%%
which  satisfy  \cite{Ramgoolam2002,JabbariTorabian2005} 
%%%%%%%%%%%%%%%%%%%%%%%%
\be
\sum_{\mu=1}^4 {\Gamma_{\mu}^{(I)}}^2=\frac{1}{2}(I+1)(I+3)\bs{1}_{\frac{(I+3)(I+1)}{2}}.  ~~~(I=1,3,5,\cdots) 
\label{gammamatrisum}
\ee
%%%%%%%%%%%%%%%%%%%%%%%%%%%
(\ref{s3matrixelexmu}) is now obtained as   
%%%%%%%%%%%%%%%%%%%%%%%%%%%%%
\be
X_{\mu}(n)=\frac{I+1}{(2n+I+1)(2n+I+3)}\Gamma_{\mu}^{(2n+I)}. 
\ee
%%%%%%%%%%%%%%%%%%%%%%%%%%%%%%
%Thus, we have shown that for any LL, only for $s=\pm 1/2$, generate the geometry of the fuzzy three-sphere. 

To summarize, the non-trivial matrix geometry appears only in the subband $|s|=1/2$ for odd $I$: 
%%%%%%%%%%%%%%%%%%
\be
X_{\mu}(n)= 
\begin{pmatrix}
\bs{0} _{d(n,1/2)} &     Y_{\mu}^{(I)}(n)  \\ 
{Y_{\mu}^{(I)}(n)}^{\dagger} & \bs{0} _{d(n,1/2)} 
\end{pmatrix} = \frac{I+1}{(2n+I+1)(2n+I+3)}\Gamma_{\mu}^{(2n+I)}.  
\label{tosummazmu}
\ee
%%%%%%%%%%%%%%%%%%%  
In particular, for the lowest Landau level ($n=0$),  
%%%%%%%%%%%%%%%%%%%
\be
 X_{\mu}(n=0)=\frac{1}{I+3}~\Gamma_{\mu}^{(I)}.~~~~(I=1,3,5,\cdots.) 
\label{n0xmugammarel}
\ee
%%%%%%%%%%%%%%%%%%%
$X_{\mu}$ (\ref{tosummazmu}) satisfy the relation 
%%%%%%%%%%%%%%%%%%%%%%%%%%%%
\be
\sum_{\mu=1}^4 X_{\mu}(n)X_{\mu}(n)= {R_n^{(I)}}^2 \bs{1}_{ \frac{1}{2}{(2n+I+1)(2n+I+3)} }
\label{gammamatrisumx}
\ee
%%%%%%%%%%%%%%%%%%%%%%%%%%%%%
with 
%%%%%%%%%%%%%%%%%%%%%%%
\be
R_n^{(I)} \equiv \frac{I+1}{\sqrt{2(2n+I+1)(2n+I+3)}}. 
\label{llindexdepradius}
\ee
%%%%%%%%%%%%%%%%%%%%%%%%
The relation (\ref{gammamatrisumx}) is invariant under the $SO(4)$ rotations and is a non-commutative counterpart of the definition of three-sphere. We thus find that $X_{\mu}$ denote the matrix  coordinates of  fuzzy three-sphere. 
The radius decreases as the Landau level increases. A similar behavior is observed  in the fuzzy two-sphere of the $SO(3)$ Landau model \cite{Hasebe-2015}. 
%Usually, the level projection is realized in the strong magnetic field limit  $I\rightarrow \infty$. Meanwhile in this case, the  subband energy interval does not depend on the magnetic field, and it is not probable to project the physics onto the $|s|=1/2$ subbands solely.  
 %However, 
In each Landau level, only the $|s|=1/2$ subband realizes the fuzzy three-sphere geometry and  each Landau level accommodates just one fuzzy three-sphere.  
%In this sense, the fuzzy sphere geometry is realized in each Landau level. 

\begin{figure}[tbph]
\center
%\hspace{5cm}
\includegraphics*[width=50mm]{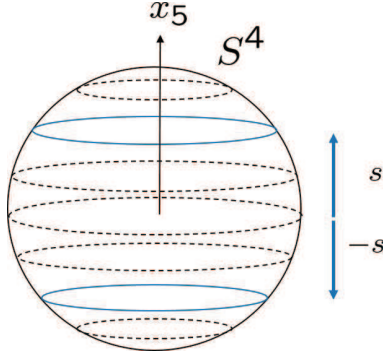}
\caption{$SO(4)$ Landau level subbands as the two latitudes on the virtual fuzzy four-sphere }
\label{TwoLatitudes.fig}
%\vspace{-3mm}
\end{figure}

The fuzzy three-sphere geometry of the $SO(4)$ Landau model is naturally understood as a subspace embedded in the one dimension higher fuzzy four-sphere of the $SO(5)$ Landau model \cite{Hasebe-2014-2, JabbariTorabian2005}.    
If we regard the chirality parameter $s$ as an extra fifth coordinate $x_5$,   
 $X_{\mu}$ can be interpreted as  the coordinates of two $S^3$-latitudes with $x_5=\pm s$  on the virtual fuzzy $S^4$ (Fig.\ref{TwoLatitudes.fig}), as suggested by the equation 
%%%%%%%%%%%%%%%%%%%%%%%%%%%%%%
\be
\sum_{s=-I/2}^{I/2} d_{n=0}(s) =\frac{1}{6}(I+1)(I+2)(I+3)=d^{SO(5)}_{n=0}, 
\ee
%%%%%%%%%%%%%%%%%%%%%%%%%%%%%%% 
where $d_{n=0}(s)$ (\ref{dimso4irrpsns}) and $d^{SO(5)}_{n=0}$ respectively denote the lowest Landau level degeneracy of the $SO(4)$ and $SO(5)$ Landau models.  
%%%%%%%%%%%%%%%%%%%%%%%%%%%%%%%%%%%%%%%%%%%%%%%%%%%%%%%%%%%%
%%%%%%%%%%%%%%%%%%%%%%%%%%%%%%%%%%%%%%%%%%%%%%%%%%%%%%%%%%%%%% 

%In the thermodynamic limit, (\ref{gammamatrisumx}) gives 
%%%%%%%%%%%%%%%%%%
%\be
%\sum_{\mu=1}^4 X_{\mu}(n) X_{\mu}(n) ~\rightarrow ~\frac{1}{2}. ~~~~~~~(I\rightarrow \infty)
%\ee
%%%%%%%%%%%%%%%%%%
%This $1/2$ comes from the right-hand side of the properties of the $SO(4)$ gamma matrices (\ref{gammamatrisum}).  
%Notice that in the thermodynamic limit, the matrix elements of $x_{\mu}$ do not reproduce the classical result: 
%%%%%%%%%%%%%%%%%%%%%
%\be
%\sum_{\mu=1}^4 x_{\mu}x_{\mu}=1. 
%\ee
%%%%%%%%%%%%%%%%%%%%%%%

%In the case of $S_F^2$ \cite{Hasebe-2015}, the $n$th LL matrix elements of $x_i$ $(i=1,2,3)$ are given by 
%%%%%%%%%%%%%%%%%%%%
%\be
%X_i(n)= \frac{I}{(2n+I)(2n+I+2)}{\Gamma}_i^{(2n+I)} , 
%\ee
%%%%%%%%%%%%%%%%%%%%%
%where $\frac{1}{2}\Gamma_i^{(I)}$  denotes the $SU(2)$ spin matrix of the spin magnitude $\frac{I}{2}$ \footnote{$\sum_{i=1}^3 {\Gamma}_i^{(I)}{\Gamma}_i^{(I)}=I(I+2)\bs{1}_{I+1}.$}, and then  
%%%%%%%%%%%%%%%%%%%%%%%%
%\be
%\sum_{i=1}^3 X_i(n) X_i(n) = \frac{I^2}{(2n+I)(2n+I+2)} \bs{1}_{2n+I+1} ,
%\ee
%%%%%%%%%%%%%%%%%%%%%%%%
%which reproduces the classical result in the thermodynamic limit: 
%%%%%%%%%%%%%%%%%%%%
%\be
%\sum_{i=1}^3 X_i(n) X_i(n)  ~~\rightarrow ~~1.~~~~(I~\rightarrow~\infty) 
%\ee
%%%%%%%%%%%%%%%%%%%%%%

%%%%%%%%%%%%%%%%%%%%%%%%%%%%%%%%%%%%%%%%%%%%%
\subsection{Spinor Landau model}\label{subsec:llprojectionspinorll}
%%%%%%%%%%%%%%%%%%%%%%%%%%%%%%%%%%%%%%%%%%%%%%%

As discussed in Sec.\ref{sec:spinorlandau}, the spinor Landau model consists of two independent Landau models: 
%%%%%%%%%%%%%%%%%%%%%%%%%%%%%%%
\begin{align}
&\text{Spinor~Landau~model~with~monopole~charge}~\frac{I}{2}\nn\\
&~~~~~~=~\text{Landau~models~with~}~J^+=\frac{I+1}{2}~\oplus~J^-=\frac{I-1}{2}.
\end{align}
%%%%%%%%%%%%%%%%%%%%%%%%%%%%%%%%
$s=\pm 1/2$ can be realized for even $I$. From the upper figure of 
Fig.\ref{Spinorspectra.fig}, we find that the non-trivial cases occur in\footnote{Even if there was not  external magnetic field (\ref{smalls31}), the $SU(2)$  connection of the holonomy would act as a fictitious magnetic field to generate fuzzy three-sphere geometry.}      
%%%%%%%%%%%%%%%%%%%%%%%%%%%%%%
\begin{subequations}
\begin{align}
&I=0,~~n=~1,2,\cdots ~~~~~~~~~\rightarrow~S_F^3(n, J^+=\frac{1}{2}, |s|=\frac{1}{2}), \label{smalls31} \\
&I=2,4,\cdots~,~ n=1,2,\cdots~\rightarrow~S_F^3(n, J^-=\frac{I-1}{2}, |s|=\frac{1}{2})\oplus S_F^3(n-1, J^+=\frac{I+1}{2}, |s|=\frac{1}{2}), \label{usuals3f}  \\
&I=2,4,\cdots~,~n=0~~~~~~~~~\rightarrow~S_F^3(n=0, J^-=\frac{I-1}{2}, |s|=\frac{1}{2}).
\label{smalls32}
\end{align}
\end{subequations}
%%%%%%%%%%%%%%%%%%%%%%%%%%%%%  
The radii of the corresponding fuzzy three-spheres are respectively given by 
%%%%%%%%%%%%%%%%%%%%
\begin{subequations}
\begin{align}
&R_n^{(1)}=\frac{1}{\sqrt{2(n+1)(n+2)}}, \\
&R_n^{(I-1)}=\frac{I}{\sqrt{2(2n+I)(2n+I+2)}}, ~~~~R_{n-1}^{(I+1)}=\frac{I+2}{\sqrt{2(2n+I)(2n+I+2)}}, \\
&R_{n=0}^{(I-1)} =\sqrt{\frac{I}{2(I+2)}},   
\end{align}
\end{subequations}
%%%%%%%%%%%%%%%%%%
and the matrix sizes of $X_{\mu}$ are 
%%%%%%%%%%%%%%%%%%%%%%%%%
\begin{subequations}
\begin{align}
&2(n+1)n, \\
&\frac{1}{2}(2n+I+2)(2n+I), ~~\frac{1}{2}(2n+I+2)(2n+I)  ,\\
&\frac{1}{2}I({I}+2).
\end{align}
\end{subequations}
%%%%%%%%%%%%%%%%%%%%%%%%%

%%%%%%%%%%%%%%%%%%%%%%%%%%%%%%%%%%%%%%%%%%%%%
\subsection{Relativistic Landau models}\label{subsec:llprojectionrelll}
%%%%%%%%%%%%%%%%%%%%%%%%%%%%%%%%%%%%%%%%%%%%%%%

The Weyl-Landau model is a ``square root'' of the spinor Landau model and the non-trivial matrix geometry can occur for even $I$.   In the spectrum of the Weyl-Landau model,  
the degeneracies of $s=1/2$ and $-1/2$ corresponding to (\ref{smalls31}) and (\ref{smalls32})   are completely lifted (see the left-figure of Fig.\ref{schepDLspectra.fig}),   and only  the half of the degeneracies in (\ref{usuals3f})  survives in the Weyl-Landau model to generate the fuzzy three-sphere. 
From the left figure of Fig.\ref{schepDLspectra.fig}, we can see  
%%%%%%%%%%%%%%%%%%%%%%%%%%%%%%
\be
I=2,4,\cdots~,~ \pm n=\pm 1, \pm 2,\cdots~\rightarrow~ S_F^3. \label{usuals3fdirac}  
\ee
%%%%%%%%%%%%%%%%%%%%%%%%%%%%%  
In the Weyl-Landau level, $-i\fsl{\mathcal{D}}=+\lambda_n(s)$ for $n=1,2,\cdots$\footnote{In the following, we discuss the positive  relativistic Landau levels. The extension to the negative Landau level is obvious.}, the degenerate eigenstates with $s=\pm 1/2$ are given by (\ref{secondeigenstate})
%%%%%%%%%%%%%%%%%%
\begin{subequations}
\begin{align}
&\Psi_{s=+1/2} =U\begin{pmatrix}
\alpha_+ \Phi^{(n-1,  s=1/2, J^+)} \\
\beta_+ \Phi^{(n,  s=1/2, J^-)}
\end{pmatrix}, ~~~~~~({\alpha_+}^2+{\beta_+}^2=1)   \\
&\Psi_{s=-1/2} =U\begin{pmatrix}
\alpha_- \Phi^{(n-1,  s=-1/2, J^+)} \\
\beta_- \Phi^{(n,  s=-1/2, J^-)}
\end{pmatrix}, ~~~~({\alpha_-}^2+{\beta_-}^2=1)  
\end{align} \label{linearcombdiraclandau}
\end{subequations}
%%%%%%%%%%%%%%%%%%%%% 
and the matrix elements  
%%%%%%%%%%%%%%%%%%%%
\be
\mathbf{X}_{\mu}(n, {I}/{2}) =\begin{pmatrix}
\langle \Psi_{+1/2}|x_{\mu}|\Psi_{+1/2}\rangle & \langle \Psi_{+1/2}|x_{\mu}|\Psi_{-1/2}\rangle \\
\langle \Psi_{-1/2}|x_{\mu}|\Psi_{+1/2}\rangle & \langle \Psi_{-1/2}|x_{\mu}|\Psi_{-1/2}\rangle \\
\end{pmatrix} 
\label{diraclandauthreespherefuzzy}
\ee
%%%%%%%%%%%%%%%%%%%%%%%
are evaluated as 
%%%%%%%%%%%%%%%%%%%%%%%%%%%%%%%%%%%%%%%%%%%
\be
\mathbf{X}_{\mu}(n, {I}/{2})=\alpha_+\alpha_- X_{\mu}(n-1, J^+) + \beta_+\beta_- X_{\mu}(n, J^-), 
\label{calxmurel}
\ee
%%%%%%%%%%%%%%%%%%%%%%%%%%%%%%%%%%%%%%%
where $X_{\mu}(n, J)$ denote (\ref{tosummazmu}) with the replacement of $I$ with $2J$: 
%%%%%%%%%%%%%%%%%%%%%%%%%%%
\be
X_{\mu}(n, J)=\frac{2J+1}{(2n+2J+1)(2n+2J+3)} ~\Gamma^{(2n+2J)}_{\mu}. 
\ee
%%%%%%%%%%%%%%%%%%%%%%%%%%%
Consequently, $\mathbf{X}_{\mu}$ (\ref{calxmurel}) are  given by 
%%%%%%%%%%%%%%%%%%%%%%%%%%%%
\be
\mathbf{X}_{\mu}(n, {I}/{2})=\frac{1}{2n+I}\biggl( \alpha_+\alpha_- \frac{I+2}{2n+I+2} +  \beta_+\beta_- \frac{I}{2n+I-2} \biggr)~\Gamma_{\mu}^{(2n+I-1)}, 
\label{weyllfuzz}
\ee
%%%%%%%%%%%%%%%%%%%%%%%%%%%%% 
whose  radius becomes 
%%%%%%%%%%%%%%%%%%%%%%%%%%%%%%%%%%
\be
\mathbf{R}_n^{(I)}=\sqrt{\frac{2n+I+2}{2(2n+I)}}~ (\alpha_+\alpha_- \frac{I+2}{2n+I+2}+\beta_+\beta_-\frac{I}{2n+I-2}). 
\ee
%%%%%%%%%%%%%%%%%%%%%%%%%%%%%%%%%%%

Next, we investigate the case of the Dirac-Landau model. In the massless case, the Dirac-Landau operator becomes a simple direct sum of the two Weyl-Landau operators, and there exist two identical fuzzy three-spheres, each of which originates from each Weyl-Landau operator. Since $\Gamma^4$  is a off-diagonal block matrix (\ref{so4diragamm}), the mass term brings ``interaction'' between such two Weyl-Landau sectors.  
For $-i\fsl{{D}}+M\Gamma^4=\Lambda_n(s=1/2)$,   we have  degenerate eigenstates ,  $\bs{\Xi}^{(1)}_{+\Lambda_n(s=1/2)}$, $\bs{\Xi}^{(1)}_{+\Lambda_n(s=-1/2)}$ and $\bs{\Xi}^{(2)}_{+\Lambda_n(s=1/2)}$, $\bs{\Xi}^{(2)}_{+\Lambda_n(s=-1/2)}$ (\ref{+lambdandiracmass}). Taking the matrix elements of $x_{\mu}$ between them, we obtain   
%%%%%%%%%%%%%%%%%
\be
\mathbb{X}_{\mu}(M)%(n, {I}/{2})
%&=\sqrt{\frac{2\Lambda_n}{\Lambda_n+\lambda_n}}~
%\left(
%\begin{array}{cc:cc}
%0 & Y_{\mu}  & 0 & Z^{(+)}_{\mu}(M) \\
%Y_{\mu}^{\dagger} & 0 & Z^{(-)}_{\mu}(M)^{\dagger}   & 0 \\
%\hdashline %
%0 &  {Z}_{\mu}^{(-)}(M)& 0  & Y^{\mu} \\
% Z_{\mu}^{(+)}(M)^{\dagger} & 0 & Y_{\mu}^{\dagger} & 0 
%\end{array}
%\right)\nn\\
%&
= \left(
\begin{array}{c:c}
\mathbf{X}_{\mu}(n, I/2) & 0 \\
\hdashline
0 & \mathbf{X}_{\mu}(n, I/2)
\end{array} 
\right)
+\frac{M}{\Lambda_n} \left(
\begin{array}{cc:cc}
0 & 0  & 0 & Z^{(+)}_{\mu} \\
0 & 0 & {Z^{(-)}_{\mu}}^{\dagger}   & 0 \\
\hdashline %
0 &  {Z}_{\mu}^{(-)} & 0  & 0 \\
 {Z_{\mu}^{(+)}}^{\dagger} & 0 & 0 & 0 
\end{array}
\right), 
\ee
%%%%%%%%%%%%%%%%%%%
where $\Lambda_n\equiv \Lambda_n(s=1/2)$, $\lambda_n\equiv \lambda_n(s=1/2)$,   and 
%%%%%%%%%%%%%%%%%
%\be
%X_{\mu}=
%\begin{pmatrix}
%0 & Y_{\mu} \\
%Y_{\mu}^{\dagger} & 0 
%\end{pmatrix}\equiv \frac{1}{(2n+I+2)}\biggl( \alpha_+\alpha_- \frac{I+2}{2n+I+4} +  \beta_+\beta_- \frac{I}{2n+I} \biggr)~\Gamma_{\mu}^{(2n+I+1)}, 
%\ee
%%%%%%%%%%%%%%%%%
%(which is (\ref{weyllfuzz})) and 
%%%%%%%%%%%%%%%%%%
\begin{align}
\mathbf{W}_{\mu}^{(+)}&=
\begin{pmatrix}
0 & Z^{(+)}_{\mu}\\
{Z^{(+)}_{\mu}}^{\dagger} &0
\end{pmatrix} \equiv ~\frac{  I\alpha_{-}\beta_{+}   -(I+2)\alpha_{+}\beta_{-}}{(2n+I-1)(2n+I+2)}~\Gamma_{\mu}^{(2n+I-1)}, \nn\\
\mathbf{W}_{\mu}^{(-)}&=
\begin{pmatrix}
0 & Z^{(-)}_{\mu}\\
{Z^{(-)}_{\mu}}^{\dagger} &0 
\end{pmatrix} \equiv ~\frac{  I\alpha_{+}\beta_{-}   -(I+2)\alpha_{-}\beta_{+}}{(2n+I-1)(2n+I+2)}~{\Gamma_{\mu}^{(2n+I-1)}}. 
\end{align}
%%%%%%%%%%%%%%%%%%%%%%%%%
The square of $\mathbb{X}_{\mu}$ is derived as  
%%%%%%%%%%%%%%%%%%%%%%%%%%%%%%%%%%%%%%%%%
\begin{align}
{\mathbb{X}_{\mu}}(M)^2=\left(\begin{array}{c:c}
{\mathbf{X}_{\mu}}^2    &  0 \\
\hdashline 
0 &  {\mathbf{X}_{\mu}}^2  
\end{array}\right)&+\biggl(\frac{M}{\Lambda_n}\biggr)^2 \left(\begin{array}{c:c}
 \mathbf{W}^{(+)}_{\mu} \mathbf{W}^{(+)}_{\mu}   & 0 \\
 \hdashline 
0 &   \mathbf{W}^{(-)}_{\mu} \mathbf{W}^{(-)}_{\mu}   
\end{array}\right) \nn\\
&+ {\frac{M}{\Lambda_n}} \left(\begin{array}{c:c}
 0   & \mathbf{W}_{\mu}^{(+)} \mathbf{X}_{\mu} +\mathbf{X}_{\mu}\mathbf{W}_{\mu}^{(-)} \\
 \hdashline 
\mathbf{W}_{\mu}^{(-)} \mathbf{X}_{\mu} +\mathbf{X}_{\mu}\mathbf{W}_{\mu}^{(+)} &   0  
\end{array}\right)  , 
%&={\frac{2\Lambda_n}{\Lambda_n+\lambda_n}} \begin{pmatrix}
%X_{\mu}X_{\mu} + W^{(+)}_{\mu} W^{(+)}_{\mu}   & 0 \\
%0 &  X_{\mu}X_{\mu} + W^{(+)}_{\mu} W^{(+)}_{\mu}   
%\end{pmatrix} \bb\\
%&+\frac{2\Lambda_n}{\Lambda_n+\lambda_n}
%\begin{pmatrix}
%0 & W_{\mu}^{(+)} X_{\mu} +X_{\mu}W_{\mu}^{(-)} \\
%W_{\mu}^{(-)} X_{\mu} +X_{\mu}W_{\mu}^{(+)} & 0 
%\end{pmatrix}.
\end{align}
%%%%%%%%%%%%%%%%%%%%%%%%%%%%%%%%%%%%%%%%%%%%%%
where we interchanged the second and fourth columns and rows.  
In the massless limit, the off-diagonal blocks of $\mathbb{X}_{\mu}(M)$ vanish to yield $\mathbb{X}_{\mu}(M) ~\overset{M\rightarrow 0}{\longrightarrow} ~\begin{pmatrix} \mathbf{X}_{\mu} & 0 \\
0 &  \mathbf{X}_{\mu} \end{pmatrix}$ that actually represents the two identical non-interacting fuzzy three-spheres. When the  mass term is turned on, the off-diagonal block matrices appear to bring interactions between the two fuzzy three-spheres. $M/\Lambda_n$ can be interpreted as the coupling of the interaction.   
Meanwhile, for each of the cases of (\ref{otherdegenerateeigenstatemassive}), the degenerate subbands with $s=\pm 1/2$   appear for even $I$, and the fuzzy three-spheres are respectively realized as 
%%%%%%%%%%%%%%%%%%%%%%%%%%%%%%
\begin{subequations}
\begin{align}
&{\mathbb{X}_{\mu}}(M) =\frac{M}{\Lambda_n} \mathbf{X}_{\mu}(n, I/2)|_{(n, I/2=0, s=1/2)}=\frac{M}{\sqrt{(n+\frac{1}{2})^2+M^2}}X_{\mu}(n, \frac{I}{2}=\frac{1}{2} ),~~(n=1,2,\cdots) \\
&{\mathbb{X}_{\mu}}(M) =\frac{M}{\Lambda_{n}} \mathbf{X}_{\mu}(n, I/2)|_{(n=0, {I}/{2}, s=1/2)}=\frac{M}{\sqrt{M^2+\frac{1}{4}}}X_{\mu}(n=0, \frac{I-1}{2}).  
\end{align}
\end{subequations}
%%%%%%%%%%%%%%%%%%%%%%%%%%%%%%%%
In the massless limit,  the fuzzy geometries  collapse in either case. 

In the supersymmetric Landau model, the Hamiltonian (\ref{susyhamildef}) consists of two independent spinor Landau Hamiltonians and  accommodates twice the number of fuzzy three-spheres of the spinor Landau  model.  
The mass term in the SUSY Landau model does not bring any particular effect to  fuzzy geometry, as the mass term just shifts the energy levels uniformly (\ref{susyhammass}). 

%%%%%%%%%%%%%%%%%%%%%%%%%%%%%%%%
\section{Summary and Discussions}\label{sec:summary}
%%%%%%%%%%%%%%%%%%%%%%%%%%%%%%%%

 We throughly investigated the $SO(4)$ Landau models based on  the  Dirac and the Schwinger gauges.   
 The gauge fixing enabled us  to elaborate the previous works about the $SO(4)$ Landau models and  bring the new observations, such as  the properties of the $SO(4)$ monopole harmonics and 
 the $SO(4)$ symmetry of the  relativistic operators. 
%We constructed the $SO(4)$ monopole harmonics  by using the fully symmetric products of the chiral Hopf spinors, and derived the coordinate 
 % representation of the Landau level wavefunctions. It was confirmed that 
 In the present analysis, we took into account the spin connection of three-sphere to construct the relativistic Landau operators. 
 With the synthesized connection of the spin connection and gauge field,  we  solved the eigenvalue problem of the spinor Landau operator and of the Weyl-Landau operator also. 
The obtained results are confirmed to reproduce the known results of the  $SO(4)$ spherical harmonics and the free Dirac operator in the free background limit.   The eigenvalue problems of the massive Dirac-Landau  and the supersymmetric Landau model are analyzed too. 
 We applied  the  level projection method to the $SO(4)$ Landau models and derived  the odd dimensional matrix geometry  for the first time.     
 It was shown that, for  each of the (non-relativistic) Landau levels,  the fuzzy three-sphere geometry appears only in  the lowest energy $|s|=1/2$ subband, and the size of fuzzy three-sphere depends on the Landau level.     
 We also clarified realizations of the fuzzy three-sphere geometry in the relativistic Landau models. In particular, we designated that the mass term of the Dirac-Landau model induces interaction between two fuzzy spheres realized in the relativistic Landau level.  
 
 As the $SO(3)$ Landau model has a wide range of  applications, we expect that the $SO(4)$ Landau model may also find its playing fields in many branches of physics. Even if limited to  condensed matter physics, one may conceive its possible applications to Weyl/Dirac semi-metal, three-dimensional quantum Hall effect and chiral topological insulator. We have clarified the emergent fuzzy three-sphere geometry in the Landau physics. 
  It is interesting to see  that such an exotic geometry realizes ``inside''  the physical models, and the dynamics of the fuzzy spaces can be controlled by a physical parameter which in principle can be controlled by experiment.

%%%%%%%%%%%%%%%%%%%%%%%%%%%%%%%%%%%%%%%%%%%%%%%%%%%%%%%%%%%%%%%%%%%%%%%%%%%%%%%%%%%%%%%%%%%%%%
%%%%%%%%%%%%%%%%%%%%%%%%%%%%%%%%%%%%%%%%%%%%%%%%%%%%%%%%%%%%%%%%%%%%%%%%%%%%%%%%%%%%%%%%%%%%%%
\section*{Acknowledgment}
%%%%%%%%%%%%%%%%%%%%%%%%%%%%%%%%%%%%%%%%%%%%%%%%%%%%%%%%%%%%%%%%%%%%%%%%%%%%%%%%%%%%%%%%%%%%%%
%%%%%%%%%%%%%%%%%%%%%%%%%%%%%%%%%%%%%%%%%%%%%%%%%%%%%%%%%%%%%%%%%%%%%%%%%%%%%%%%%%%%%%%%%%%%%%

 This work was supported by JSPS KAKENHI Grant No. 16K05334 and No. 16K05138. 
 
%%%%%%%%%%%%%%%%%%%%%%%%%%%%%%%%%%%%%%%%%%%%%%%%%%%
\appendix
%%%%%%%%%%%%%%%%%%%%%%%%%%%%%%%%%%%%%%%%%%%%%%%%%%

%%%%%%%%%%%%%%%%%%%%%%%%%%%%%%%%
\section{Geometric quantities of three-sphere: component method}\label{append:threespheregeo}
%%%%%%%%%%%%%%%%%%%%%%%%%%%%%%%%%

%%%%%%%%%%%%%%%%%%%%%%%%%%%%%%%%%%%
\subsection{Metric and curvature}
%%%%%%%%%%%%%%%%%%%%%%%%%%%%%%%%%%

The metric on $S^3$ is given by (\ref{metricthreesphere}):   
%%%%%%%%%%%%%%%%%%%
\be
g_{\alpha\beta}=
\begin{pmatrix}
g_{\chi\chi} & g_{\chi\theta} & g_{\chi\phi} \\
g_{\theta\chi} & g_{\theta\theta} & g_{\theta\phi} \\
g_{\phi\chi} & g_{\phi\theta} & g_{\phi\phi} 
\end{pmatrix}
=\begin{pmatrix}
1 & 0 & 0 \\
0 & \sin^2\chi & 0 \\
0 & 0 & \sin^2\chi\sin^2\theta
\end{pmatrix}. \label{metrics3}
\ee
%%%%%%%%%%%%%%%%%%%%%%%%%%%%%%%%%%%%%%%%%%%%%%%%%%%%%%%%%%%%%%%%%%
From (\ref{metrics3}), the non-zero components of Christoffel symbol,  $\Gamma^{\alpha}_{~~\beta\gamma}=\Gamma^{\alpha}_{~~\gamma\beta} =\frac{1}{2}g^{\alpha\delta}(\partial_{\beta}g_{\gamma\delta} +\partial_{\gamma}g_{\beta\delta}-\partial_{\delta}g_{\beta\gamma})$,  are derived as 
%%%%%%%%%%%%%%%%%%%%
\begin{align}
&\Gamma^{\chi}_{~~\theta\theta}=-\sin\chi\cos\chi,~~~~~\Gamma^{\chi}_{~~\phi\phi}=-\sin\chi\cos\chi\sin^2\theta,\nn\\
&\Gamma^{\theta}_{~~\chi\theta}=\Gamma^{\theta}_{~~\theta\chi}=\cot\chi,~~~~\Gamma^{\theta}_{~~\phi\phi}=-\sin\theta\cos\theta, \nn\\
&\Gamma^{\phi}_{~~\chi\phi}=\Gamma^{\phi}_{~~\phi\chi}=\cot\chi,~~~~\Gamma^{\phi}_{~~\theta\phi}=\Gamma^{\phi}_{~~\phi\theta}=\cot\theta, 
\end{align}
%%%%%%%%%%%%%%%%%%%%%%
and the non-zero components of curvature, $R^{\alpha}_{~~\beta\gamma\delta}= \partial_{\gamma }\Gamma^{\alpha}_{~~\beta\delta} - \partial_{\delta}\Gamma^{\alpha}_{~~\gamma\beta} +\Gamma_{~~\gamma\epsilon}^{\alpha}\Gamma^{\epsilon}_{~~\beta\delta}-\Gamma_{~~\delta\epsilon}^{\alpha}\Gamma^{\epsilon}_{~~\beta\gamma}$, are derived as 
%%%%%%%%%%%%%%%%%%%%%%%
\begin{align}
&R^{\chi}_{~~\theta\chi\theta}=\sin^2\chi,~~R^{\chi}_{~~\phi\chi\phi}=\sin^2\chi\sin^2\theta, ~~~R^{\theta}_{~~\chi\chi\theta}=-1,~~~R^{\theta}_{~~\phi\theta\phi}=\sin^2\chi\sin^2\theta, \nn\\
&R^{\phi}_{~~\chi\chi\phi}=-1,~~~~~~R^{\phi}_{~~~\theta\theta\phi}=-\sin^2\chi. 
\label{explicitcurvaturetensors}
\end{align}
%%%%%%%%%%%%%%%%%%%%%%%%%%
The non-zero components of Ricci tensor, $\mathcal{R}_{\mu\nu}\equiv R^{\rho}_{~~\mu\rho\nu}$,  are given by 
%%%%%%%%%%%%%%%%%%%%%%%%%
\be
\mathcal{R}_{\chi\chi}=2,~~~\mathcal{R}_{\theta\theta}=2\sin^2\chi, ~~~\mathcal{R}_{\phi\phi}=2\sin^2\chi\sin^2\theta, 
\label{mathcarmunu}
\ee
%%%%%%%%%%%%%%%%%%%%%%%%%%
and the Ricci scalar, $R=\mathcal{R}^{\mu}_{~~\mu}$, is  
%%%%%%%%%%%%%%%%%
\be
R=2\times 3=6. 
\ee
%%%%%%%%%%%%%%%%%%

%%%%%%%%%%%%%%%%%%%%%%%%%%%%%%%%%%%%%%%%%%%%%%%%%%%%%%%%%%%
\subsection{Spin connection}
%%%%%%%%%%%%%%%%%%%%%%%%%%%%%%%%%%%%%%%%%%%%%%%%%%%%%%%%%%%

From $e^a=e^a_{~~\alpha}dx^{\alpha}$,  we can read off the components of the dreibein (\ref{dreibeinschwing}): 
%%%%%%%%%%%%%%%%%%%
\be
e^a_{~~\alpha}=\begin{pmatrix}
1 & 0 & 0 \\
0 & \sin\chi & 0 \\
0 & 0 & \sin\chi\sin\theta 
\end{pmatrix}  .
\ee
%%%%%%%%%%%%%%%%%%%
The inverse is 
%%%%%%%%%%%%%%%%%%%
\be
e_a^{~~\alpha}=\begin{pmatrix}
1 & 0 & 0 \\
0 & \sin^{-1}\chi & 0  \\
0 & 0 & \sin^{-1}\chi\sin^{-1}\theta
\end{pmatrix}. 
\ee
%%%%%%%%%%%%%%%%%%%
The components of the spin-connection are obtained by the formula \cite{Boulware-1975}
%%%%%%%%%%%%%%%%%%%%%%
\be
\omega_{ab\alpha}=-e_{b\beta}\nabla_{\alpha}e_{a}^{~~\beta}=-e_{b\beta}(\partial_{\alpha}e_{a}^{~~\beta}+\Gamma_{\alpha \gamma}^{\beta}e_{a}^{~~ \gamma}), 
\ee
%%%%%%%%%%%%%%%%%%%%
as 
%%%%%%%%%%%%%%%%%%%%
\begin{align}
&\omega_{12\alpha}(=-\omega_{21\alpha})=\{\omega_{12\chi},~\omega_{12\theta},~ \omega_{12\phi} \}=\{0, -\cos\chi, 0\}, \nn\\
&\omega_{31\alpha}(=-\omega_{13\alpha})=\{\omega_{31\chi},~\omega_{31\theta},~ \omega_{31\phi} \}=\{0, 0,  \sin\theta\cos\chi\}, \nn\\
&\omega_{23\alpha}(=-\omega_{32\alpha})=\{\omega_{23\chi},~\omega_{23\theta},~ \omega_{23\phi} \}=\{0,  0, -\cos\theta\}, 
\label{componentspinconns3}
\end{align}
%%%%%%%%%%%%%%%%%%%
which are consistent with  (\ref{spinconnes3oneform}).  The matrix form of the spin connection is given by 
%%%%%%%%%%%%%%%%%%%%%%%%
\be
\omega_{\alpha}=\sum_{a<b =1,2,3}\omega_{ab\alpha}\sigma^{ab},     
\ee 
%%%%%%%%%%%%%%%%%%%%%%%
with 
%%%%%%%%%%%%%%%%%%%%%%%
\be
\omega_{\chi}=0, ~~~~\omega_{\theta}=-\cos\chi \sigma^{12}, ~~~~\omega_{\phi}=\sin\theta \cos\chi \sigma^{31}-\cos\theta \sigma^{23},  
\label{spinconnes3free}
\ee
%%%%%%%%%%%%%%%%%%%%%%%%%
or 
%%%%%%%%%%%%%%%%%%%%%%%%%%%%%%%%%%%
\be
\omega=\omega_{\alpha}dx^{\alpha}=-\frac{1}{2} (\gamma^1 \cos\theta -\gamma_2 \cos\chi\sin\theta) d\phi-\frac{1}{2}\gamma_3 \cos\chi d\theta , 
\label{su2gaugefieldfromspin}
\ee
%%%%%%%%%%%%%%%%%%%%%%%%%%%%%%%%%%%%
where we used 
%%%%%%%%%%%%%%%%
\be
\sigma^{ab}=-i\frac{1}{4}[\gamma^a, \gamma^b]=\frac{1}{2}\epsilon^{abc}\gamma^c.
\ee
%%%%%%%%%%%%%%%%%%
The curvature is obtained as  
%%%%%%%%%%%%%%%%%%%%%%%%%%%%
\begin{align}
f&=d\omega+i\omega^2\nn\\
&=\frac{1}{2}\gamma_1 \sin^2\chi \sin\theta ~d\theta \wedge d\phi + \frac{1}{2}\gamma_2 \sin\chi\sin\theta ~d\phi \wedge d\chi + \frac{1}{2}\gamma_3 \sin\chi  ~d\chi \wedge d\theta   \nn\\
&=\frac{1}{2}\epsilon_{abc} e^ b\wedge e^c ~\frac{1}{2}\gamma_{a}.   
\label{totalspinfieldst}
\end{align}
%%%%%%%%%%%%%%%%%%%%%%%%%%%%%%
%which is exactly equal to the field strength of the $SU(2)$  monopole with minimal charge $I/2=1/2$.  

%%%%%%%%%%%%%%%%%%%%%%%%%%%%%%%%%%%%%%%%%%%%%%%
\section{The $SO(4)$ spherical harmonics and free Dirac operator}
%%%%%%%%%%%%%%%%%%%%%%%%%%%%%%%%%%%%%%%%%%%%%%%%%

%%%%%%%%%%%%%%%%%%%%%%%%%%%%%%%
\subsection{The $SO(4)$ spherical harmonics}\label{sphericalharmoso4}
%%%%%%%%%%%%%%%%%%%%%%%%%%%%%%%%

In the polar coordinates, the $SO(4)$  free angular momentum operators 
%%%%%%%%%%%%%%%%%%%%%%
\be
l_{\mu\nu}=-ix_{\mu}\partial_{\nu}+ix_{\nu}\partial_{\mu}
\ee
%%%%%%%%%%%%%%%%%%%%%%
are given by 
%%%%%%%%%%%%%%%%%%%%
\begin{align}
&l_{12}=-i\partial_{\phi}, \nn\\
&l_{13}=i(\cos\phi\partial_{\theta}-\cot\theta \sin\phi\partial_{\phi}),\nn\\
&l_{14}=i(\sin\theta\cos\phi\partial_{\chi}+\cot\chi\cos\theta\cos\phi\partial_{\theta}-\cot\chi\frac{1}{\sin\theta}\sin\phi \partial_{\phi}),\nn\\
&l_{23}=i(\sin\phi\partial_{\theta}+\cot\theta \cos\phi\partial_{\phi}),\nn\\
&l_{24}=i(\sin\theta\sin\phi\partial_{\chi}+\cot\chi\cos\theta\sin\phi\partial_{\theta}+\cot\chi\frac{1}{\sin\theta} \cos\phi\partial_{\phi}),\nn\\
&l_{34}=i(\cos\theta\partial_{\chi}-\cot\chi \sin\theta\partial_{\theta}),  
\end{align}
%%%%%%%%%%%%%%%%
and the $SO(4)$ Casimir is derived as 
%%%%%%%%%%%%%%%%%%%%%%%%%%%%%%%%%%%
\be
\sum_{\mu<\nu}^4 {l_{\mu\nu}}^2=-\frac{1}{\sin^2\chi} \partial_{\chi}(\sin^2\chi\partial_{\chi}) -\frac{1}{\sin^2\chi}\frac{1}{\sin\theta}\biggl(\partial_{\theta}(\sin\theta\partial_{\theta}) +\frac{1}{\sin\theta}\partial^2_{\phi}\biggr)=-\Delta_{S^3}, 
\label{freeangularsqaurecasi}
\ee
%%%%%%%%%%%%%%%%%%%%%%%%%%%%%%%%%%%%%%
where 
%%%%%%%%%%%%%%%%%%%%%%%%%%%%%%%%%%
\be
\Delta_{S^3}=\frac{1}{\sqrt{g}}\partial_{\mu}(\sqrt{g}g^{\mu\nu}\partial_{\nu})=\frac{1}{\sin^2\chi} \partial_{\chi}(\sin^2\chi\partial_{\chi}) +\frac{1}{\sin^2\chi\sin\theta}\partial_{\theta}(\sin\theta\partial_{\theta}) +\frac{1}{\sin^2\chi\sin^2\theta}\partial^2_{\phi}. 
\ee
%%%%%%%%%%%%%%%%%%%%%%%%%%%%%%%%%
The $SO(4)$ spherical harmonics that satisfy\footnote{From the general formula, we have 
%%%%%%%%%%%%%%%%%%%%%%
\be
\sum_{\mu <\nu} {l^{(0)}_{\mu\nu}}^2 =2(L(L+1)+R(R+1))_{L=R=\frac{n}{2}} = n(n+2).  
\ee
%%%%%%%%%%%%%%%%%%%%%%%%% 
}
%%%%%%%%%%%%%%%%%%%%%%%%%%%%%
\be
\sum_{\mu<\nu=1}^4{l_{\mu\nu}}^2 \Phi_{m_L, m_R}^{(n,0,0)} (x)  =n(n+2) \Phi_{m_L, m_R}^{(n,0,0)} (x) ,  ~~~(n=0,1,2,\cdots)
\ee
%%%%%%%%%%%%%%%%%%%%%%%%%%%%%%%
are usually denoted as \cite{Biedenharn1961,Domokos-1967,Hochstadt-book}
%%%%%%%%%%%%%%%%%%%
\begin{align}
Y_{n l m}(x) &=2^l l!\sqrt{\frac{2(n+1)(n-l)!}{\pi(n+l+1)!}}~\sin^l(\chi)~C_{n-l}^{l+1}(\cos\chi) \cdot Y_{lm}(\theta,\phi). \\
&~~~~~~~~~~~~~(l=0,1,2,\cdots, n),~~(m=-l, -l+1, \cdots, l) \nn
\end{align}
%%%%%%%%%%%%%%%%%%%%
Here $Y_{lm}$ are the $SO(3)$ spherical harmonics, and $C_{n-l}^{l+1}$ are the Gegenbauer polynomials: 
%%%%%%%%%%%%%%%%%%%%%%%%
\be
C_{n}^{\alpha} (x) \equiv \frac{(2\alpha)_n}{(\alpha+\frac{1}{2})_n}P_n^{(\alpha-\frac{1}{2}, \alpha-\frac{1}{2})}(x)= \frac{(-2)^n}{n!} \frac{\Gamma(n+\alpha) \Gamma(n+2\alpha)}{\Gamma(\alpha)\Gamma(2n+2\alpha)} (1-x^2)^{-\alpha+\frac{1}{2}}\frac{d^n}{dx^n} [(1-x^2)^{n+\alpha-\frac{1}{2}}], 
\ee
%%%%%%%%%%%%%%%%%%%%%%%%
with  $P_n^{(\alpha-\frac{1}{2}, \alpha-\frac{1}{2})}(x)$ being Jacobi polynomials and $(\alpha)_n\equiv \alpha(\alpha+1)(\alpha+2)\cdots (\alpha+n-1)$.  
 The degeneracy is 
%%%%%%%%%%%%%%%%%%%%%%%
\be
\sum_{l=0}^n (2l+1) =(n+1)^2.  
\ee
%%%%%%%%%%%%%%%%%%%%%%

%%%%%%%%%%%%%%%%%%%%%%%%%%%%%%%%%%%%%%%%%%%%%%%
\subsection{The free Weyl operator and eigenstates}\label{subsec:freeweyl}
%%%%%%%%%%%%%%%%%%%%%%%%%%%%%%%%%%%%%%%%%%%%%%%%%

The eigenvalue problems of  free Dirac operators on arbitrary dimensional spheres  are generally solved in \cite{CamporesiHiguchi1996, Tranutman1995,Tranutman1993}. Here, we apply the results to the $S^3$ case. For  spinor particle, the covariant derivatives on $S^3$ are given by 
%%%%%%%%%%%%%%%%%%%%
%\be
%-i\nabla_{\alpha}\bs{\psi}=(-i\partial_{\alpha}+\omega_{\alpha})\bs{\psi}, 
%\ee
%%%%%%%%%%%%%%%%%%%%
%or 
%%%%%%%%%%%%%%%%%%%%
\begin{align}
&-i\nabla_{\chi}=-i\partial_{\chi}, \nn\\
&-i\nabla_{\theta}=-i\partial_{\theta}-\frac{1}{2}\gamma_3\cos\chi ,\nn\\
&-i\nabla_{\phi}=-i\partial_{\phi}-\frac{1}{2}\gamma_1\cos\theta+\frac{1}{2}\gamma_2\cos\chi\sin\theta , 
\label{eachderidiracop-fin}
\end{align}
%%%%%%%%%%%%%%%%%%%%%%%
where we used the spin-connection (\ref{spinconnes3free}).  
(\ref{eachderidiracop-fin}) is formally equivalent to the  covariant derivatives of $\it{spinless}$ particle in the  $SU(2)$ monopole background with  $I/2=1/2$.  
We construct the free Weyl operator  as\footnote{
(\ref{diracopspherecoord-conc}) can be written as 
%%%%%%%%%%%%%%%%%%%%%%%%%
\be
-i\fsl{\nabla}_{S^3}=-i\gamma_1 (\partial_{\chi}+\cot \chi)-i\frac{1}{\sin\chi}\fsl{\nabla}_{S^2}, 
\label{fslnablaori}
\ee
%%%%%%%%%%%%%%%%%%%%%%%%%%
where 
%%%%%%%%%%%%%%%%%%%%%%%%%
\be
\fsl{\nabla}_{S^2}\equiv \gamma^2 (\partial_{\theta}+\frac{1}{2}\cot\theta)+\gamma_3 \frac{1}{\sin\theta}\partial_{\phi}. 
\ee
%%%%%%%%%%%%%%%%%%%%%%%%%%
(\ref{fslnablaori}) is consistent with the general relation in Sec.3.1 of \cite{CamporesiHiguchi1996}:  %that  gives the relation : 
%%%%%%%%%%%%%%%%%%%%%%%%%%} 
\be
-i\fsl{\nabla}_{S^3}=-i\gamma_1 (\partial_{\chi}+\frac{N-1}{2}|_{N=3}\cot \chi)-i\frac{1}{\sin\chi}\fsl{\nabla}_{S^2}. 
\label{fslnablacomporese}
\ee
%%%%%%%%%%%%%%%%%%%%%%%%%%%%%%%
} 
%%%%%%%%%%%%%%%%%%%%%
\begin{align}
-i\fsl{\nabla}_{S^3}&=-ie_{a}^{~~\alpha} \gamma^{a}\nabla_{\alpha}=
-i \gamma_{1}\nabla_{\chi} -i\frac{1}{\sin\chi}\gamma_2\nabla_{\theta} -i\frac{1}{\sin\chi\sin\theta}\gamma_3\nabla_{\phi} 
\nn\\
&=-i\gamma_1 (\partial_{\chi}+\cot \chi)-i\gamma_2\frac{1}{\sin\chi} (\partial_{\theta}+\frac{1}{2}\cot\theta)-i\gamma_3 \frac{1}{\sin\chi\sin\theta}\partial_{\phi},  
\label{diracopspherecoord-conc}
\end{align}
%%%%%%%%%%%%%%%%%%%%%%%
or  
%%%%%%%%%%%%%%%%%%%%%%%%%%%%
\begin{align}
-i\fsl{\nabla}_{S^3}&
=-i\sigma_3 (\partial_{\chi}+\cot \chi)-i\sigma_1\frac{1}{\sin\chi} (\partial_{\theta}+\frac{1}{2}\cot\theta)-i\sigma_2 \frac{1}{\sin\chi\sin\theta}\partial_{\phi}
\nn\\&=\begin{pmatrix}
-i (\partial_{\chi}+\cot \chi)   & \frac{1}{\sin\chi}\biggl(-i\partial_{\theta}-\frac{1}{\sin\theta}(\partial_{\phi}+i\frac{1}{2}\cos\theta) \biggr) \\
  \frac{1}{\sin\chi}\biggl(-i\partial_{\theta}+\frac{1}{\sin\theta}(\partial_{\phi}-i\frac{1}{2}\cos\theta) \biggr)  & i (\partial_{\chi}+\cot \chi) 
\end{pmatrix}.  
\label{s3diracoperatorinspcho}
\end{align}
%%%%%%%%%%%%%%%%%%%%%%%%%%%%
The eigenvalue problem of the Weyl operator (\ref{s3diracoperatorinspcho}) is expressed as 
%%%%%%%%%%%%%%%%%%%%%%%%%%
\be
-i\fsl{\nabla}\bs{\psi}^{(\pm)}_{n, l, m}(x)=\pm (\frac{3}{2}+n)\bs{\psi}^{(\pm)}_{n, l, m}(x), 
\ee
%%%%%%%%%%%%%%%%%%%%%%%%%
with 
%%%%%%%%%%%%%%%%%%%%%%%%%%%
\be
n=0, 1, 2, \cdots, ~~~~~~l=0, 1, 2, \cdots, n, ~~~~~m=0, 1, 2, \cdots, l.  
\ee
%%%%%%%%%%%%%%%%%%%%%%%%%%%%
The corresponding eigenstates are 
%%%%%%%%%%%%%%%%%%%%%%%%%%%%%%%%%%
\begin{align}
&-i\fsl{\nabla}=+(\frac{3}{2}+n)~:\nn\\
&\bs{\psi}^{(+)}_{n, l, m, \pm}(x) =\frac{1}{A} \biggl(\cos\frac{\chi}{2}\cdot \sin\frac{\chi}{2}\biggr)^l  \biggl(\cos\frac{\theta}{2}\cdot \sin\frac{\theta}{2}\biggr)^m \nn\\
&~~~~~~~~~~~\cdot \begin{pmatrix}
\biggl(\cos\frac{\chi}{2} ~P_{n-l}^{l+\frac{1}{2}, l+\frac{3}{2}}(\cos\chi) +i\sin\frac{\chi}{2} ~P_{n-l}^{l+\frac{3}{2}, l+\frac{1}{2}}(\cos\chi)\biggr)~\sin\frac{\theta}{2}~P_{l-m}^{m+\frac{1\pm 1}{2}, m+\frac{1\mp 1}{2}}(\cos\theta) \\
\pm \biggl(\cos\frac{\chi}{2} ~P_{n-l}^{l+\frac{1}{2}, l+\frac{3}{2}}(\cos\chi) -i\sin\frac{\chi}{2} ~P_{n-l}^{l+\frac{3}{2}, l+\frac{1}{2}}(\cos\chi)\biggr)~\cos\frac{\theta}{2}~P_{l-m}^{m+\frac{1\mp 1}{2}, m+\frac{1\pm 1}{2}}(\cos\theta)
\end{pmatrix} e^{\mp i(m+\frac{1}{2})\phi}, \nn\\
&-i\fsl{\nabla}=-(\frac{3}{2}+n)~:\nn\\
&\bs{\psi}^{(-)}_{n, l, m, \pm }(x) =\frac{1}{A} \biggl(\cos\frac{\chi}{2}\cdot \sin\frac{\chi}{2}\biggr)^l  \biggl(\cos\frac{\theta}{2}\cdot \sin\frac{\theta}{2}\biggr)^m \nn\\
&~~~~~~~~~~\cdot \begin{pmatrix}
\biggl(\cos\frac{\chi}{2} ~P_{n-l}^{l+\frac{1}{2}, l+\frac{3}{2}}(\cos\chi) -i\sin\frac{\chi}{2} ~P_{n-l}^{l+\frac{3}{2}, l+\frac{1}{2}}(\cos\chi)\biggr)~\sin\frac{\theta}{2}~P_{l-m}^{m+\frac{1\pm 1}{2}, m+\frac{1\mp 1}{2}}(\cos\theta) \\
\pm \biggl(\cos\frac{\chi}{2} ~P_{n-l}^{l+\frac{1}{2}, l+\frac{3}{2}}(\cos\chi) +i\sin\frac{\chi}{2} ~P_{n-l}^{l+\frac{3}{2}, l+\frac{1}{2}}(\cos\chi)\biggr)~\cos\frac{\theta}{2}~P_{l-m}^{m+\frac{1\mp 1}{2}, m+\frac{1\pm 1}{2}}(\cos\theta)
\end{pmatrix} e^{\mp i(m+\frac{1}{2})\phi}, 
\label{eigenstatess3diracexplicitrepr}
\end{align}
%%%%%%%%%%%%%%%%%%%%%%%%%%%%%%%%%%
where\footnote{
The normalization constant $A$ is determined so as to satisfy 
%%%%%%%%%%%%%%%%%%%%%%%%%
\be
\int_0^{\pi} d\chi ~\sin^2\chi \int_0^{\pi} d\theta~\sin\theta \int_0^{2\pi} ~ d\phi~ \bs{\psi}^{(\pm)}_{n, l, m, \pm}(x)^{\dagger}~\bs{\psi}^{(\pm)}_{n, l, m, \pm}(x)=1.
\ee
%%%%%%%%%%%%%%%%%%%%%%%
}  
%%%%%%%%%%%%%%%%%%%%%%%%
\be
A%=A_2 A_3 
=\frac{l!~(2n+1)!!}{2^{n-1}}\frac{\pi}{\sqrt{(n-l)!(n+l+2)!(l-m)!(l+m+1)!}}. 
\ee
%%%%%%%%%%%%%%%%%%%%%%%%%%
The number of each of the  eigenstates (\ref{eigenstatess3diracexplicitrepr}) is 
%%%%%%%%%%%%%%%%%%%%%
\be
\sum_{l=0}^n (l+1)=\frac{1}{2}(n+1)(n+2).  
\ee
%%%%%%%%%%%%%%%%%%%%%%%
Hence  for   $-i\fsl{\nabla} =+(\frac{3}{2}+n)$, the degeneracy  is 
%%%%%%%%%%%%%%%%%%%%%%%%
\be
2\times \frac{1}{2}(n+1)(n+2)=(n+1)(n+2),  
\ee
%%%%%%%%%%%%%%%%%%%%%%%%%%
and for $-i\fsl{\nabla} =-(\frac{3}{2}+n)$, 
%%%%%%%%%%%%%%%%%%%%%%%%
\be
2\times \frac{1}{2}(n+1)(n+2)=(n+1)(n+2). 
\ee
%%%%%%%%%%%%%%%%%%%%%%%%%%
From the covariant derivatives (\ref{eachderidiracop-fin}), we construct the $SO(4)$ angular momentum operators as  
%%%%%%%%%%%%%%%%%%%%%%%%%
\be
J_{\mu\nu} =-ix_{\mu}\nabla_{\nu} + ix_{\nu}\nabla_{\nu} +f_{\mu\nu}, \label{angularspinconne}
\ee
%%%%%%%%%%%%%%%%%%%%%
where $\nabla_{\mu}=\partial_{\mu}+i\omega_{\mu}$ and $f_{\mu\nu}=-i[\nabla_{\mu}, \nabla_{\nu}]$ (\ref{totalspinfieldst}). $J_{\mu\nu}$ (\ref{angularspinconne}) is formally equivalent to  the non-relativistic angular momentum with minimal monopole charge $I/2=1/2$. 
The Weyl operator $-i\fsl{\nabla}$ is invariant under the $SO(4)$ transformation generated by $J_{\mu\nu}$: 
% $-i\fsl{\nabla}_{S^3}$ is invariant under the $SO(4)$ transformation: 
%%%%%%%%%%%%%%%%%%%%%
\be
[-i\fsl{\nabla}_{S^3}, J_{\mu\nu}]=0. 
\ee
%%%%%%%%%%%%%%%%%%%%%%%%
The Weyl operator eigenstates (\ref{eigenstatess3diracexplicitrepr}) correspond to the $SO(4)\simeq SU(2)\otimes SU(2)$ irreducible representations: 
%%%%%%%%%%%%%%%%%%%%%
\begin{align}
&-i\fsl{\nabla} =+(\frac{3}{2}+n) ~\leftrightarrow~(L, R)=(\frac{n}{2}, \frac{n}{2}+\frac{1}{2}), \nn\\
&-i\fsl{\nabla} =-(\frac{3}{2}+n) ~\leftrightarrow~(L, R)=(\frac{n}{2}+\frac{1}{2}, \frac{n}{2}). 
\end{align}
%%%%%%%%%%%%%%%%%%%%%
In particular for $n=0$,  (\ref{eigenstatess3diracexplicitrepr}) becomes  
%%%%%%%%%%%%%%%%%%%%%%
\begin{align}
&-i\fsl{\nabla} =+\frac{3}{2}~~:~~ \bs{\psi}_{n=0,l=0,m=0 , +}^{(+)}=\frac{1}{\sqrt{2}\pi}\begin{pmatrix}
\sin\frac{\theta}{2}e^{\frac{i}{2}(\chi-\phi)} \\
\cos\frac{\theta}{2}e^{-\frac{i}{2}(\chi+\phi)}
\end{pmatrix},~~~\bs{\psi}_{0,0,0 , -}^{(+)}=\frac{1}{\sqrt{2}\pi}\begin{pmatrix}
\cos\frac{\theta}{2}e^{\frac{i}{2}(\chi+\phi)} \\
-\sin\frac{\theta}{2}e^{-\frac{i}{2}(\chi-\phi)}
\end{pmatrix} , \nn\\
&-i\fsl{\nabla} =-\frac{3}{2}~~:~~ \bs{\psi}_{n=0,l=0,m=0 , +}^{(-)}=\frac{1}{\sqrt{2}\pi}\begin{pmatrix}
\sin\frac{\theta}{2}e^{-\frac{i}{2}(\chi+\phi)} \\
\cos\frac{\theta}{2}e^{\frac{i}{2}(\chi-\phi)}
\end{pmatrix} ,~~~\bs{\psi}_{0,0,0 , -}^{(-)}=\frac{1}{\sqrt{2}\pi}\begin{pmatrix}
\cos\frac{\theta}{2}e^{-\frac{i}{2}(\chi-\phi)} \\
-\sin\frac{\theta}{2}e^{\frac{i}{2}(\chi+\phi)}
\end{pmatrix}.
\end{align}
%%%%%%%%%%%%%%%%%%%%%%%%
Meanwhile for the $SO(4)$ monopole harmonics (\ref{vectorrepsorep}), we have  
%%%%%%%%%%%%%%%%%%%%%
\begin{align}
&s=+\frac{1}{2}~:~\bs{\Phi}^{(0, 1/2, 1/2)}_{1/2,0}=\frac{1}{\sqrt{2}\pi}\begin{pmatrix}
\cos\frac{\theta}{2}e^{-\frac{i}{2}(\chi-\phi)} \\
-\sin\frac{\theta}{2}e^{\frac{i}{2}(\chi+\phi)}
\end{pmatrix} ,~~~~~\bs{\Phi}^{(0, 1/2, 1/2)}_{-1/2, 0}=\frac{1}{\sqrt{2}\pi}\begin{pmatrix}
\sin\frac{\theta}{2}e^{-\frac{i}{2}(\chi+\phi)} \\
\cos\frac{\theta}{2}e^{\frac{i}{2}(\chi-\phi)}
\end{pmatrix} , \nn\\
&s=-\frac{1}{2}~:~\bs{\Phi}^{(0, -1/2, 1/2)}_{0, 1/2}=\frac{1}{\sqrt{2}\pi}\begin{pmatrix}
\cos\frac{\theta}{2}e^{\frac{i}{2}(\chi+\phi)} \\
-\sin\frac{\theta}{2}e^{-\frac{i}{2}(\chi-\phi)}
\end{pmatrix} ,~~\bs{\Phi}^{(0, -1/2, 1/2)}_{0, -1/2}=\frac{1}{\sqrt{2}\pi}\begin{pmatrix}
\sin\frac{\theta}{2}e^{\frac{i}{2}(\chi-\phi)} \\
\cos\frac{\theta}{2}e^{-\frac{i}{2}(\chi+\phi)}
\end{pmatrix} .
\end{align}
%%%%%%%%%%%%%%%%%%%%%%%
They are related as 
%%%%%%%%%%%%%%%%%%%%%%%%%%%%%%%%%%
\begin{align}
&\bs{\psi}_{0,0,0 , +}^{(+)}=\bs{\Phi}^{(0, -1/2, 1/2)}_{0, -1/2},~~~\bs{\psi}_{0,0,0 , -}^{(+)}=\bs{\Phi}^{(0, -1/2, 1/2)}_{0, +1/2},\nn\\
&\bs{\psi}_{0,0,0 , +}^{(-)}=\bs{\Phi}^{(0, 1/2, 1/2)}_{-1/2, 0},~~~~~\bs{\psi}_{0,0,0 , -}^{(-)}=\bs{\Phi}^{(0, 1/2, 1/2)}_{+1/2, 0}. 
\label{reltionn0free} 
\end{align}
%%%%%%%%%%%%%%%%%%%%%%%%%%%%%%%%%%
(\ref{reltionn0free}) is a special case of (\ref{freediraceigenrelations}).

%%%%%%%%%%%%%%%%%%%%%%%%%%%%%%%%%%
\subsection{Square of the free Weyl operator}\label{subsec:freeweylsquare}
%%%%%%%%%%%%%%%%%%%%%%%%%%%%%%%%%

The square of the free Weyl operator on $S^3$ (\ref{fslnablaori}) is derived as 
%%%%%%%%%%%%%%%%%%%%%%%
\be
(-i\fsl{\nabla}_{S^3})^2=
-(\partial_{\chi}+\cot\chi)^2 +\frac{1}{\sin^2\chi}(-i\fsl{\nabla}_{S^2})^2 +\frac{\cot\chi}{\sin\chi}\gamma^1 \fsl{\nabla}_{S^2}, 
\ee
%%%%%%%%%%%%%%%%%%%%%%%%%%%%%%%%%%%%%%
or 
%%%%%%%%%%%%%%%%%%%%%%%%%%%%%%%%%%%
\begin{align}
(-i\fsl{\nabla}_{S^3})^2=&-(\partial_{\chi}+\cot\chi)^2 -\frac{1}{\sin^2\chi} (\partial_{\theta}+\frac{1}{2}\cot\theta)^2 -\frac{1}{\sin^2\chi\sin^2\theta}\partial_{\phi}^2 \nn\\
&+i\gamma^1 \frac{\cot\theta}{\sin^2\chi\sin\theta}\partial_{\phi}-i\gamma^2 \frac{\cot\chi}{\sin\chi\sin\theta}\partial_{\phi}+i\gamma^3\frac{\cot\chi}{\sin\chi}(\partial_{\theta}+\frac{1}{2}\cot\theta). \label{squarefreediraconerep}
\end{align}
%%%%%%%%%%%%%%%%%%%%%%%%%%
Using  (\ref{eachderidiracop-fin}), we can show
%%%%%%%%%%%%%%%%%%%%%%
\begin{align}
(-i\fsl{\nabla}_{S^3})^2&=-\frac{1}{\sin^2\chi}\nabla_{\chi}(\sin^2\chi\nabla_{\chi}) -\frac{1}{\sin^2\chi\sin\theta}\nabla_{\theta}(\sin\theta\nabla_{\theta})-\frac{1}{\sin^2\chi\sin^2\theta}{\nabla_{\phi}}^2+\frac{3}{2}\nn\\
&=-\Delta_{S^3}+\frac{R}{4}|_{R=6}  \nn\\
&=\sum_{\mu<\nu}^4 {J_{\mu\nu}}^2+\frac{R}{8}|_{R=6}, 
\label{s3diracsquanoth}
\end{align}
%%%%%%%%%%%%%%%%%%%%%%%%
where %We can derive the square of the $SO(4)$ angular momentum as  
%%%%%%%%%%%%%%%%%%%%%%%%%%%%%%%
\begin{subequations}
\begin{align}
%\ee
%%%%%%%%%%%%%%%%%%%%%%%%%%%%%%%%%
%to show
%%%%%%%%%%%%%%%%%%%%%%%%%%%%%%%%%%
%\begin{align}
&\Delta_{S^3}=\frac{1}{\sqrt{g}}\nabla_{\alpha}(\sqrt{g}g^{\alpha\beta}\nabla_{\beta})=\frac{1}{\sin^2\chi} {\nabla}_{\chi}(\sin^2\chi{\nabla}_{\chi}) +\frac{1}{\sin^2\chi}\frac{1}{\sin\theta}\biggl(\nabla_{\theta}(\sin\theta\nabla_{\theta}) +\frac{1}{\sin\theta}{\nabla}^2_{\phi}\biggr), \label{freelaplacespinors3} \\
&\sum_{\mu<\nu}^4 {J_{\mu\nu}}^2=-\frac{1}{\sin^2\chi} {\nabla}_{\chi}(\sin^2\chi{\nabla}_{\chi}) -\frac{1}{\sin^2\chi}\frac{1}{\sin\theta}\biggl(\nabla_{\theta}(\sin\theta\nabla_{\theta}) +\frac{1}{\sin\theta}{\nabla}^2_{\phi}\biggr)+\frac{3}{4}. 
\label{freeangmospinorsq} 
\end{align}
\end{subequations}
%%%%%%%%%%%%%%%%%%%%%%%%%%%%%%%%%
%%%%%%%%%%%%%%%%%%%%%%% 
%\be 
%(-i\fsl{\nabla}_{S^3})^2=\sum_{\mu<\nu}^4 {J_{\mu\nu}}^2+\frac{R}{8}|_{R=6}=-\Delta_{S^3}+\frac{R}{4}|_{R=6},  
%\label{sqaurespinornabla}
%\ee
%%%%%%%%%%%%%%%%%%%%%%
%where the Laplacian for the spinor particle is given by  
%Since $J_{\mu\nu}$ (\ref{angularspinconne}) is formally equivalent to  the non-relativistic angular momentum with minimal monopole charge $I/2=1/2$,  %$L_{\mu\nu}=-ix_{\mu}(\partial_{\nu}+iA_{\nu}^{(1/2)}) +ix_{\nu}(\partial_{\mu}+iA_{\mu}^{(1/2)})+F_{\mu\nu}^{(1/2)}$, 
(\ref{freelaplacespinors3}) and (\ref{freeangmospinorsq})   are simply related as\footnote{Also from the general formula in \cite{Dolan-2003,Hasebe-2017}, we have 
%%%%%%%%%%%%%%%%%%%%%%
\be 
\Delta_{S^3} =-\sum_{\mu<\nu}^4 {J_{\mu\nu}}^2+\frac{R}{8}|_{R=6}.   
\ee 
%%%%%%%%%%%%%%%%%%%%%%
}   
%%%%%%%%%%%%%%%%%%%%%%%%%%
\be
\Delta_{S^3} =-\sum_{\mu<\nu}^4 {J_{\mu\nu}}^2+\frac{3}{4}. 
\ee
%%%%%%%%%%%%%%%%%%%%%%%%%%%%
The eigenvalues of the $SO(4)$ Casimir  (\ref{freeangmospinorsq}) can be obtained from the non-relativistic result (\ref{so4casimireigenns}) for $I/2=|s|=1/2$: 
%%%%%%%%%%%%%%%%%%%%
\be
\sum_{\mu<\nu}^4 {J_{\mu\nu}}^2 =n^2+3n+\frac{3}{2}, 
\ee
%%%%%%%%%%%%%%%%%%
and then\footnote{(\ref{squnabeige}) is consistent with  (\ref{so4spec1}) with the replacement $(n, I) \rightarrow (n+1, 0)$.} 
%%%%%%%%%%%%%%%%%%%
\be
(-i\fsl{\nabla}_{S^3})^2 =(\frac{3}{2}+n)^2,
\label{squnabeige}
\ee
%%%%%%%%%%%%%%%%%%%
or   
%%%%%%%%%%%%%%%%%%
\be
-i\fsl{\nabla}_{S^3}=\pm (\frac{3}{2}+n).
\ee
%%%%%%%%%%%%%%%%%

%%%%%%%%%%%%%%%%%%%%%%%%%%%%%%%%%%%%%%%%%%%%%%
\section{Integral of the product of three $SO(4)$ monopole harmonics}\label{append:integral}
%%%%%%%%%%%%%%%%%%%%%%%%%%%%%%%%%%%%%%%%%%%%%%

From the explicit form of the  $SO(4)$ monopole harmonics (\ref{basiswithnormal}), we can evaluate the integral of the product of three $SO(4)$ monopole harmonics as 
\begin{align}
&\frac{1}{I+1}\int_{S^3} d\Omega_3~ (\sum_{A=-\frac{I}{2}}^{\frac{I}{2}}~\Phi_{m_L,m_R}^{[l_L,l_R,\frac{I}{2}]}(\bs{\chi})_A^*\cdot \Phi^{[\frac{p}{2}, \frac{p}{2}, 0]}_{m'_L, m'_R}(\bs{\chi})\cdot {\Phi_{m''_L,m''_R}^{[l_L,l_R,\frac{I}{2}]}(\bs{\chi})}_A)\nn\\
&=(p+1)\sqrt{\frac{(I+1)(2l_L+1)(2l_R+1)}{2\pi^2}} \begin{Bmatrix}
l_L & l_R & \frac{I}{2} \\
\frac{p}{2} & \frac{p}{2} & 0 \\
l_L & l_R & \frac{I}{2}
\end{Bmatrix} ~C_{\frac{p}{2}, m'_L~;~l_L, m''_L}^{l_L, m_L}~C_{\frac{p}{2}, m'_R~;~l_R, m''_R}^{l_R, m_R} \nn\\
&=(p+1)\sqrt{\frac{(I+1)(2l_L+1)(2l_R+1)}{2\pi^2}}(-1)^{2l_L+2l_R+I+p} \begin{Bmatrix}
l_L & l_R & \frac{I}{2} \\
l_L & l_R & \frac{I}{2}\\
\frac{p}{2} & \frac{p}{2} & 0 
\end{Bmatrix} ~C_{\frac{p}{2}, m'_L~;~l_L, m''_L}^{l_L, m_L}~C_{\frac{p}{2}, m'_R~;~l_R, m''_R}^{l_R, m_R} \nn\\
&=\sqrt{\frac{(p+1)(2l_L+1)(2l_R+1)}{2\pi^2}}(-1)^{3l_L+3l_R+\frac{3}{2}I+\frac{3}{2}p} \begin{Bmatrix}
l_L & l_R & \frac{I}{2} \\
l_R & l_L & \frac{p}{2}
\end{Bmatrix} ~C_{\frac{p}{2}, m'_L~;~l_L, m''_L}^{l_L, m_L}~C_{\frac{p}{2}, m'_R~;~l_R, m''_R}^{l_R, m_R} \nn\\
&=\sqrt{\frac{(p+1)(2l_L+1)(2l_R+1)}{2\pi^2}}~(-1)^{-(l_L+l_R+\frac{I}{2}+\frac{p}{2})}~\begin{Bmatrix}
l_L & l_R & \frac{I}{2} \\
l_R & l_L & \frac{p}{2}
\end{Bmatrix}~C_{\frac{p}{2}, m'_L~;~l_L, m''_L}^{l_L, m_L}~C_{\frac{p}{2}, m'_R~;~l_R, m''_R}^{l_R, m_R}.  \label{threeintegralso4lleigendetail}
\end{align}
%%%%%%%%%%%%%%%%%%%%%%%%%%
This is a special formula of  Eq.(3.11) in \cite{Ishiki-Takayama-Tsuchiya-2006,Ishiki-Shimasaki-Takayama-Tsuchiya-2006}. 
For $\Phi_{m_L, m_R}^{[\frac{p}{2},\frac{p}{2},0]}|_{n=m_L=m_R=0}=\frac{1}{\sqrt{2\pi^2}}$, (\ref{threeintegralso4lleigendetail})  becomes  
%%%%%%%%%%%%%%%%%%%%%
\begin{align}
&\frac{1}{\sqrt{2\pi^2}(I+1)}\int_{S^3} d\Omega ~\sum_A \Phi_{m_L,m_R}^{[l_L,l_R,\frac{I}{2}]}(\bs{\chi})_A^*\cdot {\Phi_{m''_L,m''_R}^{[l_L,l_R,\frac{I}{2}]}(\bs{\chi})}_A\nn\\
&~~~=\sqrt{\frac{(2l_L+1)(2l_R+1)}{2\pi^2}}(-1)^{-(l_L+l_R+\frac{I}{2})}
 \begin{Bmatrix}
l_L & l_R & \frac{I}{2} \\
l_R & l_L & 0 
\end{Bmatrix}~C^{l_L,m_L}_{0,0~;~l_L,m''_L}C^{l_R,m_R}_{0,0~;~l_R,m''_R}
\nn\\
&~~~=\frac{1}{\sqrt{2\pi^2}}\delta_{m_L,m''_L}\delta_{m_R,m''_R}, 
\label{normalllso4ii}
\end{align}
%%%%%%%%%%%%%%%%%%%%%%%%%%
where, in the third equation, we used 
%%%%%%%%%%%%%%%%%%%%%
\be
\begin{Bmatrix}
l_L & l_R & \frac{I}{2} \\
l_R & l_L & 0 
\end{Bmatrix} =
\frac{1}{\sqrt{(2l_L+1)(2l_R+1)}}(-1)^{l_L+l_R+\frac{I}{2}},~~~C^{l',m'}_{0,0~;~l,m}=\delta_{l,l'}\delta_{m,m'}.
\ee
%%%%%%%%%%%%%%%%%%%%%%%%%%%
(\ref{normalllso4ii}) is consistent with  (\ref{normalllso4}).

Similarly, we have 
%%%%%%%%%%%%%%%%%%%%%%%%%%%%%%%%%%%%
\begin{align}
&\frac{1}{I+1}\int_{S^3} d\Omega_3~ (\sum_{A=-\frac{I}{2}}^{\frac{I}{2}}~\Phi_{m_L,m_R}^{[l_L,l_R,\frac{I}{2}]}(\bs{\chi})_A^*\cdot \Phi^{[\frac{p}{2}, \frac{p}{2}, 0]}_{m'_L, m'_R}(\bs{\chi})\cdot {\Phi_{m''_R,m''_L}^{[l_R,l_L,\frac{I}{2}]}(\bs{\chi})}_A)\nn\\
&=(p+1)\sqrt{\frac{(I+1)(2l_L+1)(2l_R+1)}{2\pi^2}} \begin{Bmatrix}
l_L & l_R & \frac{I}{2} \\
\frac{p}{2} & \frac{p}{2} & 0 \\
l_R & l_L & \frac{I}{2}
\end{Bmatrix} ~C_{\frac{p}{2}, m'_L~;~l_R, m''_R}^{l_L, m_L}~C_{\frac{p}{2}, m'_R~;~l_L, m''_L}^{l_R, m_R} \nn\\
&=(p+1)\sqrt{\frac{(I+1)(2l_L+1)(2l_R+1)}{2\pi^2}}(-1)^{2l_L+2l_R+I+p} \begin{Bmatrix}
l_L & l_R & \frac{I}{2} \\
l_L & l_R & \frac{I}{2}\\
\frac{p}{2} & \frac{p}{2} & 0 
\end{Bmatrix}~C_{\frac{p}{2}, m'_L~;~l_R, m''_R}^{l_L, m_L}~C_{\frac{p}{2}, m'_R~;~l_L, m''_L}^{l_R, m_R} \nn\\
&=\sqrt{\frac{(p+1)(2l_L+1)(2l_R+1)}{2\pi^2}}(-1)^{2l_L+4l_R+\frac{3}{2}I+\frac{3}{2}p} \begin{Bmatrix}
l_L & l_R & \frac{I}{2} \\
l_L & l_R & \frac{p}{2}
\end{Bmatrix} ~C_{\frac{p}{2}, m'_L~;~l_R, m''_R}^{l_L, m_L}~C_{\frac{p}{2}, m'_R~;~l_L, m''_L}^{l_R, m_R} \nn\\
&=\sqrt{\frac{(p+1)(2l_L+1)(2l_R+1)}{2\pi^2}}~(-1)^{2l_L+\frac{3}{2}I+\frac{3}{2}p }~\begin{Bmatrix}
l_L & l_R & \frac{I}{2} \\
l_L & l_R & \frac{p}{2}
\end{Bmatrix}~C_{\frac{p}{2}, m'_L~;~l_R, m''_R}^{l_L, m_L}~C_{\frac{p}{2}, m'_R~;~l_L, m''_L}^{l_R, m_R}.  \label{2threeintegralso4lleigendetail}
\end{align}
%%%%%%%%%%%%%%%%%%%%%%%%%%%%%%%%%%%%%%
 
%%%%%%%%%%%%%%%%%%%%%%%%%%%%%%%%%%%%%%%%%%%%%
\section{$SU(2)$ transformation between Dirac and Schwinger gauges}\label{append:gaugediracschwingerform}
%%%%%%%%%%%%%%%%%%%%%%%%%%%%%%%%%%%%%%%%%%%%

%%%%%%%%%%%%%%%%%%%%%%%%%%%%%%%%%%%
\subsection{Curvature for the spin connection}
%%%%%%%%%%%%%%%%%%%%%%%%%%%%%%%%%%

With respect to the spin connections in the   Schwiger gauge (\ref{schexpsu2diracsphefafie}) and the Dirac gauge (\ref{omegadirac}), 
the curvatures  
%%%%%%%%%%%%%%%%%%%%%%%%%%
\be
f=d\omega+i\omega^2, 
\label{dformulac}
\ee
%%%%%%%%%%%%%%%%%%%%%%%%%
are respectively derived as 
%%%%%%%%%%%%%%%%%%%%%
\begin{align}
f_{\text{S}} 
&=\frac{1}{2}\gamma_1 \sin^2\chi \sin\theta ~d\theta \wedge d\phi + \frac{1}{2}\gamma_2 \sin\chi\sin\theta ~d\phi \wedge d\chi + \frac{1}{2}\gamma_3 \sin\chi  ~d\chi \wedge d\theta, \nn\\
f_{\text{D}}
&=\frac{1}{2}\gamma_1 \sin\chi\sin\theta(\cos\theta~ d\theta\wedge d\phi -\sin\theta ~d\phi\wedge d\chi) \nn\\
&- \frac{1}{2}\gamma_2 \sin\chi \sin\phi (d\chi\wedge d\theta -\sin\theta\cos\theta \cot\phi ~d\phi\wedge d\chi -\sin\chi \sin^2\theta \cot\phi~ d\theta\wedge d\phi) \nn\\
&+ \frac{1}{2}\gamma_3 \sin\chi \cos\phi (d\chi\wedge d\theta +\sin\theta\cos\theta \tan\phi ~d\phi\wedge d\chi +\sin\chi \sin^2\theta \tan\phi ~d\theta\wedge d\phi). 
\end{align}
%%%%%%%%%%%%%%%%%%%%%%%%
Using the dreibeins, (\ref{dreibeinschwing}) and (\ref{diracgaugedreibein}), they are concisely represented as 
%%%%%%%%%%%%%%%%%%
\be 
f_{\text{S}} 
=\frac{1}{4}\epsilon_{abc}e^a_{\text{S}}\wedge e_{\text{S}}^b~\gamma_c, ~~~~
f_{\text{D}}%=\frac{1}{2}{f_{\text{S}}}^{ij}\sigma_{ij}
=\frac{1}{4}\epsilon_{abc}e^a_{\text{D}}\wedge e_{\text{D}}^b~\gamma_c.  
\ee
%%%%%%%%%%%%%%%%%%%%%
$f_{\text{S}}$ and $f_{\text{D}}$ are related by the $SU(2)$ transformation: 
%%%%%%%%%%%%%%%%%%%%%%
\be
f_{\text{D}} 
=\frac{1}{4}\epsilon_{abc}e_{\text{D}}^{a}\wedge e_{\text{D}}^b~\gamma_c =\frac{1}{4}\epsilon_{abc}O_{ad}O_{be}{e}^{d}_{\text{S}}\wedge {e}^e_{\text{S}}~\gamma_c=\frac{1}{2}\epsilon_{cde}O_{cf} e^d_{\text{S}}\wedge e^e_{\text{S}}~\gamma_c 
= g~f_{\text{S}}~g^{\dagger}, 
\ee
%%%%%%%%%%%%%%%%%%%%%
where  
in the third equation  we used  the $SO(3)$ invariance of the Levi-Civita tensor 
%%%%%%%%%%%%%%%%%%%%%
\be
\epsilon_{ade}O_{db}O_{ec}=\epsilon_{bcd}O_{ad}, 
\label{propso3mat}
\ee
%%%%%%%%%%%%%%%%%%%%%%%%%
and in the last equation (\ref{sigmatranssu2}) .

%%%%%%%%%%%%%%%%%%%%%%%%%%%%%%%%%%%%%%%%%%%%%
\subsection{$D$ function and gauge transformation}\label{appndsec:dfuncgaugetrans}
%%%%%%%%%%%%%%%%%%%%%%%%%%%%%%%%%%%%%%%%%%%%%

We introduce two parameterizations of the $SU(2)$ group elements: 
%%%%%%%%%%%%%%%%%%%%%%%%%%
\begin{subequations}
\begin{align}
&\Psi_{\text{D}}^{(l)}(\bs{\chi}) \equiv e^{-i\chi\bs{x}\cdot \bs{S}^{(l)}}, \\
&\Psi_{\text{S}}^{(l)} (\bs{\chi}) \equiv e^{-i{\chi} S_z^{(l)}}e^{i{\theta}  S_y^{(l)}}e^{i{\phi} S_z^{(l)}}. 
\end{align}\label{twogaufuncd}
\end{subequations}
%%%%%%%%%%%%%%%%%%%%%%%%%%%
In particular for $l=1/2$,  
%%%%%%%%%%%%%%%%%%%%%%%%%%%%%
\begin{subequations}
\begin{align}
&\Psi^{(1/2)}_{\text{D}}(\bs{\chi}) =\begin{pmatrix}
\cos\frac{\chi}{2} -i\sin\frac{\chi}{2} \cos\theta &  -i\sin\frac{\chi}{2}\sin\theta e^{-i\phi}  \\
 -i\sin\frac{\chi}{2}\sin\theta e^{i\phi}   &  \cos\frac{\chi}{2} +i\sin\frac{\chi}{2} \cos\theta 
\end{pmatrix}, \\
&\Psi^{(1/2)}_{\text{S}}(\bs{\chi}) =\begin{pmatrix}
\cos\frac{\theta}{2} e^{-i\frac{1}{2}(\chi-\phi)} &  \sin\frac{\theta}{2} e^{-i\frac{1}{2}(\chi+\phi)}   \\
 -\sin\frac{\theta}{2} e^{i\frac{1}{2}(\chi+\phi)}   &  \cos\frac{\theta}{2} e^{i\frac{1}{2}(\chi-\phi)} 
\end{pmatrix}. 
\end{align}
\end{subequations}
%%%%%%%%%%%%%%%%%%%%%%%%%%%%%
With  the $D$-function 
%%%%%%%%%%%%%%%%%%%%%%%%%
\be
D^{(l)}(\chi, \theta, \phi) \equiv e^{-i\chi S^{(l)}_z} e^{-i\theta S^{(l)}_y}e^{-i\phi S^{(l)}_z}, 
\ee
%%%%%%%%%%%%%%%%%%%%%%%%%%
(\ref{twogaufuncd}) is given by 
%%%%%%%%%%%%%%%%%%%%
\begin{subequations}
\begin{align}
&\Psi_{\text{D}}^{(l)}(\bs{\chi}) =D^{(l)}(\phi,\theta,0) D^{(l)}(\chi,-\theta,-\phi), \\
&\Psi_{\text{S}}^{(l)}(\bs{\chi}) =D^{(l)}(\chi,-\theta,-\phi). 
\end{align}
\end{subequations}
%%%%%%%%%%%%%%%%%%%
For the $SO(4)$ representations, $(l_L, l_R)=(l,0)$ and $(0, l)$, we have 
%%%%%%%%%%%%
\begin{subequations}
\begin{align}
&(l_L, l_R)=(l, 0)~:~\Psi_L^{(\text{D})}\equiv \Psi_{\text{D}}(\bs{\chi}), ~~~~% \nn\\ 
(l_L, l_R)=(0, l)~:~\Psi_R^{(\text{D})}\equiv \Psi_{\text{D}}(-\bs{\chi}) , 
%\label{diracgaugeirreps} 
\\
%\ee
%%%%%%%%%%%
%and  in the Schwinger gauge: 
%%%%%%%%%%%%
%\be
&~~~~~~~~~~~~~~~~~~~~~~~~\Psi_L^{(\text{S})}\equiv \Psi_{\text{S}}^{(l)}(\bs{\chi}), ~~~~% \nn\\ 
~~~~~~~~~~~~~~~~~~~~~~~\Psi_R^{(\text{S})}\equiv \Psi_{\text{S}}^{(l)}(-\bs{\chi}).  
\end{align}
\end{subequations}
%%%%%%%%%%%
They satisfy 
%%%%%%%%%%%%%%%%
\be
{\Psi_L^{(\text{D})}}^{\dagger}=\Psi_R^{(\text{D})}, ~~~~{\Psi_L^{(\text{S})}}^{\dagger}=g^{(l)} \Psi_R^{(\text{S})} g^{(l)} , 
\label{landlschdirdaggerre}
\ee
%%%%%%%%%%%%%%%%
with  
%%%%%%%%%%%%%%%%
\be
g^{(l)} (\theta,\phi)\equiv D^{(l)}(\phi,\theta,0)=e^{-i\phi S^{(l)}_z} e^{-i\theta S^{(l)}_y}. 
\label{gaugefunctra}
\ee
%%%%%%%%%%%%%%%%%%%
For $l=1/2$, 
%%%%%%%%%%%%%%%%%%
\be
g^{(1/2)}(\theta, \phi) =\begin{pmatrix}
\cos\frac{\theta}{2} e^{-i\frac{\phi}{2}} &  -\sin\frac{\theta}{2} e^{-i\frac{\phi}{2}}  \\
 \sin\frac{\theta}{2} e^{i\frac{\phi}{2}}  & \cos\frac{\theta}{2} e^{i\frac{\phi}{2}}
 \end{pmatrix}. 
\ee
%%%%%%%%%%%%%%%%%% 
(\ref{gaugefunctra}) also acts as the gauge function between the Dirac and Schwinger gauges    
%%%%%%%%%%%%
\be
{\Psi_L^{(\text{D})}} =g^{(l)}  {\Psi_L^{(\text{S})}}, ~~~~{\Psi_R^{(\text{D})}} =g^{(l)}  {\Psi_R^{(\text{S})}}.  
\label{landlschdiracrel}
\ee
%%%%%%%%%%%
%One may find that (\ref{landlschdirdaggerre}) and (\ref{landlschdiracrel}) are consistent. 
Using (\ref{landlschdirdaggerre}), we can rewrite (\ref{landlschdiracrel}) as 
%%%%%%%%%%%%%%%%
\be
\Psi_L^{(\text{D})} ={\Psi_R^{(\text{S})}}^{\dagger}g^{\dagger},~~~\Psi_R^{(\text{D})} ={\Psi_L^{(\text{S})}}^{\dagger}g^{\dagger}. 
\label{landlschdiracreld}
\ee
%%%%%%%%%%%%%%%
(\ref{landlschdiracreld}) implies that the $SU(2)$ gauge fields in the two gauges  
%%%%%%%%%%%%%%%%
\begin{subequations}
\begin{align}
&A^{(l)} _{\text{D}}=-i\frac{1}{2}({\Psi_L^{(\text{D})}}^{\dagger}d\Psi_L^{(\text{D})}+{\Psi_R^{(\text{D})}}^{\dagger} d\Psi_R^{(\text{D})}  ),\label{afromdiragene} \\
&A^{(l)} _{\text{S}}=-i\frac{1}{2}({\Psi_L^{(\text{S})}}d{\Psi_L^{(\text{S})}}^{\dagger}+{\Psi_R^{(\text{S})}} d{\Psi_R^{(\text{S})}}^{\dagger}  ), \label{afromschgene}
\end{align}
\end{subequations}
%%%%%%%%%%%%%%%%%%
are related as 
%%%%%%%%%%%%%%%%%%%%
\be
A^{(l)} _{\text{D}}=g^{(l)} A^{(l)} _{\text{S}} {g^{(l)} }^{\dagger}-ig^{(l)} d{g^{(l)} }^{\dagger}. 
\label{gaugetradiracschw}
\ee
%%%%%%%%%%%%%%%%%%%
From  (\ref{afromdiragene}), we have     
%%%%%%%%%%%%%%
\be
A_{\text{D}}^{(I/2)}=\sum_{i=1}^3 A_{\text{D}}^i S_i^{(I/2)}, 
\ee
%%%%%%%%%%%%%%
with 
%%%%%%%%%%%%%%%%%%%%%%%
\begin{align}
&A_{\text{D}}^1=-(1-\cos\chi)(\sin\phi d\theta +\sin\theta\cos\theta \cos\phi d\phi), \nn\\
&A_{\text{D}}^2=(1-\cos\chi)(\cos\phi d\theta -\sin\theta\cos\theta \sin\phi d\phi), \nn\\
&A_{\text{D}}^3=(1-\cos\chi)\sin^2\theta d\phi. 
\label{diracggaugesu2}
\end{align}
%%%%%%%%%%%%%%%%%%%%%%%%
In the Cartesian coordinate, they are given by 
%%%%%%%%%%%%%%%%%%%%%%%%
\be
A_{\text{D}}^i=-\frac{1}{1+x_4}\epsilon_{ijk}x_jS_k^{(I/2)}, ~~~~~~{A^{\text{D}}}_4=0.  
\label{diracgaugesu2gau}
\ee
%%%%%%%%%%%%%%%%%%%%%%%%%%%
Also from (\ref{afromschgene}), we obtain  
%%%%%%%%%%%%%%
\be
A_{\text{S}}^{(I/2)}=\sum_{i=1}^3 A_{\text{S}}^i S_i^{(I/2)}, 
\ee
%%%%%%%%%%%%%%
with 
%%%%%%%%%%%%%%%%%%%%%%%
\be
A_{\text{S}}^1= \cos\chi\sin\theta  d\phi, ~~~~A_{\text{S}}^2=-\cos\chi d\theta, ~~~~A_{\text{S}}^3=-\cos\theta d\phi. 
\label{schiggaugesu2}
\ee
%%%%%%%%%%%%%%%%%%%%%%%%
With the explicit expressions (\ref{diracggaugesu2}) and (\ref{schiggaugesu2}),  it is not difficult to verify (\ref{gaugetradiracschw}).

Using the gauge field (\ref{schiggaugesu2})\footnote{One can alternatively adopt  (\ref{diracggaugesu2}) in the Dirac gauge, but the concrete calculations are rather laborious in the polar coordinates.}   we can construct the covariant derivative as  
%%%%%%%%%%%%%%%%%%%%%%%%%%%%
\be
D_{\alpha}=\partial_{\alpha}+iA_{\alpha}^i S_{i}^{(I/2)}~
~~~(\alpha=\chi,\theta,\phi), 
\ee
%%%%%%%%%%%%%%%%%%%%%%%%%%%%%%%%%%
and the covariant angular momentum operators as 
%%%%%%%%%%%%%%%%%%%%%5
\be
\Lambda_{\mu\nu}=-ix_{\mu}{D}_{\nu}+ ix_{\nu}{D}_{\mu}
\ee
%%%%%%%%%%%%%%%%%%%%%%
are expressed as 
%%%%%%%%%%%%%%%%%%%%
\begin{align}
&{\Lambda}_{12}=-i{D}_{\phi}, \nn\\
&\Lambda_{13}=i(\cos\phi{D}_{\theta}-\cot\theta \sin\phi{D}_{\phi}),\nn\\
&\Lambda_{14}=i(\sin\theta\cos\phi{D}_{\chi}+\cot\chi\cos\theta\cos\phi{D}_{\theta}-\cot\chi\frac{1}{\sin\theta}\sin\phi {D}_{\phi}),\nn\\
&\Lambda_{23}=i(\sin\phi{D}_{\theta}+\cot\theta \cos\phi{D}_{\phi}),\nn\\
&\Lambda_{24}=i(\sin\theta\sin\phi{D}_{\chi}+\cot\chi\cos\theta\sin\phi{D}_{\theta}+\cot\chi\frac{1}{\sin\theta} \cos\phi{D}_{\phi}),\nn\\
&\Lambda_{34}=i(\cos\theta{D}_{\chi}-\cot\chi \sin\theta{D}_{\theta}). 
\label{covariantangfromderiva}
\end{align}
%%%%%%%%%%%%%%%%%%%%%
The sum of their squares gives  
%%%%%%%%%%%%%%%%%%%%%%%%%%
\begin{align}
\sum_{\mu<\nu}^4 {\Lambda_{\mu\nu}}^2&=-\frac{1}{\sin^2\chi} {D}_{\chi}(\sin^2\chi{D}_{\chi}) -\frac{1}{\sin^2\chi}\frac{1}{\sin\theta}\biggl({D}_{\theta}(\sin\theta{D}_{\theta}) +\frac{1}{\sin\theta}{D}^2_{\phi}\biggr), 
\label{covangularsqaurecasi}
\end{align}
%%%%%%%%%%%%%%%%%%%%%%%%%
which is equal to (\ref{freeangularsqaurecasi}) with the  replacement of  $\partial_{\alpha}$ with $D_{\alpha}$. One can confirm that $\bs{\Phi}^{(n,s,I/2)}_{m_L,m_R}$ (\ref{basiswithnormalsch}) are the eigenstates of $\sum_{\mu<\nu}^4 {\Lambda_{\mu\nu}}^2$ (\ref{covangularsqaurecasi}) by using their explicit coordinate representation.

%%%%%%%%%%%%%%%%%%%%%%%%%%%%%%%%%%%%%%%%%%%%%%%%%%%%%%%%%%%%%
\subsection{Gauge covariance of the Weyl-Landau operator}
%%%%%%%%%%%%%%%%%%%%%%%%%%%%%%%%%%%%%%%%%%%%%%%%%%%%%%%%%%%%%%%%

From the relations %between the Dirac and Schwinger gauges,   
%%%%%%%%%%%%%%%%%%%%%%%%%%% 
$e^a_{\text{D}} =O_{ab} e^b_{\text{S}}$ and $g~\gamma_a ~g^{\dagger} =\gamma_b O_{ba}$ (see Sec.\ref{sec:three-sphereint}), we have     
%%%%%%%%%%%%%%%%%%%%%%%%%
\be 
 g^{(1/2)} ~(e^a_{\text{S}}\gamma_a) ~{g^{(1/2)}}^{\dagger} = e^a_{\text{D}}\gamma_a. 
\ee
%%%%%%%%%%%%%%%%%%%%%%%%%%%
Meanwhile, the covariant derivatives, $\nabla_{\alpha}=\partial_{\alpha}+i\omega_{\alpha}$, in the Dirac and Schwinger gauges are related as  
%%%%%%%%%%%%%%%%%%%%%%%%%%%%%%%
\be 
g^{(1/2)} ~\nabla_{\alpha}^{\text{S}}~ {g^{(1/2)}}^{\dagger} = \nabla_{\alpha}^{\text{D}} , 
\ee
%%%%%%%%%%%%%%%%%%%%%%%%%%%%%%% 
with 
%%%%%%%%%%%%%%%%%%%%%%%%%%%%%%%%%%%%%
\be
g^{(1/2)} ~\omega_{\alpha}^{\text{S}} ~{g^{(1/2)}}^{\dagger} -ig^{(1/2)}\partial_{\alpha}{g^{(1/2)}}^{\dagger}=\omega_{\alpha}^{\text{D}}. 
\ee
%%%%%%%%%%%%%%%%%%%%%%%%%%%%%%%%%%%%
Consequently, the free Weyl operators 
%%%%%%%%%%%%%%%%%%%%%%%%%%%%%%%
\be
\fsl{\nabla}^{\text{S}} =\gamma^a e_{\text{S} a}^{~~\alpha}\nabla_{\alpha}^{\text{S}} , ~~~~~\fsl{\nabla}^{\text{D}} =\gamma^a e_{\text{D}a}^{~~\alpha}\nabla_{\alpha}^{\text{D}} ,
\ee
%%%%%%%%%%%%%%%%%%%%%%%%%%%%%%%
are related as 
%%%%%%%%%%%%%%%%%%%%%%%%%%%%%%%%%%%
\be 
g^{(1/2)}~\fsl{\nabla}^{\text{S}}~
{g^{(1/2)}}^{\dagger}=\fsl{\nabla}^{\text{D}}.   
\ee
%%%%%%%%%%%%%%%%%%%%%%%%%%%%%%%%
We can verify similar relations for the Weyl-Landau operator, $\fsl{\mathcal{D}} =\gamma^a e_{ a}^{~~\alpha}\mathcal{D}_{\alpha}$. 
The synthesized gauge function 
%%%%%%%%%%%%%%%%%%%%%%%%%%%%%
\be
\mathcal{G}(\theta,\phi)= g^{(1/2)}(\theta, \phi)\otimes g^{(I/2)}(\theta, \phi), 
\ee
%%%%%%%%%%%%%%%%%%%%%%%%%%%
 generates the gauge transformation for $\mathcal{A}=\omega\otimes 1 +1\otimes A$, 
%%%%%%%%%%%%%%%%%%%%%%%%
\be
\mathcal{G} ~\mathcal{A}_{\alpha}^{\text{S}} ~\mathcal{G}^{\dagger} -i\mathcal{G}\partial_{\alpha}\mathcal{G}^{\dagger}=\mathcal{A}_{\alpha}^{\text{D}}, 
\label{ggfir}
\ee
%%%%%%%%%%%%%%%%%%%%%%%%%%%
and 
%%%%%%%%%%%%%%%%%%%%%%%%%%%%%%%%%%%%%
\be
 \mathcal{G} ~(e^a_{\text{S}}\gamma_a \otimes 1)~\mathcal{G}^{\dagger} =(g^{(1/2)} ~e^a_{\text{S}}\gamma_a ~{g^{(1/2)}}^{\dagger} ) \otimes (g^{(I/2)}{g^{(I/2)}}^{\dagger})= e^a_{\text{D}}\gamma_a \otimes 1.   
 \label{ggsec}
\ee
%%%%%%%%%%%%%%%%%%%%%%%%%%%%%%%%%%%%%%%%%
From (\ref{ggfir}) and (\ref{ggsec}), we easily see that the Weyl-Landau operator  satisfies  
%%%%%%%%%%%%%%%%%%%%%%%%%%%%%%%%%%
\be 
\mathcal{G}~\fsl{\mathcal{D}}^{\text{S}}~
\mathcal{G}^{\dagger}=\fsl{\mathcal{D}}^{\text{D}}, 
\ee
%%%%%%%%%%%%%%%%%%%%%%%%%%%%%%%%%%
and then 
%%%%%%%%%%%%%%%%%%%%%
\be
\mathcal{G} \bs{\Psi}^{\text{S}} = \bs{\Psi}^{\text{D}}. 
\ee
%%%%%%%%%%%%%%%%%%%%%

%%%%%%%%%%%%%%%%%%%%%%%%%%%%%%%%%%%%%%%%%%%%%%%%%%%%%%%%%%%%%%%%%%%%%%%%%%%%%%%%%%%%
\section{Examples of the Weyl-Landau operator eigenstates}\label{append:examplesdirac}
%%%%%%%%%%%%%%%%%%%%%%%%%%%%%%%%%%%%%%%%%%%%%%%%%%%%%%%%%%%%%%%%%%%%%%%%%%%%%%%%%%%%

For several Weyl-Landau operators, we explicitly  derive a coordinate representation of the eigenstates based on the general procedure presented in Sec.\ref{subsec:so4weyllanmod}. 

%%%%%%%%%%%%%%%%%%%%%%%%%%%%%%%%%%%%%%%%%%%%%%%%%%%%%%%%%
\subsection{${I}/{2}={1}/{2}$}\label{subsec:weyllandau1/2}
%%%%%%%%%%%%%%%%%%%%%%%%%%%%%%%%%%%%%%%%%%%%%%%%%%%%%%

\begin{figure}[tbph]
\center
\includegraphics*[width=70mm]{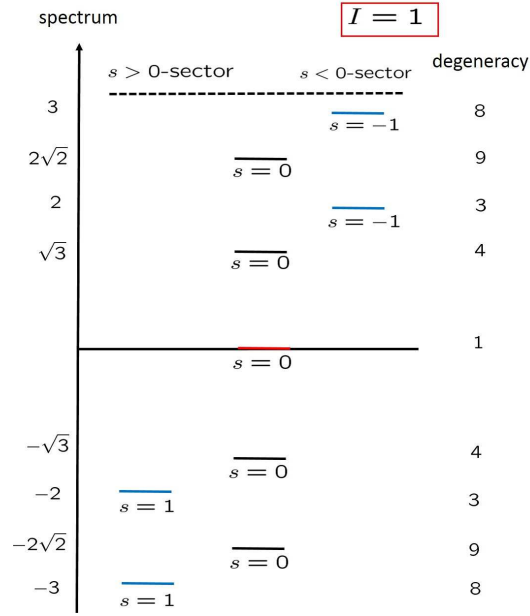}
\caption{The Weyl-Landau spectrum for $I/2=1/2$. }
\label{I1diraclandau.fig}
\end{figure}

The synthesized spin magnitudes are given by 
%%%%%%%%%%%%%%%%%%%%%%%%%%%
\be
J^+=\frac{I}{2}+\frac{1}{2}=1, ~~~J^-=\frac{I}{2}-\frac{1}{2}=0, 
\ee
%%%%%%%%%%%%%%%%%%%%%%%%%
and the Weyl-Landau operator spectrum becomes   
%%%%%%%%%%%%%%%%%
\be
\pm \lambda(n, s)=\pm \sqrt{n(n+2)+s^2}, 
\ee
%%%%%%%%%%%%%%%%%%%
with 
%%%%%%%%%%%%%%%%%%
\be
n=0:~~s=0, ~~~~~n\neq 0~:~s=1,0,-1. 
\ee
%%%%%%%%%%%%%%%%%%
For instance (Fig.\ref{I1diraclandau.fig}),
%%%%%%%%%%%%%%%%%
\be
\lambda(0,0)=0, ~~~~\lambda(1,0)=\sqrt{3}, ~~~~\lambda(1,  1)=2 , ~~~~\lambda(2, 0) = 2\sqrt{2}, ~~~\lambda(2, 1) = 3, \cdots. 
\ee
%%%%%%%%%%%%%%%%%%
%%%%%%%%%%%%%%%%%%%%%%%%%%%%%%%%%%%%%%%%%%%%%%%%%%%%%%%%%%%%
%%%%%%%%%%%%%%%%%%%%%%%%%%%%%%%%%%%%%%%%%%%%%%%%%%%%%%%%%%%%%% 
In the following, we use the notation: 
%%%%%%%%%%%%%%%%%%%%%%%%%%%%%%%%%
\be
{\Phi}^{(n,s,-)}_{M_L, M_R} \equiv  {\Phi}^{(n,s,J^-=0)}_{M_L, M_R}, ~~~\bs{\Phi}^{(n,s,+)}_{M_L, M_R} \equiv  \bs{\Phi}^{(n,J^+=1)}_{M_L, M_R}, 
\label{nonreexap1}
\ee
%%%%%%%%%%%%%%%%%%%%%%%%%%%%%
where ${\Phi}^{(-,n,s)}$ and $\bs{\Phi}^{(+,n,s)}$ respectively denote the one-component ($S=0$) and three component ($S=1$) irreducible representation of the $SU(2)$ gauge group. 
The unitary matrix for the irreducible decomposition, $1/2\otimes 1/2 = 1\oplus 0$, is given by 
%%%%%%%%%%%%%%%%%%%%%%%%%%%%%%%%%%
\be
U=\frac{1}{\sqrt{2}}
\left(
\begin{array}{ccc:c}
\sqrt{2} & 0 & 0 & 0 \\
0 & 1 & 0 & 1 \\
0 & 1 & 0 & -1 \\
0 & 0 & \sqrt{2} & 0 
\end{array}\right). 
\label{unitarymatex1}
\ee
%%%%%%%%%%%%%%%%%%%%%%%%%%%%%%%%%% 
It is straightforward to  construct the eigenstates of the spinor Landau model by acting (\ref{unitarymatex1}) to (\ref{nonreexap1}). 

 Derivation of the Weyl-Landau operator eigenstates is not difficult. We act the Weyl-Landau operator to linear combination of the eigenstates and determine the coefficients of the linear combination ($\alpha$ and $\beta$ in (\ref{secondeigenstate})) so that the linear combination to be the eigenstate of the Weyl-Landau operator. The results are as follows.

\begin{itemize}
%%%%%%%%%%%%%%%%%%%%%%%%%%%%%%%%%%%%%%%%%%%%%%%%%%%%%%%%%
\item{$\lambda=0~:~(n^-,s^-)=(0,0)$}
%%%%%%%%%%%%%%%%%%%%%%%%%%%%%%%%%%%%%%%%%%%%%%%%%%%%%%
\end{itemize}

%%%%%%%%%%%%%%%%%%%%%%%%%%%%%%%%%%%%%%%%%%
\be
\bs{\Psi}_{\lambda=0}=
U\begin{pmatrix}
0 \\
0 \\
0 \\
\Phi^{(0,0,-)}
\end{pmatrix}=
U\begin{pmatrix}
0 \\
0 \\
0 \\
1 
\end{pmatrix} =\frac{1}{\sqrt{2}} 
\begin{pmatrix}
0 \\
1 \\
-1 \\
0
\end{pmatrix}.
\ee
%%%%%%%%%%%%%%%%%%%%%%%%%%%%%%%%%%%%%%%%%%

\begin{itemize}
%%%%%%%%%%%%%%%%%%%%%%%%%%%%%%%%%%%%%%%%%%%%%%%%%%%%%%%%%
\item{$\pm\lambda=\pm\sqrt{3}~:~(n^-,s^-)=(1,0),~~(n^+,s^+)=(0, 0)$}
%%%%%%%%%%%%%%%%%%%%%%%%%%%%%%%%%%%%%%%%%%%%%%%%%%%%%%
\end{itemize}

%the  Weyl-Landau eigenstates are constructed as 
%%%%%%%%%%%%%%%%%%%%%%%%
\begin{align}
&\bs{\Psi}_{\lambda=\sqrt{3}, M_L, M_R} =\frac{1}{\sqrt{2}}( \bs{\Psi}_{M_L, M_R}^+ +\bs{\Psi}_{M_L, M_R}^-     ) =\frac{1}{\sqrt{2}}U \begin{pmatrix}
\bs{\Phi}^{(1,0,+)}_{M_L, M_R} \\
{\Phi}^{(0,0,-)}_{M_L, M_R} 
\end{pmatrix}, ~~~~~~~~(M_L, M_R=1/2, -1/2)\nn\\
&\bs{\Psi}_{-\lambda=-\sqrt{3}, M_L, M_R} =\frac{1}{\sqrt{2}}( \bs{\Psi}_{M_L, M_R}^+ -\bs{\Psi}_{M_L, M_R}^-     )=\frac{1}{\sqrt{2}}U \begin{pmatrix}
\bs{\Phi}^{(1,0,+)}_{M_L, M_R} \\
-{\Phi}^{(0,0,-)}_{M_L, M_R} 
\end{pmatrix}, ~~~(M_L, M_R=1/2, -1/2)\label{lambda=sqrt3}
\end{align}
%%%%%%%%%%%%%%%%%%%%%%%%
where 
%%%%%%%%%%%%%%%%%%%%%%%%%%
\be
\bs{\Psi}_{M_L, M_R}^- =U \begin{pmatrix}
0 \\
0 \\
0 \\
{\Phi}_{M_L, M_R}^{( 0, 0, -)} 
\end{pmatrix}, ~~~
\bs{\Psi}_{M_L, M_R}^+ =U \begin{pmatrix}
\bs{\Phi}_{M_L, M_R}^{(1,0,+)} \\
0
\end{pmatrix}. % ~~~~~(M_L, M_R=1/2, -1/2),   
\ee
%%%%%%%%%%%%%%%%%%%%%%%%%%%%

\begin{itemize}
%%%%%%%%%%%%%%%%%%%%%%%%%%%%%%%%%%%%%%%%%%%%%%%%%%%%%%%%%
\item{$\pm \lambda=\pm 2~:~(n^+,s^+)=(0,\pm 1)$}
%%%%%%%%%%%%%%%%%%%%%%%%%%%%%%%%%%%%%%%%%%%%%%%%%%%%%%
\end{itemize}

%%%%%%%%%%%%%%%%%%%%%%%%
\begin{align}
&\bs{\Psi}_{\lambda=2, M_L=0, M_R} =U \begin{pmatrix}
\bs{\Phi}_{M_L=0, M_R}^{(0, -1, +)} \\
0
\end{pmatrix},  ~~~~(M_R=+1, 0, -1)\nn\\
&\bs{\Psi}_{-\lambda=-2, M_L, M_R=0} =U  \begin{pmatrix}
\bs{\Phi}_{M_L, M_R= 0}^{(0, 1, +)} \\
0 
\end{pmatrix}.~~(M_L=+1, 0, -1)\label{lambda=2}
\end{align}
%%%%%%%%%%%%%%%%%%%%%%%%

\begin{itemize}
%%%%%%%%%%%%%%%%%%%%%%%%%%%%%%%%%%%%%%%%%%%%%%%%%%%%%%%%%
\item{$\pm \lambda=\pm 2\sqrt{2}~:~(n^-,s^-)=(2,0),~~(n^+,s^+)=(1,0)$}
%%%%%%%%%%%%%%%%%%%%%%%%%%%%%%%%%%%%%%%%%%%%%%%%%%%%%%
\end{itemize}

%In a similar manner to  (\ref{lambda=sqrt3}), 
%%%%%%%%%%%%%%%%%%%%%%%%
\begin{align}
&\bs{\Psi}_{\lambda=2\sqrt{2}, M_L, M_R} =\frac{1}{\sqrt{2}}U \begin{pmatrix}
\bs{\Phi}^{(2,0,+)}_{M_L, M_R} \\
{\Phi}^{(1,0,-)}_{M_L, M_R} 
\end{pmatrix}, ~~~~~(M_L, M_R=1,0,-1)\nn\\
&\bs{\Psi}_{-\lambda=-2\sqrt{2}, M_L, M_R}=\frac{1}{\sqrt{2}}U \begin{pmatrix}
\bs{\Phi}^{(2,0,+)}_{M_L, M_R} \\
-{\Phi}^{(1,0,-)}_{M_L, M_R} 
\end{pmatrix}. ~~~(M_L, M_R=1,0,-1)\label{lambda=2sqrt2}
\end{align}
%%%%%%%%%%%%%%%%%%%%%%%%

\begin{itemize}
%%%%%%%%%%%%%%%%%%%%%%%%%%%%%%%%%%%%%%%%%%%%%%%%%%%%%%%%%
\item{$\pm \lambda=\pm 3~:~(n^-,s^-)=(2,\pm 1),~(n^+, s^+)=(1, \pm 1)$}
%%%%%%%%%%%%%%%%%%%%%%%%%%%%%%%%%%%%%%%%%%%%%%%%%%%%%%
\end{itemize}

%%%%%%%%%%%%%%%%%%%%%%%%
\begin{align}
&\bs{\Psi}_{\lambda=3, M_L, M_R} =U \begin{pmatrix}
\bs{\Phi}_{M_L, M_R}^{(1, -1, +)} \\
0
\end{pmatrix},  ~~~~(M_L=1/2, -1/2,~~M_R=3/2, 1/2, -1/2, -3/2)\nn\\
&\bs{\Psi}_{-\lambda=-3, M_L, M_R} =U  \begin{pmatrix}
\bs{\Phi}_{M_L, M_R}^{(1, 1, +)} \\
0 
\end{pmatrix}.~~~(M_L=3/2, 1/2, -1/2, -3/2, ~~M_R=1/2, -1/2)\label{lambda=3}
\end{align}
%%%%%%%%%%%%%%%%%%%%%%%%

%%%%%%%%%%%%%%%%%%%%%%%%%%%%%%%%%%%%%%%%%%%%%%%%%%%%%%%%%
\subsection{${I}/{2}=1$}\label{subsec:weyllandau3/2}
%%%%%%%%%%%%%%%%%%%%%%%%%%%%%%%%%%%%%%%%%%%%%%%%%%%%%%

\begin{figure}[tbph]
\center
\includegraphics*[width=70mm]{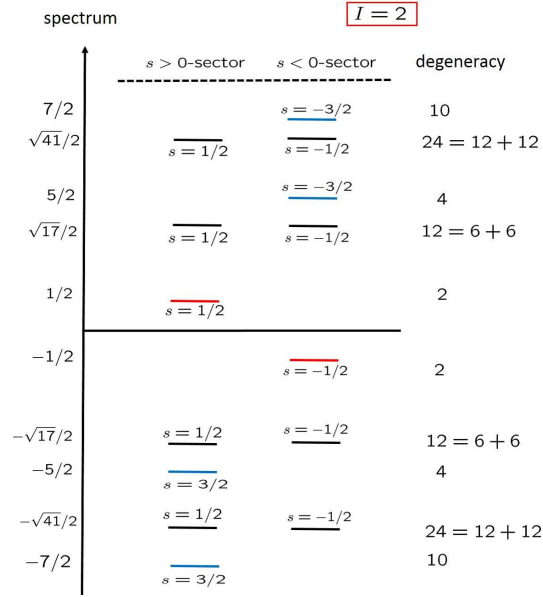}
\caption{The Weyl-Landau spectrum for $I/2=1$. }
\label{I2diraclandau.fig}
\end{figure}

The synthesized spins are 
%%%%%%%%%%%%%%%%%%%%%%%%%%%
\be
J^+=\frac{I}{2}+\frac{1}{2}=\frac{3}{2}, ~~~J^-=\frac{I}{2}-\frac{1}{2}=\frac{1}{2},   
\ee
%%%%%%%%%%%%%%%%%%%%%%%%%
and the  Weyl-Landau operator spectrum is  
%%%%%%%%%%%%%%%%%
\be
\pm \lambda(n, s)=\pm \sqrt{n(n+3)+s^2}, 
\ee
%%%%%%%%%%%%%%%%%%%
with
%%%%%%%%%%%%%%%%%%
\be
n=0:~~s=\frac{1}{2}, -\frac{1}{2},  ~~~n\neq 0~:~s=\frac{3}{2},\frac{1}{2},-\frac{1}{2}, -\frac{3}{2}. 
\ee
%%%%%%%%%%%%%%%%%%
For instance (Fig.\ref{I2diraclandau.fig}), 
%%%%%%%%%%%%%%%%%
\be
\lambda(0,1/2)=1/2, ~~~~\lambda(1,1/2)=\sqrt{17}/2, ~~~~\lambda(1,3/2)=5/2 , ~~~~\lambda(2, 1/2) = \sqrt{41}/2, ~~~\lambda(2, 3/2) = 7/2, ~~~\cdots. 
\ee
%%%%%%%%%%%%%%%%%%
%%%%%%%%%%%%%%%%%%%%%%%%%%%%%%%%%%%%%%%%%%%%%%%%%%%%%%%%%%%%
%%%%%%%%%%%%%%%%%%%%%%%%%%%%%%%%%%%%%%%%%%%%%%%%%%%%%%%%%%%%%% 

We adopt the notation: 
%%%%%%%%%%%%%%%%%%%%%%%%%%%%%%%%%
\be
\bs{\Phi}^{(n,s,-)}_{M_L, M_R} \equiv  \bs{\Phi}^{(n,s,J^-=1/2)}_{M_L, M_R}, ~~~\bs{\Phi}^{(n,s,+)} _{M_L, M_R}\equiv  \bs{\Phi}^{(n,s,J^+=3/2)}_{M_L, M_R}, 
\ee
%%%%%%%%%%%%%%%%%%%%%%%%%%%%%
where $\bs{\Phi}^{(-,n,s)}$ and $\bs{\Phi}^{(+,n,s)}$ respectively denote the two-component ($S=1/2$)  and four-component ($S=3/2$) irreducible representations of the $SU(2)$ gauge group. 
The unitary matrix for the decomposition $1/2\otimes 1=3/2\oplus 1/2$ is given by 
%%%%%%%%%%%%%%%%%%%%%%%%%%%%%%%%%%
\be
U=\frac{1}{\sqrt{3}}\left(
\begin{array}{cccc:cc}
\sqrt{3} & 0 & 0 & 0 & 0 & 0  \\
0 & 1 & 0 & 0 & -\sqrt{2} & 0  \\
0 & \sqrt{2} & 0 & 0 &  1 & 0  \\
0 & 0 & \sqrt{2} & 0 & 0 & -1 \\
0 & 0 & 1 & 0 & 0 & \sqrt{2} \\
0 & 0 & 0  & \sqrt{3} & 0 & 0  
\end{array}
\right).
\ee
%%%%%%%%%%%%%%%%%%%%%%%%%%%%%%%%%%  
%Based on a similar procedure to Appendix \ref{subsec:weyllandau1/2},  the eigenstates of the present  Weyl-Landau operator are derived as follows.   

\begin{itemize}
%%%%%%%%%%%%%%%%%%%%%%%%%%%%%%%%%%%%%%%%%%%%%%%%%%%%%%%%%
\item{$\pm \lambda=\pm 1/2 ~:~(n^-,s^-)=(0,\pm 1/2)$}
%%%%%%%%%%%%%%%%%%%%%%%%%%%%%%%%%%%%%%%%%%%%%%%%%%%%%%
\end{itemize}

%%%%%%%%%%%%%%%%%%%%%%%%%%%%%%%%%%%%%%%%%%
\begin{align}
&\bs{\Psi}_{+\lambda=+{1}/{2}, M_L, M_R=0}=
U\begin{pmatrix}
\bs{0} \\
\bs{\Phi}^{(0,1/2,-)}_{M_L, M_R=0}
\end{pmatrix}, ~~~~(M_L=1/2, -1/2) \nn\\
&\bs{\Psi}_{-\lambda=-{1}/{2}, M_L=0, M_R}=
U\begin{pmatrix}
\bs{0} \\
\bs{\Phi}^{(0,-1/2,-)}_{M_L=0, M_R}
\end{pmatrix}.~~~~(M_R=1/2, -1/2)
\end{align}
%%%%%%%%%%%%%%%%%%%%%%%%%%%%%%%%%%%%%%%%%%

\begin{itemize}
%%%%%%%%%%%%%%%%%%%%%%%%%%%%%%%%%%%%%%%%%%%%%%%%%%%%%%%%%
\item{$\pm\lambda=\pm\sqrt{17}/2~:~(n^-,s^-)=(1,\pm 1/2),~~(n^+,s^+)=(0, \pm 1/2)$}
%%%%%%%%%%%%%%%%%%%%%%%%%%%%%%%%%%%%%%%%%%%%%%%%%%%%%%
\end{itemize} 

For $s=+1/2$ sector, 
%%%%%%%%%%%%%%%%%%%%%%%%
\begin{align}
&\bs{\Psi}_{+\lambda=+\sqrt{17}/2, M_L, M_R}  =\frac{1}{\sqrt{6(51+5\sqrt{17})}}U \begin{pmatrix}
8\sqrt{2}\bs{\Phi}^{(0,1/2,+)}_{M_L, M_R} \\
-(5+3\sqrt{17})\bs{\Phi}^{(1,1/2,-)}_{M_L, M_R} 
\end{pmatrix}, ~~~~(M_L=  1, 0, -1, ~~M_R=1/2, -1/2      )\nn\\
&\bs{\Psi}_{-\lambda=-\sqrt{17}/2, M_L, M_R} =\frac{1}{\sqrt{6(51+5\sqrt{17})}}U \begin{pmatrix}
(5+3\sqrt{17})\sqrt{2}\bs{\Phi}^{(0,1/2,+)}_{M_L, M_R} \\
8\sqrt{2}\bs{\Phi}^{(1,1/2,-)}_{M_L, M_R} 
\end{pmatrix}.~~~~(M_L=  1, 0, -1, ~~M_R=1/2, -1/2      )
\end{align}
%%%%%%%%%%%%%%%%%%%%%%%%
For $s=-1/2$ sector,  
%%%%%%%%%%%%%%%%%%%%%%%%%%%%%%%%%%%
\begin{align}
&\bs{\Psi}_{+\lambda=+\sqrt{17}/2, M_L, M_R}  = \frac{1}{\sqrt{6(51-5\sqrt{17})}} U \begin{pmatrix}
8\sqrt{2}\bs{\Phi}^{(0,-1/2,+)}_{M_L, M_R} \\
(5-3\sqrt{17})\bs{\Phi}^{(1,-1/2,-)}_{M_L, M_R} 
\end{pmatrix}, ~~~~(M_L=  1/2,  -1/2, ~~M_R=1, 0, -1      )\nn\\
&\bs{\Psi}_{-\lambda=-\sqrt{17}/2, M_L, M_R} =\frac{1}{\sqrt{6(51-5\sqrt{17})}} U \begin{pmatrix}
(5-3\sqrt{17})\sqrt{2}\bs{\Phi}^{(0,-1/2,+)}_{M_L, M_R} \\
-8\sqrt{2}\bs{\Phi}^{(1,-1/2,-)}_{M_L, M_R} 
\end{pmatrix}.~~~~(M_L=  1/2,  -1/2, ~~M_R=1, 0,  -1      )
\end{align}
%%%%%%%%%%%%%%%%%%%%%%%%

\begin{itemize}
%%%%%%%%%%%%%%%%%%%%%%%%%%%%%%%%%%%%%%%%%%%%%%%%%%%%%%%%%
\item{$\pm \lambda=\pm 5/2~:~(n^+,s^+)=(0,\pm 3/2)$}
%%%%%%%%%%%%%%%%%%%%%%%%%%%%%%%%%%%%%%%%%%%%%%%%%%%%%%
\end{itemize}

%%%%%%%%%%%%%%%%%%%%%%%%
\begin{align}
&\bs{\Psi}_{\lambda=5/2, M_L=0, M_R} =U \begin{pmatrix}
\bs{\Phi}_{M_L=0, M_R}^{(0, -3/2, +)} \\
\bs{0}
\end{pmatrix},  ~~~(M_R=3/2, 1/2, -1/2, -3/2)\nn\\
&\bs{\Psi}_{-\lambda=-5/2, M_L, M_R=0} =U  \begin{pmatrix}
\bs{\Phi}_{M_L, M_R= 0}^{(0, 3/2, +)} \\
\bs{0} 
\end{pmatrix}.~~~(M_L=3/2, 1/2, -1/2, -3/2)
\end{align}
%%%%%%%%%%%%%%%%%%%%%%%%

%%%%%%%%%%%%%%%%%%%%%%%%%%%%%%%%

%%%%%%%%%%%%%%%%%%%%%%%%%%%%%%%%%%%%%%%

%%%%%%%%%%%%%%%%%%%%%%%%%%%%%%%%%%%%%%%%%%%%%%%%%%%%%%%%%%%%

\end{document}